\documentclass[useAMS,usenatbib]{mn2e}
\usepackage{rotating}
\usepackage{amsmath}
\usepackage{txfonts}
\usepackage{graphics}
\usepackage{graphicx}
\usepackage{times}
\usepackage{subfigure}
\usepackage{pdflscape}
\usepackage{array}
\usepackage[T1]{fontenc}
\usepackage{aecompl}
\begin{document}
\voffset-1.25cm
\title[Small-scale Redshift Space Distortions]{A 2.5\% measurement of the growth rate from small-scale redshift space clustering of SDSS-III CMASS galaxies}
\author[Reid et al.]{
\parbox{\textwidth}{
Beth A. Reid$^{1,2,3}$\thanks{E-mail: beth.ann.reid@gmail.com}, Hee-Jong Seo$^{3,4}$, Alexie Leauthaud$^{5}$, Jeremy L. Tinker$^{6}$, Martin White$^{1,7}$
}
\vspace*{4pt} \\
$^{1}$ Lawrence Berkeley National Laboratory, 1 Cyclotron Road, Berkeley, CA 94720, USA \\
$^{2}$ Hubble Fellow \\
$^{3}$ Berkeley Center for Cosmological Physics, LBL and Department of Physics, University of California, Berkeley, CA, 94720, USA \\
$^{4}$ Center for Cosmology and Astro-Particle Physics, Ohio State University, Columbus, OH 43210, USA \\
$^{5}$ Kavli Institute for the Physics and Mathematics of the Universe, Todai Institutes for Advanced Study, the University of Tokyo, Kashiwa, Japan 277-8583 \\
$^{6}$ Center for Cosmology and Particle Physics, Department of Physics New York University, USA \\
$^{7}$ Departments of Physics and Astronomy, University of California, Berkeley, CA, 94720, USA
}
\date{\today}
\pagerange{\pageref{firstpage}--\pageref{firstpage}}
\maketitle
\label{firstpage}
\begin{abstract}
We perform the first fit to the anisotropic clustering of SDSS-III CMASS DR10
galaxies on scales of $\sim 0.8 - 32$ $h^{-1}$ Mpc.  A standard halo occupation
distribution model evaluated near the best fit Planck $\Lambda$CDM cosmology
provides a good fit to the observed anisotropic clustering, and implies a
normalization for the peculiar velocity field of $M \sim 2 \times 10^{13}$
$h^{-1}$ $M_{\sun}$ halos of $f\sigma_8(z=0.57) = 0.450 \pm 0.011$.  Since this
constraint includes both quasi-linear and non-linear scales, it should severely
constrain modified gravity models that enhance pairwise infall velocities on
these scales.  Though model dependent, our measurement represents a factor of
2.5 improvement in precision over the analysis of DR11 on large scales,
$f\sigma_8(z=0.57) = 0.447 \pm 0.028$, and is the tightest single constraint on
the growth rate of cosmic structure to date.  Our measurement is consistent with
the Planck $\Lambda$CDM prediction of $0.480 \pm 0.010$ at the $\sim 1.9\sigma$
level.  Assuming a halo mass function evaluated at the best fit Planck
cosmology, we also find that $10\%$ of CMASS galaxies are satellites in halos of
mass $M \sim 6 \times 10^{13}$ $h^{-1}$ $M_{\sun}$.  While none of our tests and
model generalizations indicate systematic errors due to an insufficiently
detailed model of the galaxy-halo connection, the precision of these first
results warrant further investigation into the modeling uncertainties and
degeneracies with cosmological parameters. 
\end{abstract}


\section{Introduction} \label{sec:intro} The clustering of galaxies on small
scales provides important constraints on the relationship between galaxies and
the underlying dark matter distribution.  This relation is of interest in itself
as a constraint on galaxy formation and evolution, as well as for quantifying
the impact of galaxy-formation scale physics on larger scale clustering measures
used for cosmological parameter constraints.  Modern approaches to modeling the
relationship between galaxies and the underlying dark matter distribution rely
on the basic tenet that galaxy formation requires a gravitationally-bound dark
matter halo or sub-halo to accumulate and condense gas
\citep{Peacock00,Seljak00,Benson00,White01,Berlind02,CooraySheth02,Yang03}.  In
their simplest form, such ``halo models'' contain one dominant variable that
determines the probability that a (sub-)halo hosts a galaxy of interest.  In the
halo occupation distribution (HOD) formalism adopted in this paper, halo mass is
the dominant variable and halos are permitted to host more than one galaxy.  In
the sub-halo abundance matching (``SHAM'') formalism, the maximum circular
velocity at accretion is often used \citep{Marinoni02,Vale06,Conroy06}.  The
primary advantage of SHAM is that each sub-halo hosts only a single galaxy, thus
requiring fewer free parameters to specify the model but assuming a specific but
physically motivated relation between central and satellite galaxies.  The
practical disadvantage is that $N$-body simulations require higher resolution to
resolve sub-halos.  In principle both of these approaches could be generalized
to include additional secondary variables such as halo formation time, with
observable consequences \citep{Gao05,Wang13,Zentner13,Cohn13}.

There are a host of observables available to constrain halo models as a function
of galaxy properties: one-point statistics like number density or luminosity
functions, two- or three-point galaxy clustering \citep{Zehavi11,Marin11},
marked statistics \citep{Sheth05,White09} and direct measurements the galaxy
group multiplicity function \citep{Yang09,Reid09}.  
The most widely used observable is the projected correlation function, $w_p$, 
which removes sensitivity to redshift space distortions by integrating over the 
line-of-sight separation.
While redshift space distortion effects are
more difficult to model, they do provide complementary constraints both on the
velocity distribution of galaxies relative to their host dark matter halos
\citep{vdB05} and on cosmological parameters \citep{Yang04}.  

The primary goal of the present paper is to use the information in the
anisotropy of the galaxy correlation function on scales $\sim 0.8 - 32$ $h^{-1}$ Mpc
to simultaneously constrain the HOD and growth rate of cosmic structure through
the pairwise infall of galaxies caused by their mutual gravitational attraction
\citep{Kaiser87}.  Constraints on gravitational infall on these scales is of
particular interest in searching for signatures of modified gravity: for
instance, an $f(R)$ model with $|f_{R0}| = 10^{-4}$ predicts a $\sim 25\%$
increase in the amplitude of pairwise infall velocities on scales of 10-30 Mpc
\citep{Keisler13,Zu13,Lam12}.  Alternatively, the non-linear regime is also a
promising avenue for constraining dark sector coupling \citep{Piloyan14}.  We
can also use the constraints on the HOD to infer the nuisance parameter
$\sigma^2_{\rm FOG}$ employed in our analysis on larger scales
\citep{Reid12,Samushia13} to account for the velocity dispersions of galaxies
relative to their host dark matter halos.  See \citet{Hikage14} for a similar
concept applied to the power spectrum multipoles. 

In this paper we focus on the CMASS sample from the SDSS-III Baryon Oscillation
Spectroscopic Survey (BOSS).  This sample has thus far been the focus of several
cosmological analyses, most recently providing a one per-cent absolute distance
measurement using the baryon acoustic oscillation (BAO) standard ruler
\citep{Anderson13} and a 6\% constraint on the growth rate of cosmic structure
\citep{Samushia13,Beutler13,Sanchez13,Chuang13}.  The projected correlation
function of these galaxies has also been used to constrain halo models
using both the HOD \citep{White11,Guo14} and SHAM \citep{Nuza13} formalisms;
this work represents the first quantitative comparison to the small-scale
anisotropic clustering of the CMASS galaxies.

The layout of the paper is as follows.  In Sec.~\ref{sec:pre} we describe the
basic conceptual elements of our analysis.  Sec.~\ref{sec:data} details our
dataset, while Sec.~\ref{sec:FB} focuses on mitigating the impact of fiber
collisions in our spectroscopic galaxy sample.  Sec.~\ref{sec:meas} presents
fiber-collision corrected measurements and uncertainties.
Sec.~\ref{sec:HODmodel} presents the details of our $N$-body simulation based
HOD model that we use to fit the observed anistropic CMASS galaxy clustering.
The principal results of a simultaneous fit to the HOD parameters and
$f\sigma_8$ are presented in Sec.~\ref{sec:results}.  In Sec.~\ref{sec:modgrav}
we discuss the implications of our results for constraining modified gravity
models, and in Sec.~\ref{sec:conc} we discuss future prospects for this
technique. 

\section{Preliminaries}
\label{sec:pre}
\subsection{Analysis Road Map}
Many components of our analysis are interdependent, so a strictly linear
presentation is impossible.  Therefore we provide an overview of the full
analysis here.  The first new product of this work is an unbiased estimate of
the small-scale aniostropic clustering of the CMASS sample from the BOSS survey.
Second, we implement a new algorithm to quickly and accurately predict two-point
clustering statistics as a function of HOD parameters.  The algorithm is based
directly on measuring the clustering of halo catalogs derived from $N$-body
simulations; it uses no analytic approximations or fits for the one-halo or
two-halo terms.  We combine these two products to constrain both HOD and growth
rate parameters.

For the measurements, the primary source of systematic uncertainty, referred to
as ``fiber collisions'', is the instrumental constraint that spectroscopic
fibers cannot be placed closer than 62'' during a single observation.
Therefore, the galaxies that do not receive a spectroscopic fiber are a
non-random subset of the targets; they preferentially reside in regions of
higher target density.  Moreover, the positioning of spectroscopic tiles depends
on the angular density of targets, so that regions in which plates overlap (and
therefore fiber collisions can be resolved) are not representative of the full
survey.  Ignoring these effects would substantially bias our clustering
measurements.  In Sec.~\ref{sec:FB} we consider two fiber collision correction
methods previously introduced in the literature: nearest neighbor redshift
assignment and angular upweighting.  Neither correction is exact, so we apply
the BOSS tiling pipeline to a mock galaxy catalog to determine which fiber
collision correction method is best on which scales, which potential observable
is the least affected, and what the residual biases are.  Below we define the
set of clustering measures we will consider for our final analysis.

We make use of halo catalogs derived from different $N$-body simulations for
three distinct purposes: \begin{itemize} \item to evaluate theoretical models
for parameter estimation \item to estimate the uncertainty (theory covariance
matrix) due to the finite volume of simulations used to compute the theoretical
models \item to generate the mock galaxy catalog to which we applied the BOSS
tiling algorithm in order to study the effects of fiber collisions.
\end{itemize} Below we specify the set of $N$-body simulations we use for these
purposes as well as explain how the halo catalogs are derived from the
simulation outputs.

The primary goal of our analysis is to constrain the growth rate of cosmic
structure using redshift space distortions, and so we review the basic physics
first.
\subsection{Redshift Space Distortions (RSD)} \label{sec:RSDpre} Because 
cosmological flows are non-relativistic, the
spectrosopically observed redshift of a galaxy can be expressed as the sum of
two components: \begin{equation} z_{\rm spec} = z_{\rm cosmo} + \frac{v_{\rm
p}^{\rm LOS}}{ac} \label{zspec} \end{equation} where $z_{\rm cosmo}$ is the
redshift expected if the universe were homogeneous, while the second term
accounts for the component of the physical ``peculiar velocity'' along the
line-of-sight (LOS), i.e., the proper motion of an object due to its local
gravitational potential.  Here $a = 1/(1+z_{\rm cosmo})$ is the scale factor of
the universe and c is the speed of light.  ``Redshift space distortions'' (RSD)
is the generic term referring to distortions in the observed galaxy density
field due to the $v_{\rm p}^{\rm LOS}$ contribution to the observed redshift
coordinate.  Throughout this work we will quote velocities in units of distance,
with the relation between peculiar velocity $v_{\rm p}$ and apparent
line-of-sight comoving distance shift $\Delta s$ for a galaxy observed at $a$ 
given by \begin{equation} \Delta s = \frac{v_{\rm p}^{\rm LOS}}{aH(a)},
\end{equation} where $H(a) = \dot{a}/a$ is the expansion rate at $a$.  

On large scales where linear perturbation theory applies, the peculiar velocity
field ${\bf v}_{\rm p}$ is simply related to the underlying matter density
fluctuations, $\delta_m$: \begin{equation} \nabla \cdot {\bf v}_{\rm p} =
-aHf\delta_{\rm m} \label{eq:pecv} \end{equation} where $f = {\rm d} \, ln \,
D/{\rm d} \, ln \, a$ is the logarithmic growth rate, and $D(a)$ is the linear
growth function that specifies the amplitude of fluctuations as a function of
$a$, relative to some initial fluctuation amplitude:  $\delta_m(a) \propto D(a)
\delta_{\rm m,i}$.  Therefore, in the linear regime, a measure of the amplitude
of the peculiar velocity field through RSD provides a constraint on $f$ times
the amplitude of matter fluctuations on some scale; often this scale is taken to
be 8 $h^{-1}$ Mpc, so that linear redshift space distortions measure
$f\sigma_8$.  Because the scale-dependence of the matter power spectrum is
extremely well-constrained by the CMB, the specified scale is not important for
many applications \citep[see Section 5.1 of][]{Reid12}.  The measurement of the
amplitude of the peculiar velocity field is typically made using the variation
of the amplitude of galaxy clustering as a function of orientation with respect
to the line of sight caused by redshift space distortions.  On large scales,
Eq.~\ref{eq:pecv} implies \citep{Kaiser87} \begin{equation} \delta_g^s({\bf k})
= (b + f\mu^2) \delta_m^r({\bf k}).  \label{eq:kaiser} \end{equation} Here
$\delta_g^s$ is the observed (in ``redshift space'') galaxy density fluctuation
for wavevector ${\bf k}$, $b$ is the real space linear galaxy bias, and
$\delta_m^r({\bf k})$ is the true underlying matter density fluctuation (i.e.,
in ``real space'', without velocity perturbations included in the redshift
direction coordinate).  The parameter $\mu$ is the cosine of the angle between
${\bf k}$ and the LOS, and the known $\mu$ dependence allows a measurement of
$f\sigma_8$ after marginalizing over the unknown galaxy bias.  In the present
work, we work strictly in configuration space; see \citet{Fisher95} for the
configuration space equivalent of Eq.~\ref{eq:kaiser}.

On smaller scales investigated in the present work, nonlinearities become
important and the relationship between ${\bf v}$ and $\delta_m$ becomes
substantially more complicated.  A detailed description of many distinct
physical effects that impact the observed redshift space galaxy clustering on
small scales is given in \citet{Tinker07}.  Because of the complexity of the
modeling and the high statistical precision of our data, we resort to $N$-body
simulations to provide predictions for our observables, which we describe below.

\subsection{Two-dimensional correlation function $\xi(r_{\sigma}, r_{\pi})$}
\label{sec:xi2dpre} 
\begin{figure} \includegraphics[width=85mm]{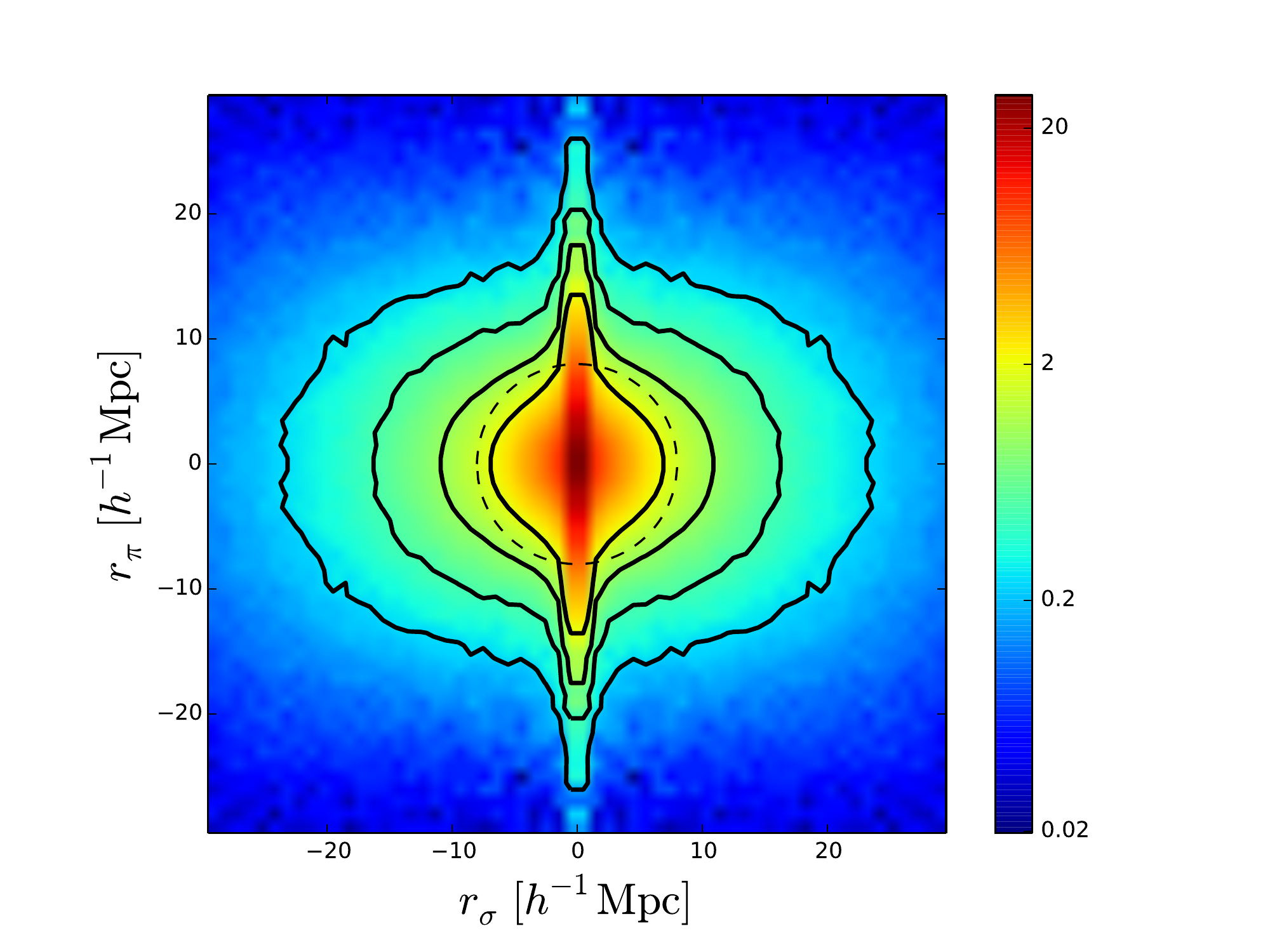} \caption{The two-dimensional
correlation function $\xi(r_{\sigma}, r_{\pi})$ of SDSS-III CMASS galaxies.  The
perturbations of the observed redshifts about the Hubble flow due to peculiar
velocities introduce anistropy in the correlation strength with respect to the
line of sight (y-axis in the figure).  In this plot fiber collisions have been
corrected using the angular upweighting method.  The dashed circle indicates the
separation scale ($\sim 8$ h$^{-1}$ Mpc) at which the observed quadrupole
transitions from positive (dominated by Finger-of-God velocities) to negative
(dominated by large scale Kaiser infall velocities).  Contours at $\xi =
[2,1,0.5,0.25]$ are shown with solid black curves.} \label{fig:xibutterfly}
\end{figure}

Because RSD effects only distort the observed coordinates
(or pair separations) in the LOS direction, the two-point correlation function
$\xi$ is fundamentally a function of two variables.  In
Fig.~\ref{fig:xibutterfly} we choose as coordinates the LOS separation,
$r_{\pi}$, and the separation transverse to the LOS, $r_{\sigma}$ to display our
measurement from the galaxy sample analysed in the present work.  This
measurement uses the angular upweighting method described in
Sec.~\ref{sec:FBang} to correct for fiber collisions.  Two primary features are
apparent: on large scales ($\sim 8$ $h^{-1}$ Mpc and above), contours of
constant $\xi$ are ``squashed'' in the LOS direction.  The correlation between
the density and velocity field described by Eq.~\ref{eq:pecv} on average reduces
the apparent separation between pairs of galaxies along the line of sight.  On
smaller scales where Eq.~\ref{eq:pecv} breaks down, the contours are instead
stretched along the LOS.  Galaxies orbiting in the potential of a
gravitationally bound dark matter halo have a virial-like velocity component.
As we will see, the SDSS-III CMASS galaxies shown here occupy massive dark
matter halos with large virial velocities.  The prominent feature in $\xi$ along
the LOS (i.e., at $r_{\sigma} < 1$ $h^{-1}$ Mpc) is due to these motions, often
called ``fingers-of-god'' (FOGs) \citep{Jackson72}; note that these virial-like
velocities distort $\xi$ at all separations, and their impact must be mitigated
even in analysis of relatively large scales \citep[e.g.,][]{Reid12}.

In this work we choose not to analyse $\xi(r_{\sigma}, r_{\pi})$ directly, since
information is spread over a large number of bins.  As described in
Sec.~\ref{sec:bootstrap}, we estimate measurement errors by bootstrapping the
survey, and therefore need to reduce the number of measurements to well below
the number of bootstrap regions, which are limited in number since each region
must span scales larger than we include in our analysis.  In this section we present the 
observables we will estimate from $\xi(r_{\sigma}, r_{\pi})$ and compare with 
theoretical models directly.

The most widely used
observable in studies of small-scale galaxy clustering is $w_p(r_{\sigma})$, 
which quantifies the clustering as a function
of transverse pair separation $r_{\sigma}$.  All pairs with line-of-sight
separations smaller than $\pi_{max}$ contribute to $w_p$: 
\begin{equation}
w_p(r_{\sigma}) = 2 \int_{0}^{\pi_{\rm max}} dr_{\pi} \xi(r_{\sigma}, r_{\pi}).
\label{eq:wp} 
\end{equation} 
$\pi_{max}$ is traditionally chosen to be large (80 $h^{-1}$ Mpc
in this work) so that the sensitivity of $w_p$ to redshift space distortions is
minimal \citep[but see][]{vdB13}.  

On large scales and
for the highly biased tracers we consider here, the majority of redshift space
information is available by measuring the first two even multipoles ($\ell =
0,2$) of $\xi$: \begin{equation} \xi_{\ell}(s_i) = \frac{2\ell+1}{2}\int d\mu_s\
\xi(s_i, \mu_s) L_{\ell}(\mu_s), \label{eq:moments} \end{equation} where
redshift space separation $s$ is defined by $s^2 = r_{\sigma}^2 + r_{\pi}^2$ and
$\mu_s = r_{\pi}/s$ is the cosine of the angle of the galaxy pair with respect
to the line of sight.  Here $L_\ell$ is the Legendre polynomial of order $\ell$.
Both our measurement and theoretical estimates of $\xi_{0,2}$ are computed by
replacing the integral with a direct sum over bins of width $d\mu_s = 0.1$.
Each bin in redshift space separation $s_i$ is averaged over a finite band of
separations.

To mitigate the effect of fiber collisions, our primary analysis uses the
statistic $\hat{\xi}_{0,2}$ which approaches $\xi_{0,2}$ on large scales, but
eliminates all bins that include pairs with $r_{\sigma} < 0.534$ $h^{-1}$ Mpc.
This choice corresponds to pairs separated by the fiber collision radius 62'' at
the maximum redshift included in our analysis, $z=0.7$.  Heuristically, we
estimate \begin{equation} \hat{\xi}_{\ell}(s_i) =
\frac{2\ell+1}{2}\int_0^{\mu_{\rm max}(s_i)} d\mu_s \xi(s, \mu_s)
L_{\ell}(\mu_s).  \label{eq:hatmoments} \end{equation} In practice, our
implementation is slightly more complicated, but we emphasize that the
measurement and theoretical predictions are computed with exactly the same
algorithm, and so the details are irrelevant for the comparison of the two.  We
start with relatively fine logarithmic binning in $s$ (${\rm d} \, log_{10} \, s
= 0.035$) and $\mu$ (${\rm d} \mu_s = 0.005$) to compute $\xi(s, \mu)$.  We then
aggregate pair counts in the small $s$ bins into larger bins for which we report our
measurements.  In the case where some of the small bins have $r_{\sigma}$ larger
than the cutoff, we estimate $\xi$ in the larger $s$ bin from only that subset
of small bins.  If none of the bins have large enough $r_{\sigma}$ in the
$\mu$-bin, we set $\xi$ for that bin to 0 before integrating over $\mu$ to
estimate $\hat{\xi}_{0,2}$; this is equivalent to only integrating up to a
$\mu_{\rm max}$ that is different for each fine $s$ bin.  The previous step
ensures that no pairs with $r_{\sigma}$ smaller than the fiber collision scale
are included.  The exact $s$ and $\mu$ boundaries for our final bins are listed
in Table \ref{tab:ximeas}.

\subsection{$N$-body simulation Halo Catalogs}
\label{sec:Nbody}
\begin{table}
\centering
\caption{Cosmological and simulation parameters for the $N$-body simulations used in this paper.}
\label{tab:simparams}
\begin{tabular}{llll}
\hline
Parameter & LowRes & MedRes & HiRes \\
\hline
$L_{\rm box}$ ($h^{-1}$ Mpc) & 2750 & 1380 & 677.7 \\
$N_p$ & $3000^3$ & $2048^3$ & $2048^3$ \\
$m_p$ ($h^{-1}$ $M_{\sun}$) & $5.86 \times 10^{10}$ & $2.5 \times 10^{10}$ & $3.10 \times 10^9$ \\
$\Omega_m$ & 0.274 & 0.292 & 0.30851 \\
$\Omega_b h^2$ & 0.0224 & 0.022 & 0.022161 \\
$h$ & 0.7 & 0.69 & 0.6777 \\
$n_s$ & 0.95 & 0.965 & 0.9611\\
$\sigma_8$ & 0.8 & 0.82 & 0.8288\\
$z_{\rm box}$ & 0.550 & 0.550 & 0.547 \\ 
$f\sigma_8(z_{\rm box})$ & 0.455 & 0.472 & 0.482 \\
\end{tabular}
\end{table}
We make use of three periodic $N$-body simulation sets throughout this paper.  
We have a single realization for the LowRes and HiRes cases, and three independent 
realizations (labelled 0,1,2) in the MedRes case.
The simulation parameters are listed in Table \ref{tab:simparams}.  The LowRes
box has parameters favored by WMAP7 \citep{WMAP7}, while the HiRes box adopts
the ``Planck+WP+highL+BAO'' constraints from the Planck analysis
\citep{PlanckXVI}; all $N$-body simulations assume massless neutrinos, while the
parameter constraints from Planck assume $\sum m_{\nu} = 0.06$ eV.  The Planck
best fit and HiRes cosmologies will therefore have slightly different expansion
and structure growth histories which should be negligible for the present
application.  The MedRes cosmology is ``between'' LowRes and HiRes.  For each
simulation we generate spherical overdensity (SO) halo catalogs using an
overdensity of $\Delta_m = 200$ relative to the mean matter density $\rho_m$ to
define the halo virial radius $r_{\rm vir}$.  Our catalogs extend down to fifty
particles per halo where necessary.  We use the \citet{TinkerSO} implementation
of the SO halo finder, which allows halo virial radii to overlap, as long as the
center of one halo is not within the virial radius of another halo; this choice
alters the halo-halo clustering on scales near $r_{\rm vir}$ compared with a
friends-of-friends halo catalog, in which two such halos would be bridged into a
single halo \citep[see figure 9 of][]{Reid09}.  

More specifically, halos are identified around pseudo-peaks in the density field, which may or may
not be located on the true density peak of the host halo.  A radius $R_{\rm
halo}$ is computed for each pseudo-peak in the density field within which the
density is 200$\rho_m$.  Starting at a radius of 1/3 the initial $R_{\rm halo}$,
the center of mass is computed within this restricted radius and iterated to
convergence. If the pseudo -peak lies on a subhalo, this procedure migrates the
halo center to the true host halo density peak. Once convergence is reached, the
top-hat radius is incrementally reduced and the center of mass is recomputed
until the top-hat radius is $R_{\rm halo}$/15.9 or the number of particles
within the top-hat radius drops below 20.  The center of mass is again computed
iteratively at the tophat radius of $R_{\rm halo}$/15.9. For halos above the 20
particle limit, this algorithm averages over $\sim 3.7\%$ of all halo members.
This algorithm was originally refined to accurately locate the halo center; we
verified that it recovers the position of the potential minimum within the halo
to within 0.01-0.02 $h^{-1}$ Mpc.  We denote the mean velocity of these densest
particles as ${\bf v}_{\rm DENS}$, and is our fiducial choice for the velocity
of each halo's central galaxy.  This choice is by no means unique, and Appendix
\ref{sec:vcen} shows that while there is strong evidence that the dense
central region of the halo does have a bulk velocity with respect to the halo
members, the rms offset between the ``central'' velocity and the center-of-mass
velocity depends on the radius over which the average is taken.  The effective
radius for our ${\bf v}_{\rm DENS}$ definition ranges from 0.04 - 0.08 $h^{-1}$
Mpc for halos with $M < 1.2 \times 10^{14}$ $h^{-1}$ $M_{\odot}$ in our MedRes
simulation; this mass
range hosts 90\% of the central galaxies in our sample for our best fit halo
occupation distribution model.  The median seeing-corrected effective radius of
CMASS targets is 1.2'' \citep{Masters11}, or 0.0087 $h^{-1}$ Mpc.  For a de
Vaucouleurs profile, 0.04 (0.08) $h^{-1}$ Mpc would contain 87\% (96\%) of the
light.  Therefore, our choice of central velocity definition is reasonably well
matched to the typical extent of our target galaxies.  Of course, since our
simulations contain only dark matter, any impact of baryonic physics on the
central dynamics has been neglected.

The LowRes box provides a volume much larger than the survey we analyse.  To
match the observed clustering strength using the LowRes box, we require halos
below the SO halo catalog threshhold; for this purpose, we use a
friends-of-friends (FOF) algorithm with linking length 0.168 to identify halos
\citep{Davis85}, and then compute their masses in a spherical aperture at
$\Delta_{180}$.  We further rescaled the FOF-derived halo masses by a factor of
0.975 to approximate $\Delta_{200}$ masses used in the SO catalogs.  The FOF
catalog extends down to $5.4 \times 10^{11}$ $h^{-1}$ $M_{\sun}$.  Typically
only $\sim 5\%$ of the mock galaxies are assigned to FOF-derived halos, so the
impact of these details should be minimal.  This hybrid catalog did show
evidence for numerical artefacts in the halo clustering, which led us to adopt
higher resolution simulations for our primary parameter constraints.  The HiRes
box provides more than sufficient mass resolution but its small volume made the
theoretical predictions somewhat noisy.  Because this box is smaller than our
survey size, we have to add a theoretical error budget to the observational one.
We use mock catalogs derived from the LowRes box for two applications.  First,
we generate a mock catalog to which we apply the BOSS tiling algorithm in order
to investigate the fiber collision effects on clustering (see
Sec.~\ref{sec:FB}).  Second, we subdivide the LowRes box into 64 subboxes to
estimate the theoretical uncertainty due to the finite volume of the HiRes box.

\section{Data} \label{sec:data} In this paper we analyse data included in data
release 10 (DR10) of the Sloan Digital Sky Survey
\citep[SDSS;][]{Gunn98,Yor00,Gunn06,Eis11}.  We refer the reader to
\citet{Ahn14} for the full details about the DR10 dataset.  Briefly, BOSS uses
imaging data available in SDSS data release 8 \citep{DR8} to target quasars and
two classes of galaxies for spectroscopy.  The larger CMASS sample analysed in
this work covers $0.4 < z < 0.8$, but as in \citet{Anderson13} we restrict our
analysis to the subsample falling within $0.43 < z < 0.7$.  The LOWZ galaxy
targets primarly have $z < 0.4$ and so we treat them as uncorrelated with the
CMASS targets.   

Errors in the redshift fitting to the BOSS spectra propagate as random errors in
the LOS redshift space position of each galaxy.  The median redshift error
reported by the BOSS spectroscopic pipeline is 0.00014.  \citet{Bolton12}
estimate that these errors are underestimated by $\sim 20\%$.  While the
redshift error increases with $z$ (by 40\% between the lower and upper third of
the sample), the conversion to distance errors partially cancels this.  We
therefore expect a typical line-of-sight offset of $0.44$ $h^{-1}$ Mpc due to
redshift errors.  We incorporate redshift errors into our theoretical model by
adding a random LOS offset to each mock galaxy that is drawn from a Gaussian
distribution of width $\sigma = 0.44$ $h^{-1}$ Mpc.

Throughout we treat the North and South galactic cap regions of the survey
separately.  Photometric calibration across the hemispheres is subject to
systematic error, and may result in slightly different populations of galaxies
being targeted in the two hemispheres; see \citet{Ross12} for a more detailed
analysis of potential differences.  We will address the consistency of the north
and south for each statistic we compute, and combine them before performing
fits.  The majority of our sample is contained in the north: 409365 of the
521958 galaxies in the final sample. 

In \citet{Anderson13} we upweighted nearest neighbor galaxies to account for two
potentially non-random sources of missing redshifts: fiber collisions and
failure to obtain redshifts for targets assigned a spectroscopic fiber.  We
adopt the same weighting procedure for redshift failures in present work.  For
fiber collisions, we assign a nearest neighbor redshift to each collided galaxy,
drawn from another CMASS target within the fiber collision group provided by the
tiling algorithm.  In total, we corrected for 7043 (2829) redshift failures in
the north (south) and 24648 (6297) fiber collided galaxies.  The survey
completeness is treated as uniform in each sector, which is defined as a unique
intersection of spectrosopic tiles.  The angular mask is defined in the same way
as in \citet{Anderson13}, except that for the purposes of fiber collision
correction uniformity, we only retain sectors where all planned spectroscopic
tiles have been observed.   We track two values of completeness in each sector.
For $c_{\rm NN}$, collided galaxies are assigned a nearest-neighbor redshift and
treated as observed when estimating completeness.  To define the completeness
used in the angular upweighting method (Sec.~\ref{sec:FBang}), fiber collided
galaxies are instead treated as a random subsample of the spectroscopic sample
that lack redshifts, so they reduce the completeness of their sector.  The
area-weighted completenesses are $c_{\rm NN} = 0.988 (0.984)$ and $c_{\rm ang} =
0.936 (0.939)$ in the north (south).  Finally, we note that we neglect both the
FKP and systematics weights adopted in \citet{Anderson13} so that galaxies
receive equal weight in both the angular and three-dimensional clustering
measurements.  The systematics weights primarily affect clustering on very large
scales, and can therefore be neglected on the small scales of interest here.

In order to place the observed angular and redshift coordinates of each galaxy
on a comoving grid, we adopt the same fiducial cosmological model as in
\citet{Anderson13}: $\Omega_m = 0.274$.  We compute $\xi$ from the Landy-Szalay
estimator \citep{LanSza93}, which depends on the data-data ($DD$), data-random
($DR$), and random-random ($RR$) pair counts in each separation bin of interest:
\begin{equation} \xi(\Delta {\bf r}_i) = \frac{DD(\Delta {\bf r}_i) -2DR(\Delta
{\bf r}_i) +RR(\Delta {\bf r}_i)}{RR(\Delta {\bf r}_i)}.  \label{eq:LS}
\end{equation} Random galaxies are an unclustered sample of points within the
survey mask with the same sector-by-sector completeness as the data and with a
radial selection function matched to the data by drawing the random galaxy
redshifts from the observed ones (i.e., the ``shuffle'' method advocated in
\citealt{Ross12}).  In contrast to some other works that use $C^{-1}$ weighting,
we combine clustering statistics estimated separately in the two hemispheres
using a simple weighted average with weights proportional to the total galaxy
weight in each hemisphere.

\section{Fiber collision corrections} \label{sec:FB} The tiling algorithm
\citep{Tiling} determines the location of the $3^\circ$ spectroscopic tiles and
allocates the available fibers among the galaxy and quasar targets.  A physical
constraint of the instrument is that fibers may not be closer than 62'' on a
given spectroscopic tile.  The algorithm divides target galaxies into
friends-of-friends groups with a linking length of 62'', and then assigns fibers
to the groups in a way that maximizes the number of targets with fibers.  The
choice of which galaxies are assigned the fibers is otherwise random.  The
algorithm adapts to the density of targets on the sky, with a net result being
that regions covered by more than one tile have a larger than average number
density.
For the DR10 sample studied in this work, 42\% (52\%) of the area in the north
(south) is covered by multiple tiles, and the number density is larger by 4.6\% 
(3.1\%) in those regions.  For the final survey footprint, the enhancement is
5.1\% in both the north and south.  Fiber collisions are partially
resolved only in the multiple tile regions, 
and therefore may not be representative of the unresolved fiber
collisions in lower target density regions.  For this reason we do not adopt the
method of \citet{Guo12}, which uses the overlap regions to correct fiber
collisions in the single tile regions.  

\begin{figure}
\includegraphics[]{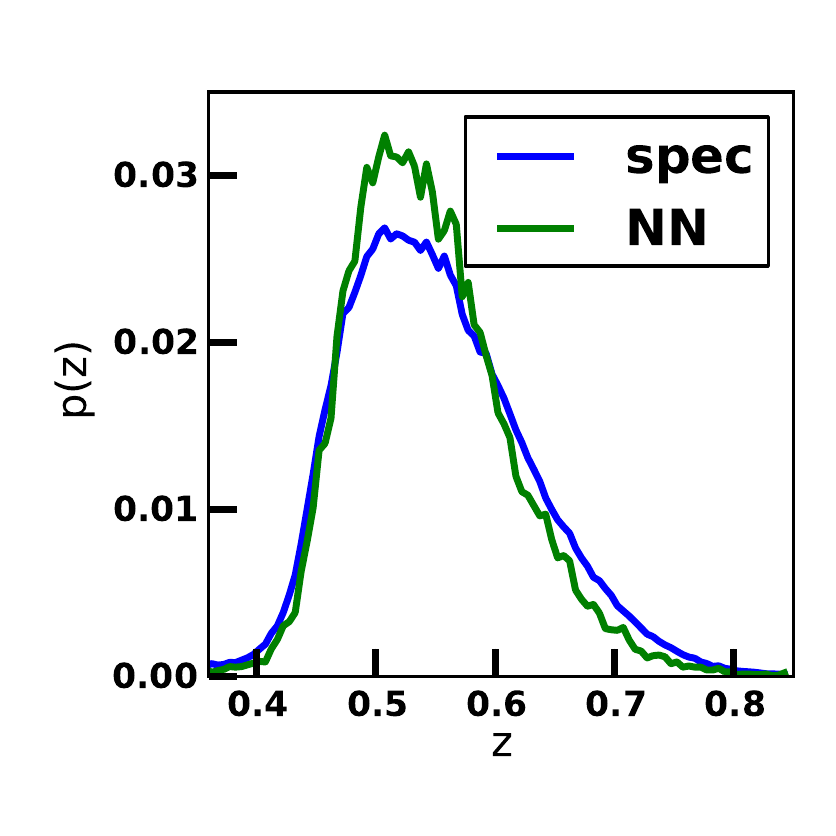} \caption{The normalized redshift probability
distribution for CMASS targets that were assigned fibers (blue) and fiber collided galaxies
(green).  For collided galaxies, we use the nearest neighbor redshifts as a
proxy; since the galaxy in a fiber collision pair that receives the fiber is
randomly chosen, this is an unbiased estimate of the redshift distribution for
objects without a fiber due to a fiber collision.} \label{fig:pofz} \end{figure}

Fiber-collided galaxies cannot
simply be accounted for by reducing the completeness of their sector, since they
are a non-random subset of targets (conditioned to have another target within
62'').  This is evident in Fig.~\ref{fig:pofz}, where we compare the redshift
distribution of all good CMASS redshifts with the redshift distribution of the
nearest neighbor of fiber collided galaxies; the latter preferentially reside
near the peak of $p(z)$, where $\bar{n}(z)$ is also the largest.  In this
section we will apply the tiling algorithm to a mock galaxy catalog in order to
assess the robustness of our fiber collision correction methods to the existence
of realistic tile position-target density correlations.  Finally, we note that
fiber collisions also occur between CMASS targets and targets from other
classes.  Since quasar targets are given higher priority, we account for them by
simply including a 62'' veto mask around each high priority quasar target.
CMASS targets that are fiber collided with LOWZ galaxies are treated as random
losses; the completeness of their sectors is reduced accordingly, rather than
assigning them a nearest neighbor redshift.  
\subsection{The angular upweighting method}
\label{sec:FBang}
The comparison of
the angular clustering of the target galaxies with the angular clustering of the
subsample of targets for which spectra were obtained quantifies the number of
pairs lost as a function of angular separation due to fiber collisions.  One
common method to use this information to correct measurements of $\xi$ is to
first treat the fiber-collided missing galaxies as if they were a random
subsample of targets (i.e., adjust the completeness of the sector based on the
fraction of targets without spectra), and then to upweight $DD$ pairs in the
Landy-Szalay estimator (Eq.~\ref{eq:LS}) using the following weight
\citep{Hawkins03}: 
\begin{equation} 
\label{eq:angweight}
{\rm w}_{\rm pair}(\theta) = \frac{1+w_s(\theta)}{1+w_t(\theta)}.
\end{equation} 
Here $w_s$
is the angular correlation function of galaxies drawn from the ``spectroscopic''
sample for which you obtained redshifts and want to estimate $\xi$.  $w_t$ is
the angular correlation function of the targets from which the spectroscopic
sample is drawn.  

As we will show using our tiled mock catalog, this method performs quite well at
recovering the true clustering down to scales well below the BOSS fiber
collision radius, and has been successfully applied in a number of other surveys
\citep{Hawkins03,Li06,Ross07,White11,delaTorre13}.  However, there are a number
of open issues and limitations of the method:
\begin{itemize} 
\item The method
assumes that the distribution of LOS separations of the fiber-collided pairs is
the same as the distribution for non-collided pairs; this will not be true in
detail, since fiber-collided pairs may occupy a different distribution of halo
masses (and therefore have a different large-scale bias) compared to the full
target sample.  Since ${\rm w}_{\rm pair}$ quickly approaches 1 on large scales,
the method will not properly account for the possibility of fiber-collided
galaxies having a larger bias.  For this reason we prefer the next method
(nearest neighbor redshift assignment) on scales well above the fiber collision
scale.
\item The weight given in Eq.~\ref{eq:angweight} is not easily generalizable to
``spectroscopic'' samples that are subsets of the full set of fibers (in our
case, cutting out stars and applying redshift boundaries), at least not without
additional assumptions regarding the redshift dependence of the target galaxy
clustering.  
\end{itemize}
The Limber approximation \citep{Limber54} allows us
to relate the angular and real-space clustering given the probability
distribution $p(\chi)$ that an object in the sample is at comoving distance
$\chi$.  \begin{equation} \label{eq:limberang} w(\theta) = \int_0^{\infty} d\chi
p^2(\chi) \int_{-\infty}^{+\infty} dr_z \xi(\sqrt{\chi^2 \theta^2 + r_z^2},
\chi) \end{equation} Here $r_z$ is the LOS separation between a pair.  The
explicit $\chi$ dependence of $\xi$ in Eq.~\ref{eq:limberang} indicates that
$\xi$ may evolve with redshift.  We can only observe $w(\theta)$ for the full
target sample.  However, if we adopt a particular model for $\xi(r,\chi)$, then
in principle we can fit for its parameters and infer the expected $w(\theta)$
for a subsample of targets occupying a narrower range of $\chi$ (i.e., different
$p(\chi)$).  The simplest choice is a power law correlation function, $\xi(r)
= (r/r_0)^{-\gamma}$.  In that case, Eq.~\ref{eq:limberang} gives $w(\theta) =
A_w \theta^{1-\gamma}$ with \begin{equation} \label{eq:wthetapowerlaw} A_w =
\sqrt{\pi} r_0^{\gamma} \frac{\Gamma(\gamma/2-1/2)}{\Gamma(\gamma/2)}
\int_0^{\infty} d\chi p^2(\chi) \chi^{1-\gamma}.  \end{equation} For a more
complicated $\xi(r)$, $w(\theta)$ in different redshift slices must generically
have a different $\theta$ dependence that depends on $p(\chi)$.  To relate the
observed $w_t$ to the one for the subsample entering Eq.~\ref{eq:angweight},
\citet{Hawkins03} employed the Limber approximation, assuming a power law form
for $\xi$ and that it does not evolve with redshift.
We make the same assumption in Sec.~\ref{sec:angweight} but propagate the 
uncertainty in this step to our final clustering measurements.

\subsection{Nearest neighbor redshift (NN) and \citet{Anderson13} weighting
schemes} \label{sec:NNz} An alternative fiber collision correction method is to
simply assign a fiber-collided galaxy the redshift of its nearest neighbor
(labelled NN in subsequent figures).  In the limit of separations large compared
to 62'', this is equivalent to the method employed in our large-scale analyses,
and described in detail in \citet{Anderson13}.  In that work we assign the
weight of a fiber collided galaxy to its nearest neighbor, which is propagated
into the redshift distribution and correlation function calculations.  A nice
property of this method is that it is guaranteed to recover the correct large
scale clustering amplitude of the full sample, at least in the limit of fiber
collision {\em pairs} rather than groups.   While the large scale bias is
considered a nuisance parameter in galaxy auto-correlation analyses, it is used
in cosmological parameter analyses when comparing galaxy clustering with
galaxy-galaxy lensing, for example in the $E_G$ test \citep{Reyes10}.  This
method will not recover accurate statistics on small scales, since the LOS
separation between the collided and nearest neighbor galaxy will be artificially
set to 0, thus suppressing true FOGs in the galaxy sample.

\subsection{Solution using a tiled mock catalog} \label{sec:mocksoln} We
generate a mock catalog covering the northern galactic cap portion of the
SDSS-III BOSS survey using the LowRes simulation box listed in
Table~\ref{tab:simparams}.  The HOD parameters for this mock catalog were chosen
based on a preliminary fit to preliminary measurements of $\xi_{\ell}$: 
$M_{\rm min} = 8.62 \times 10^{12}$
$h^{-1}$ $M_{\sun}$, $M_{1} = 2.16 \times 10^{14}$ $h^{-1}$ $M_{\sun}$, $M_{\rm
cut} = -2.53 \times 10^{12}$ $h^{-1}$ $M_{\sun}$, $\sigma_{log M} = 0.43$, and
$\alpha = 1.00$.  The central galaxy velocities were defined using ${\bf v}_{\rm
COMV}$, the center-of-mass velocity of the host halo.  We also set $\gamma_{\rm
IHV} = 1.27$ and $\gamma_{\rm cenv} = 0.48$ in order to get a reasonable fit to
the observed clustering.  All of these HOD parameters are described in detail in
Sec.~\ref{sec:HODmodel}.  In the following section we compare the clustering in
the mock catalog with the data; the agreement is not perfect but sufficiently
good for our purpose, which is to compare various fiber collision correction
methods against the true clustering in the mock catalog.

We choose the number density of the HOD to be $\bar{n} = 4.2 \times 10^{-4}$
($h^{-1}$ Mpc)$^{-3}$, larger than the maximum value of the CMASS sample and
first generate a uniform mock catalog using a single redshift output.  We then
apply the angular mask of the NGC for the final BOSS survey, and randomly
downsample the mock galaxy catalog to match the observed CMASS $\bar{n}(z)$.  We
keep galaxies between $0.3 < z < 0.8$ in the mock catalog to account for the
difference between the angular clustering of the full target sample and the
angular clustering of the targets after applying the redshift cut $0.43 < z <
0.7$.  Note that this mock catalog therefore meets the ``constant clustering''
assumption discussed in Sec.~\ref{sec:FBang}.  This broader redshift coverage
incorporates the vast majority of CMASS targets except for the 3\% of targets
that are stars.  We make use of the approximate independence of the LOWZ and
CMASS galaxy distributions, and input the positions of the true LOWZ and quasar
targets before applying the BOSS tiling algorithm to the mock catalog.  We
measure the $w(\theta)$ for the full mock target sample and $w(\theta)$ for the
mock target sample after applying redshift cuts, and find that on all scales of
interest their ratio remains within 1\% of 1.14, the value expected from
Eq.~\ref{eq:wthetapowerlaw}.  We also find that the angular weight that
estimates $w_s$ from the set of all targets assigned to fibers (and $w_t$
corresponds to the full target sample) returns the same weight as our fiducial
approach with redshift cuts, within 1\%.

As we quantify in more detail in Sec.~\ref{sec:meas} and
Fig.~\ref{fig:mockscompareall}, our tiled mock catalog shows that the angular
upweighting method recovers the true $w_p$ within 1.5\% on all scales, while the
same is true above the fiber collision scale when using the nearest neighbor
redshift method.  Similarly for the $\hat{\xi}_{0,2}$ statistics, the nearest
neighbor method almost perfectly recovers the true clustering above $s=2$
$h^{-1}$ Mpc.  The angular method is nearly unbiased on scales below $s=2$
$h^{-1}$ Mpc, but shows systematic $\sim 2\sigma$ differences on intermediate
scales for both $\hat{\xi}_2$ and $\xi_2$.  Our final clustering estimators will
use the nearest neighbor method on large scales and angular upweighting on small
scales to infer the clustering of the CMASS sample in absence of fiber
collisions, with systematic uncertainty below the statistical one.  Therefore,
we are not compelled to introduce any more complicated fiber collision
correction schemes, such as the one recently proposed in \citet{Guo12}.
However, we note that their tests corroborate our findings that angular
upweighting performs well on scales below the fiber collision scale, and the
nearest neighbor method is unbiased on large scales.  We note that their 
method has not yet been verified on a mock catalog with realistic tile-density 
correlations.
 
\section{Measurements and Covariances} \label{sec:meas}

\subsection{Survey Bootstrap Regions} \label{sec:bootstrap} \begin{figure}
\includegraphics[width=85mm]{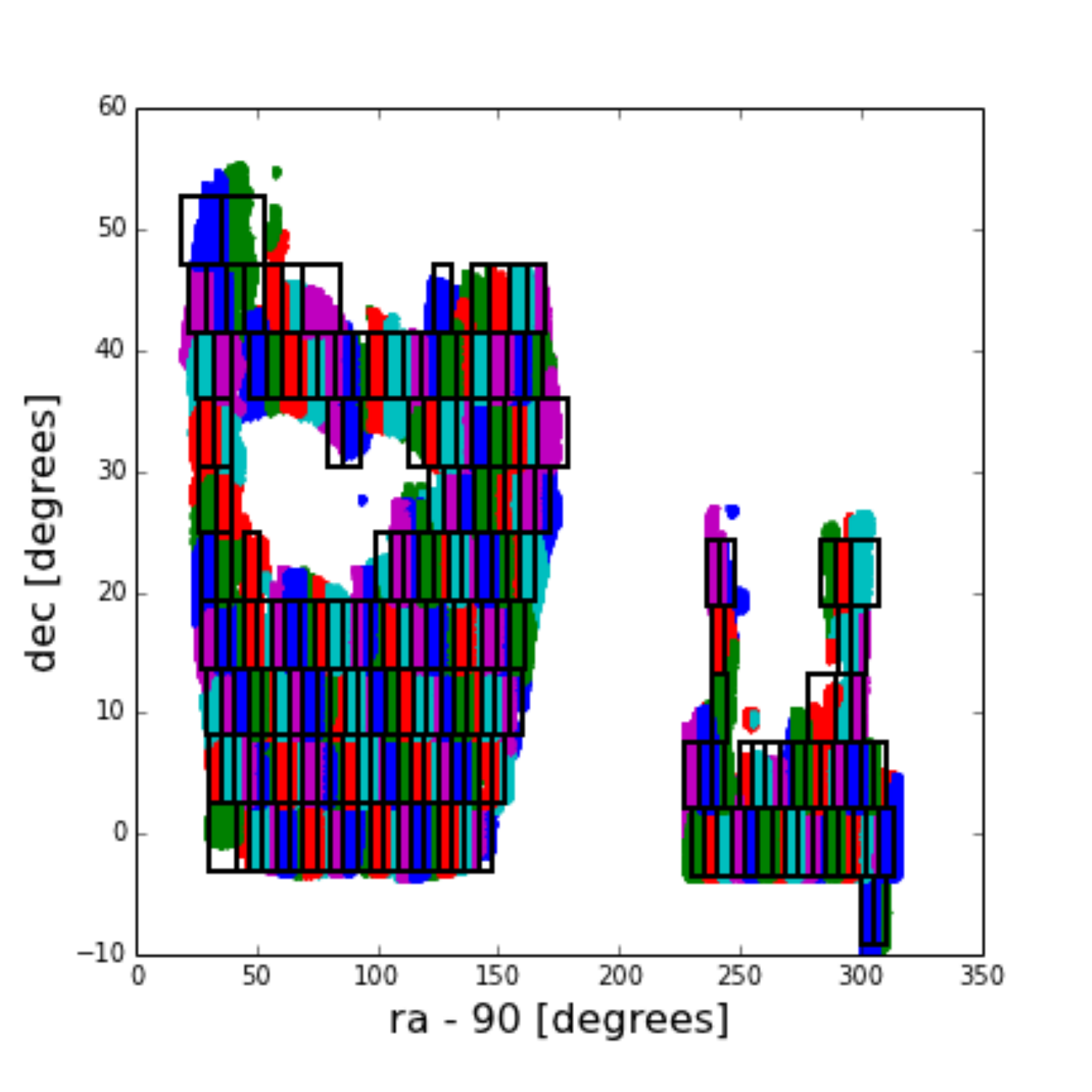} \caption{Two hundred bootstrap regions used to
estimate the covariance matrix of observables from the survey.  The individual
subregions are squares (or a union of squares) in the coordinates $\Delta$dec,
$cos({\rm dec}) \Delta$ra.} \label{fig:bootregions} \end{figure} We derive
statistical covariances on our measurements by dividing the survey into 200
subregions, roughly equal in size and shape.  Fig.~\ref{fig:bootregions} shows
the regions.  To define them, we first distributed square regions across the
survey footprint with sidelength 5.56 degrees using $\Delta$dec, $cos({\rm dec})
\Delta$ra coordinates.  At $z=0.57$ in our fiducial cosmology this corresponds
to 145 $h^{-1}$ Mpc on a side and the redshift extent from $z=0.43$ to $z=0.7$
translates to 608 $h^{-1}$ Mpc.  For comparison, the largest separations
included in our $\xi_{\ell}$ measurements is 38 $h^{-1}$ Mpc, and our $w_p$
measurements extend along the LOS to 80 $h^{-1}$ Mpc.  Along each row of
constant dec, we shifted the square centers in the ra direction to maximize the
number of ``nearly full'' squares.  We then grouped neighboring squares together
in order to homogenize the number of galaxies per bootstrap region.  Finally,
galaxies and randoms outside any of the 200 bootstrap region were assigned to
the nearest regions.  The final distribution of randoms per bootstrap region had
a one-$\sigma$ scatter of $\sim 17\%$, accounting for both survey footprint
incompleteness in the regions and variations in completeness for regions within
the survey footprint.  To derive bootstrap errors, we compute each observable
separately in each subregion, excluding pairs that cross subregion boundaries.
We use a single normalization between the data and random counts (though
different for the north and south) that enters the Landy-Szalay estimator; this
choice uses information from the entire survey to constrain the expected number
density of galaxies as a function of redshift.  The bootstrap covariance is then
estimated as \begin{equation} \label{eq:cov} C_{\rm boot} =
\frac{1}{M-1}\sum_{k=1}^{M} (x_i^k-\bar{x}_i) (x_j^k - \bar{x}_j) \end{equation}
and we set $M=5000000$.  Here $x_i^{k}$ is the $k^{th}$ mean of $N$ randomly
selected (with replacement) $x_i$ from the $N=200$ subregion measurements.

Following \citet{Hartlap07}, we estimate the inverse covariance matrix as
\begin{equation} \hat{C}^{-1} = \frac{n_{\rm boot} - p -2}{n_{\rm boot} -1}
C^{-1}_{\rm boot}; \label{eq:Hartlapfac} \end{equation} but see discussions in
\citet{Krause13} and \citet{Eifler08}.  In our case $n_{\rm boot} = 200$, and
$p$ = 27 for our default analysis, which jointly fits $w_p(r_\sigma<2 \, h^{-1}
{\rm Mpc})$ and $\hat{\xi}_{0,2}$.  We verified that covariances derived by
dividing the survey into a smaller number of boostrap regions (50 or 100) gave
similar correlation structure, and diagonal errors reassuringly agreed at the
$\sim 10\%$ level for $w_p$.
 
\subsection{Angular clustering and fiber collision angular weights} 
\label{sec:angweight}
\begin{figure} \includegraphics[width=85mm]{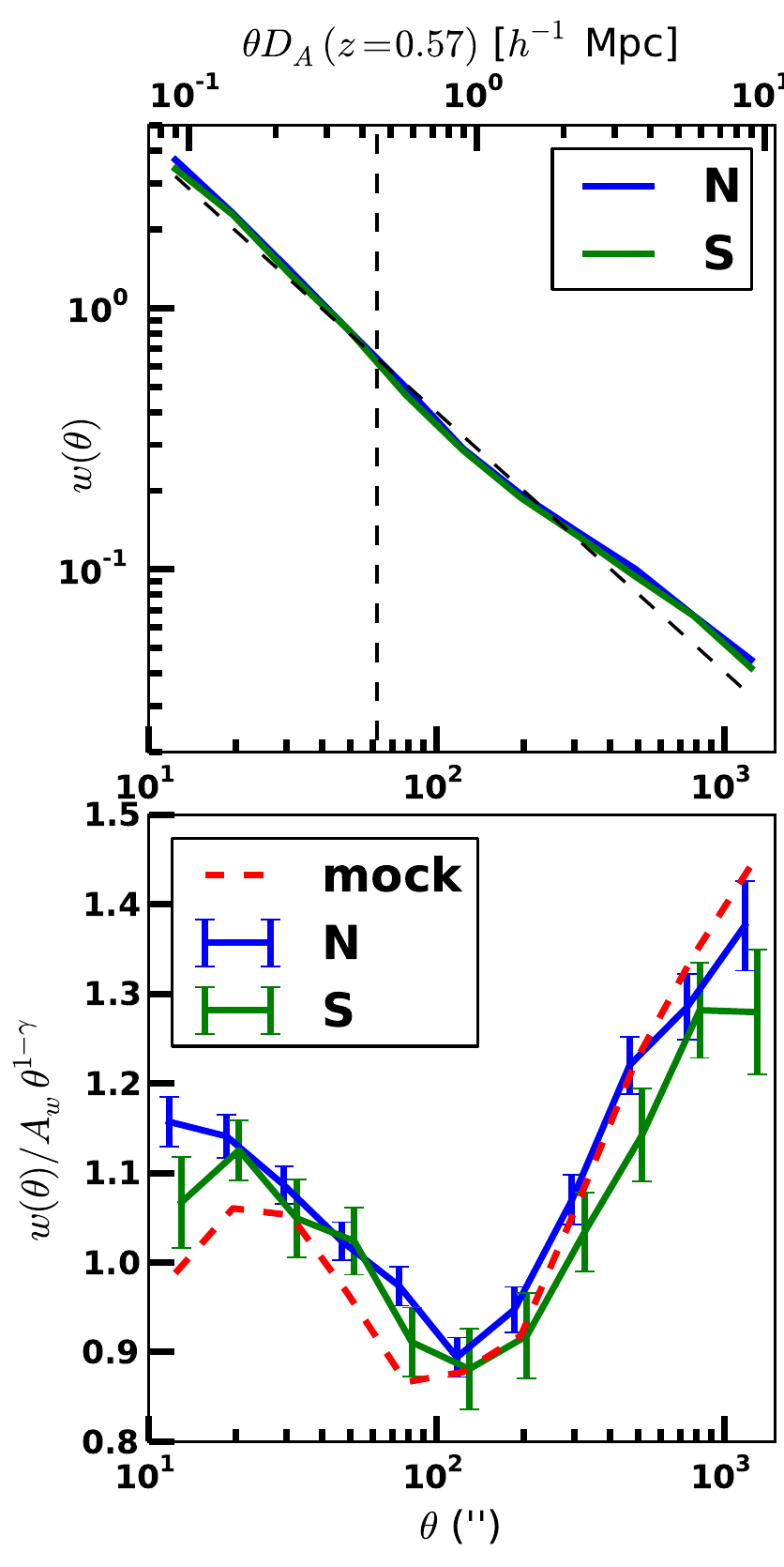} \caption{The clustering of
CMASS targets in the north and south as a function of angular separation
$\theta$ in arcseconds.  The top axis translates angular scales to comoving
separations at the effective redshift of our sample, $z_{\rm eff} = 0.57$.  The
fiber collision scale, 62'', is highlighted with the dashed vertical line in the
upper panel.  The dash-dot line shows the best fit power law.  The lower panel
highlights the $\sim 20\%$ deviation of the observed clustering from the best
fit power law and also compares $w(\theta)$ measured in the northern and
southern hemispheres to $w(\theta)$ for targets in tiled mock catalog (red). We
show the diagonal elements of the bootstrap errors derived separately for the N
and S, offset by $\pm 5$ per-cent in $\theta$ for clarity.} \label{fig:wtheta} \end{figure}

The angular
clustering of CMASS targets in the northern and southern galactic caps is shown
in Fig.~\ref{fig:wtheta}.  A power law $w(\theta) = (\theta/39.75'')^{-0.99}$ is
a reasonable description of the overall behavior, though $\sim 20\%$ level
deviations are clearly detectable.  We find $\chi^2 = 450$ for a power-law fit
with 11 degrees of freedom.  At $\sim 100''$ ($\sim$ 1 $h^{-1}$ Mpc), there is a
dip that corresponds to the one-halo to two-halo transition region in the halo
model.  We find that the difference between the north and south measurements of
$w(\theta)$ is consistent with our bootstrap errors, so there is no indication
in this statistic that the galaxies in the two hemispheres we select cluster
differently. 

\begin{figure} \includegraphics[width=85mm]{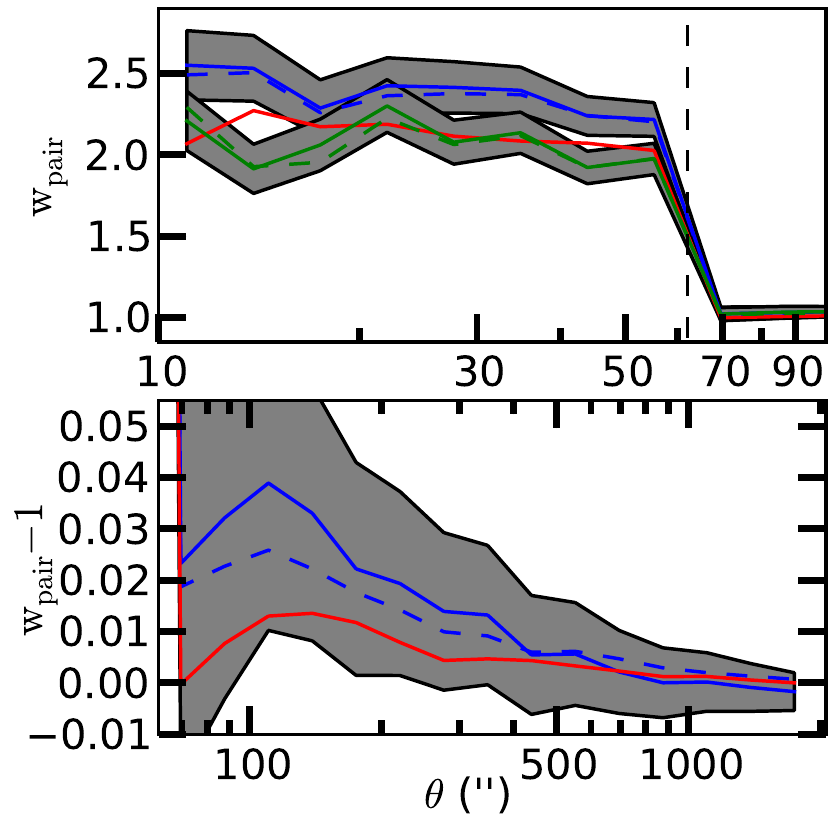} \caption{The angular weights
as a function of pair separation (solid) in the North (blue) and South (green).
The upper panel compares the ${\rm w}_{\rm pair}$ below the fiber collision
radius (dashed vertical line).  The plate density in the south is slightly
higher, so the angular weight is smaller on scales below the fiber collision
radius.  The red solid line is the angular weight derived from the tiled mock
catalog.  In the bottom panel, we show ${\rm w}_{\rm pair} - 1$ for the north
(blue) and tiled mocks (red).  The angular weight for the south is similar.  For
scales above the fiber collision radius, $|{\rm w}_{\rm pair} - 1|$ is smaller
than 4\% on all scales.  To illustrate the level of correction to transform the
full sample target clustering to the redshift cut subsample (i.e., $w_t$
entering Eq.~\ref{eq:angweight}), we also show as dashed lines ${\rm w}_{\rm pair}$ 
corresponding to $w_s$ measured from all targets assigned fibers and $w_t$ 
measured from the full target catalog.  In this case star targets and galaxies
outside our redshift cuts are included.  These two schemes produce nearly
identical angular weights.  The grey bands indicate the uncertainty in $w_t$
corresponding to the spectroscopic subsample that we propagate to our final
estimates of $w_p$, $\hat{\xi}_{0,2}$, and $\xi_{0,2}$.} \label{fig:angcorrect}
\end{figure}

In this paper we analyze a single redshift-selected subsample of CMASS targets,
$0.43 < z < 0.7$.  8\% (9\%) of targets in the north (south) are galaxies that
fall outside this redshift range, with $p^2(\chi)$ in Eq.~\ref{eq:limberang} of
these discarded targets peaking in a relatively narrow range in $\chi$ near both
redshift boundaries.  Another 2.6\% (3.1\%) of targets are stars (neglected in
the mock tiling), and their angular clustering is weak enough to be completely
neglected on these scales.  Following \citet{Hawkins03} and \citet{White11}, we
use Eq.~\ref{eq:wthetapowerlaw} to relate the observed clustering of the full
target sample (shown in Fig.~\ref{fig:wtheta}) to the expected target clustering
after the redshift cuts.  We find $w_{\rm t,subsample}/w_{\rm t,full}$ is 1.18
(1.2) in the north (south) under the ``constant clustering'' assumption that
targets inside and outside the redshift cuts have identical real-space
clustering.  If we instead assume that $\xi = 0$ outside our redshift cuts, this
factor would be reduced by 6\% (8\%).  Measurement of $w_p$ for the galaxies
outside our redshift cuts is noisy, but indicates that the constant clustering
assumption is correct, within a factor of $\sim 2$.  Therefore we assign a
systematic uncertainty to $w_{\rm t}$ in Eq.~\ref{eq:angweight} of 10\%, shown
as the grey bands in Fig.~\ref{fig:angcorrect}.   Finally, unlike \citet{Hawkins03}, we do not
systematically shift the angular coordinate of $w_t$ in the full sample when
estimating $w_t$ for the redshift subsample; a $p^2(\chi)$ weighted mean of
$\chi$ yields a difference of only 0.1\% in the subsampled case.  Our final
estimates of the angular correction $w_{\rm pair}(\theta)$ for the north (blue solid line) and south
(green solid line) are shown in Fig.~\ref{fig:angcorrect}.  The plate density in
the south is slightly higher, so the angular weight is smaller than in the north
on scales below the fiber collision radius.  We also show the angular weight
derived in the same manner using the tiled mock catalog (red).  In the tiled
mock catalog, the constant clustering assumption is accurate by construction.
In all three cases, the weights quickly approach 1 above the fiber collision
scale.  For comparison, the dashed lines show angular weights derived setting
$w_s$ to the full sample of targets that were assigned fibers, and $w_t$ to the
full target sample, as opposed to our fiducial method of estimating both after
applying redshift cuts.  Within our 10\% uncertainty in $w_t$, the two schemes
are identical.  We propagate the uncertainty in the angular pair weight by
computing the statistics of interest ($w_p$, $\xi_{0,2}$, and $\hat{\xi}_{0,2}$)
with our best estimate of ${\rm w}_{\rm pair}$ as well as ${\rm w}_{\rm pair}$
derived from $\pm 10\%$ changes in $w_t$ in Eq.~\ref{eq:angweight}.

\subsection{Projected correlation function $w_p$} 
\begin{figure} \includegraphics[width=85mm]{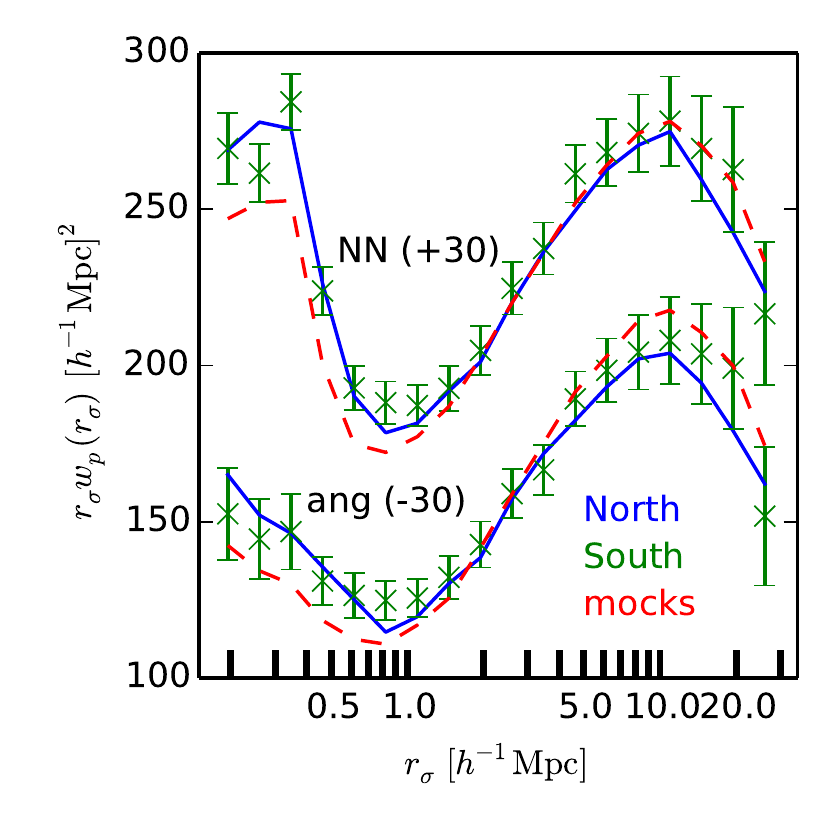} \caption{The projected
correlation function $w_p(r_{\sigma})$ in the north (blue), south (green), and
tiled mock catalog (red).  The upper curves use nearest neighbor (``NN'')
redshifts to correct for fiber collisions, while the lower ones use angular
upweighting (``ang''); the curves have been offset in $r_\sigma w_p$ by $\pm 30$
for visualization.  On the south measurements, we show the square root of the
diagonal elements of the bootstrap covariance matrix, multiplied by the inverse square
root of the fraction of total galaxy weight in the south (i.e., a factor of 2
over the N+S bootstrap errors).  The observed difference between the north and
south are consistent with random realizations of the same underlying $w_p$.  The
mock catalog show fairly good agreement with the observations as well.}
\label{fig:wpNS} \end{figure}

In Fig.~\ref{fig:wpNS} we show
our measurements of $w_p$ in the northern (blue) and southern (green)
hemispheres of the DR10 CMASS sample, compared with the tiled mock catalog
(red).  Errors on the south measurements are square root of the diagonal
elements of the bootstrap covariance matrix after dividing by the fraction of total
galaxies in the south. To assess the consistency of the two observations, we
compute \begin{equation} \Delta_{NS} = (w_{p,N} - w_{p,S})_i C^{-1}_{{\rm
boot},ij} (w_{p,N} - w_{p,S})_j. %N_{\rm gal,S}/(N_{\rm gal,S} + N_{\rm gal,N})
\label{eq:deltaNS} \end{equation} 
Assuming that the bootstrap covariance matrix
can be rescaled using the total galaxy weight in the (independent) northern and
southern subsamples to adequately describe the data covariances, we would
expect \begin{equation} \left<\Delta_{NS}\right> = \left(N_{\rm gal,S} + N_{\rm
gal,N}\right)\left(N_{\rm gal,S}^{-1} + N_{\rm gal,N}^{-1}\right) n_{\rm bins},
\end{equation} 
where $N_{\rm gal,N}$ ($N_{\rm gal,S}$) are the 
total number of galaxies in our sample in the north (south).
For $w_p$, $n_{\rm bins} = 18$.  We find $\Delta_{NS} = 110$
(expected 106) for the $w_p$ estimated using nearest neighbor redshifts, and
similar results for the angular upweighting method, so the two are perfectly
consistent.  We also compute $\Delta$ between the combined N+S $w_p$ measurement
and the tiled mock result; $\Delta = 136$ (91) for the nearest neighbor and
angular upweighting statistics, respectively, while we expect 73.  The
disagreement is worst in the nearest neighbor redshift case on the smallest
scales, where we will adopt the angular upweighting result for our final
analysis.

\subsection{Anisotropic clustering measures $\hat{\xi}_{\ell}$ and $\xi_{\ell}$}
\begin{figure} \includegraphics[width=85mm]{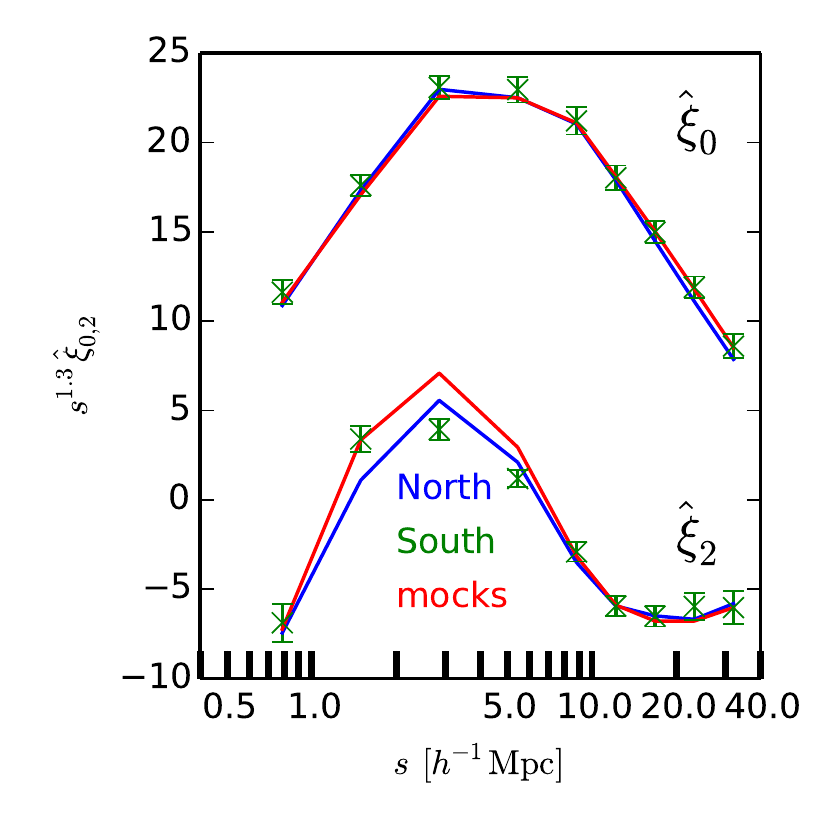} \caption{The
pseudo-multipoles $\hat{\xi}_{0,2}$ defined in Eq.~\ref{eq:hatmoments} measured
from the north (blue), south (green) and mock tiling catalog (red), as in
Fig.~\ref{fig:wpNS} using nearest neighbor redshifts to correct for fiber
collisions.} \label{fig:hatxiNS} \end{figure} Fig.~\ref{fig:hatxiNS} shows a
comparison of $\hat{\xi}_{0,2}$ measured from the north, south, and tiled mock
catalog using the nearest neighbor redshift correction; results for angular
upweighting and for $\xi_{0,2}$ are very similar.  Applying Eq.~\ref{eq:deltaNS}
to $\hat{\xi}_{0,2}$ the disagreement between the north and south subsamples is
larger: $\Delta_{NS} = 181$ (expected 106) but within the expected variation
for this quantity.  Comparing $\hat{\xi}_{0,2}$ from the tiled mock catalog to
the data we find $\sim 2\sigma$ agreement for both angular upweighting and
nearest neighbor redshift, which should be more than adequate for measuring the 
small differences between the true clustering and the clustering estimated from 
those correction methods. 

\subsection{Best estimators derived from tiled mocks and systematic
uncertainties} \label{sec:tiledmock} \begin{figure*}
\includegraphics[width=170mm]{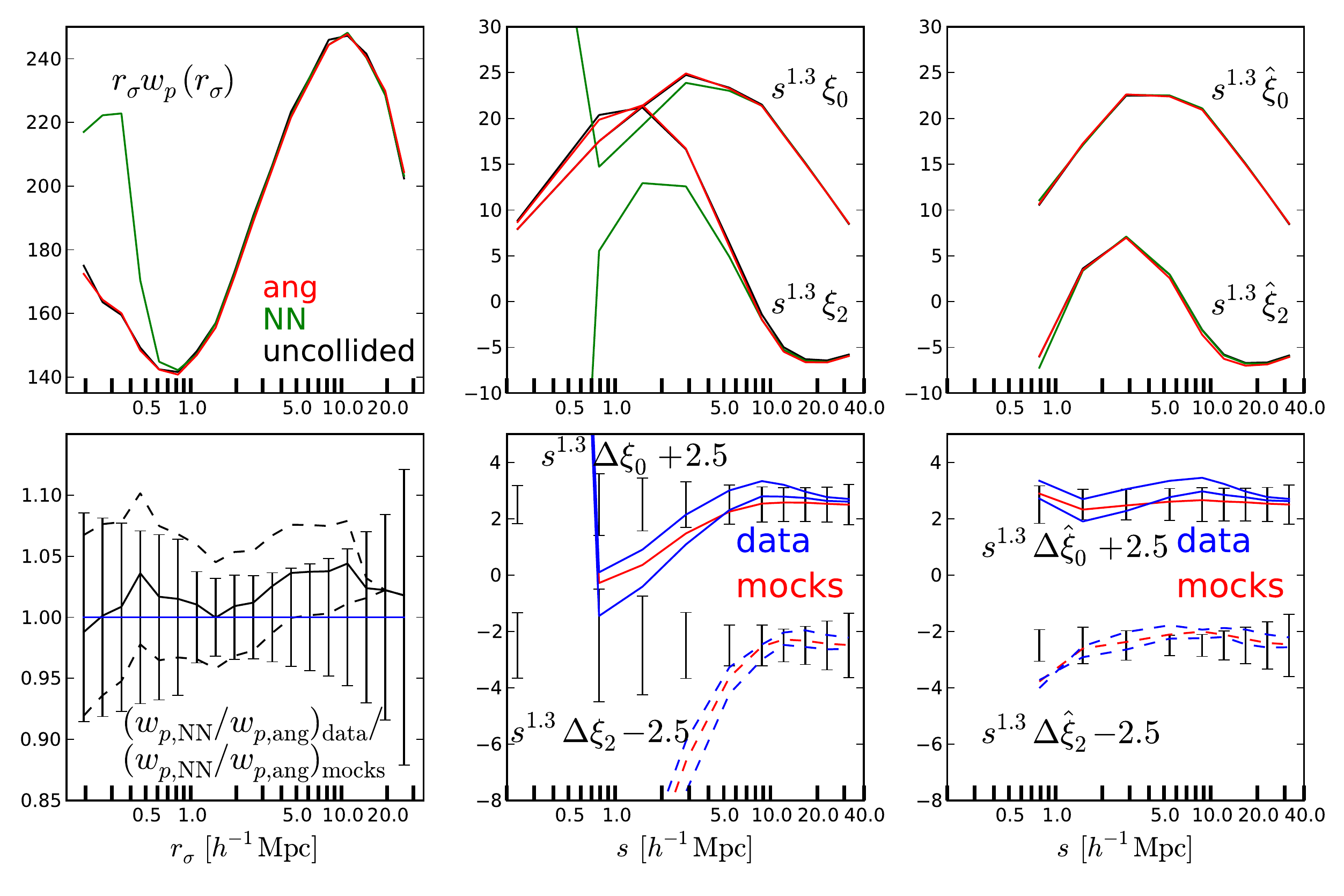} \caption{{\bf Fiber collision correction
validation}.  In the top three panels we show $w_p$, $\xi_{0,2}$, and
$\hat{\xi}_{0,2}$ as measured from our tiled mock catalog in the absence of
fiber collisions (``uncollided''; black), using the angular upweighting method
(``ang''; red), and using the nearest neighbor redshift method (``NN'',green).
The angular upweighting method is quite accurate on all scales,  but subject to
additional uncertainties for the data; we use that method on small scales and
transition to the nearest neighbor redshift method on large scales, which we
expect to be very close to unbiased.  The bottom panels compare the difference
between the angular upweighting and nearest neighbor redshift methods measured
from both the data and the mocks, allowing us to demonstrate consistency between
the data and mock galaxy catalogs.  For $w_p$ we show this comparison as a ratio
(black) along with its uncertainty from the uncertainty in the angular weights
(dashed lines).  In the middle and right bottom panels, the blue curves bracket
the measured difference between the angular upweighting and nearest neighbor
estimators for each of the four statistics $s^{1.3} \xi_{0,2}$ and $s^{1.3}
\hat{\xi}_{0,2}$.  The distance between the two blue curves for each statistic
originates from the uncertainty in deriving the angular weights for the data.
The red curves show the same differences measured from the mocks.  In all three
panels we also show diagonal errors from our final measurements + theory total
covariance matrix presented in Sec.~\ref{sec:covtot}.  We find good agreement
between the data and mocks in all three panels, lending support to our final
fiber collision correction methodology.} \label{fig:mockscompareall}
\end{figure*}

As argued in Sec.~\ref{sec:NNz}, the nearest neighbor redshift assignment method
(``NN'') should provide nearly unbiased clustering estimates on large scales, as
long as one avoids small $r_{\sigma}$ contributions, so we will rely on this
method on scales well above the fiber collision scale.  On smaller scales
angular upweighting (``ang'') method is more accurate.  In
Fig.~\ref{fig:mockscompareall} we compare the underlying ``uncollided'',
complete target catalog (black) with the nearest neighbor redshift (green) and
angular upweighting (red) methods for recovering the underlying clustering.  In
this figure we investigate the three statistics of interest, $w_p$, $\xi_{0,2}$,
and $\hat\xi_{0,2}$.  On small angular scales the angular upweighting method
does indeed recover the underlying clustering at high accuracy for all three
statistics, while the nearest neighbor method is effective only on larger
scales.  The comparison between the middle and right panels illustrates the
large contribution of the scales $r_{\sigma}/D_A(z) < 62''$ to the multipoles
$\xi_{0,2}$.  Eliminating those scales brings the nearest neighbor and angular
upweighting estimators into very good agreement for the $\hat{\xi}_{0,2}$
statistic, which is why we choose to include it rather than $\xi_{0,2}$ in our
parameter fitting.  With this comparison in hand, we define our best estimate of
these statistics from the data in the following way.  For each statistic ($w_p$,
$\xi_0$, $\xi_2$, $\hat{\xi}_{0}$, $\hat{\xi}_{2}$), we use the difference
between the two estimators and the ``truth'' to determine a transition scale at
which we switch from the angular estimator on small scales to the nearest
neighbor redshift estimator on large scales; we find (1.09, 8.8, 12.2, 1.5, 1.5)
$h^{-1}$ Mpc.  Next, we use the difference between the estimator and ``truth''
in the mock catalogs as an estimate of the bias of the observed statistics.  We
subtract this difference from our measurements, and add its square to the
diagonal elements of the bootstrap covariance matrix.  For $w_p$ this correction
is completely negligible.  For $\hat{\xi}$ it is $<0.3\sigma$ except for
$\hat{\xi}_2(1.5 \, h^{-1} {\rm Mpc})$, for which the shift is $0.7\sigma$; here
$\sigma$ refers to the diagonal element of the bootstrap covariance matrix.
This difference is slightly larger than $1\sigma$ for $\xi_2$ in the range
5.4-12.2 $h^{-1}$ Mpc.  For points using the angular upweighting method, we
translate the $\pm 10\%$ uncertainty in the angular weights, shown in
Fig.~\ref{fig:angcorrect}, into an additional uncertainty on the measured
statistics.  We add this source of uncertainty to the diagonal of the bootstrap
covariance matrices, which were computed using fixed angular pair weights.  This
increases the diagonal uncertainty on the clustering by a factor of two or more
for the points affected.  The resulting estimate of $w_p(r_\sigma)$ as well as
the diagonal uncertainty is tabulated in Table \ref{tab:wpmeas}, and
$\hat{\xi}_{0,2}$ and $\xi_{0,2}$ are listed in Table \ref{tab:ximeas}.

In the lower set of plots in Fig.~\ref{fig:mockscompareall}, we compare
differences in all three statistics using the nearest neighbor redshift and
angular upweighting method measured in both the mocks and the data; for
comparison, we show diagonal uncertainties from our final covariance matrix
presented in Sec.~\ref{sec:covtot}.  Differences between these two observables
are determined by the redshift distribution and clustering properties of the
galaxy sample as well as the tiling algorithm.  These differences seen in the
data are reproduced with good accuracy by our mock catalogs.

In the left panel we compare the ratio of $w_{p,{\rm NN}}/w_{p,{\rm ang}}$,
which is large below the fiber collision scale and approaches 1 on large scales
in the mocks.  The data behaves similarly, but with a relatively constant factor
$\sim 1.02$ offset with the mocks.  This offset is within the reported angular
upweighting uncertainty (dashed curves) and also small compared to our final
error budget.  In the middle and right bottom panels we show instead the
difference $s^{1.3} (\xi_{\ell,{\rm NN}} - \xi_{\ell,{\rm ang}})$; the same
offset in $w_p$ is also present in $\xi_0$ and $\hat{\xi}_0$.  The blue curves
in these panels show the allowed region after propagating our 10\% uncertainty
in the angular weights, and the red curves show the differences measured from
the mock catalogs.  They are consistent within the uncertainties.
\begin{table}
\centering
\begin{tabular}{lll|lll}
$r_\sigma$ & $w_p$ & $\sigma_{w_p}$ & $r_\sigma$ & $w_p$ & $\sigma_{w_p}$\\
\hline
0.195 & {\bf 1000.5} & 85.6 & 2.60 & 73.8 & 2.5 \\
0.260 & {\bf 691.7} & 56.2 & 3.46 & 59.8 & 2.2 \\
0.346 & {\bf 507.7} & 39.2 & 4.62 & 48.4 & 2.0 \\
0.462 & {\bf 358.1} & 25.3 & 6.16 & 38.0 & 1.7 \\
0.616 & {\bf 252.4} & 17.1 & 8.22 & 29.6 & 1.4 \\
0.822 & {\bf 179.8} & 11.5 & 10.96 & 22.3 & 1.2 \\
1.096 & 140.3 & 5.2 & 14.61 & 15.9 & 1.1 \\
1.461 & 111.0 & 3.5 & 19.48 & 11.2 & 0.9 \\
1.948 & 88.5 & 3.1 & 25.98 & 7.4 & 0.9 \\
\end{tabular}
\caption{Fiber collision corrected measurements of the projected correlation
function $w_p(r_\sigma)$ for $\pi_{\rm max} = 80$ $h^{-1}$ Mpc defined in
Eq.~\ref{eq:wp}.  The first 9 bins ($r_\sigma < 2$ $h^{-1}$ Mpc; left half of
the table) are included in joint fits with $\hat{\xi}_{0,2}$.  The bolded $w_p$
values are derived using the angular upweighting method while the rest use the
nearest neighbor redshift method.} \label{tab:wpmeas}
\end{table}

\begin{table*}
\centering
\begin{tabular}{lllllllllllll}
$s_{\rm cen}$ & $s_{\rm low}$ & $s_{\rm high}$ & $\mu_{\rm max,low}$ & $\mu_{\rm max,high}$ & $\xi_0$ & $\sigma_{\xi_0}$ & $\hat{\xi}_0$ & $\sigma_{\hat{\xi}_0}$ & $\xi_2$ & $\sigma_{\xi_2}$ & $\hat{\xi}_2$ & $\sigma_{\hat{\xi}_2} $\\
\hline
0.234 & 0.097 & 0.569 & 0.000 & 0.000 & {\bf 57.757} & 4.428 & - & - & {\bf 60.471} & 7.699 & - & - \\
0.785 & 0.569 & 1.084 & 0.345 & 0.845 &  {\bf 24.665} &  1.497 &  {\bf 14.339} &  0.913 &  {\bf 30.726} &  2.744 &  {\bf -8.346} &  0.767 \\
1.496 & 1.084 & 2.065 & 0.870 & 0.955 &  {\bf 12.841} &  0.555 &  10.363 &  0.317 &  {\bf 11.687} &  1.037 &  1.094 &  0.382 \\
2.851 & 2.065 & 3.936 & 0.965 & 0.985 &  {\bf 6.403} &  0.208 &  5.865 &  0.139 &  {\bf 3.811} &  0.300 &  1.309 &  0.134 \\
5.433 & 3.936 & 7.499 & 0.990 & 0.995 &  {\bf 2.559} &  0.077 &  2.503 &  0.063 &  {\bf 0.633} &  0.080 &  0.212 &  0.040 \\
8.810 & 7.499 & 10.351 & 0.995 & 0.995 &  1.273 &  0.037 &  1.244 &  0.035 &  {\bf -0.074} &  0.043 &  -0.197 &  0.023 \\
12.162 & 10.351 & 14.289 & 0.995 & 0.995 &  0.707 &  0.023 &  0.696 &  0.023 &  -0.185 &  0.023 &  -0.227 &  0.020 \\
16.788 & 14.289 & 19.724 & 0.995 & 0.995 &  0.377 &  0.015 &  0.374 &  0.015 &  -0.151 &  0.017 &  -0.164 &  0.017 \\
23.174 & 19.724 & 27.227 & 0.995 & 0.995 &  0.190 &  0.010 &  0.189 &  0.010 &  -0.103 &  0.015 &  -0.107 &  0.014 \\
31.989 & 27.227 & 37.584 & 0.995 & 0.995 &  0.088 &  0.008 &  0.088 &  0.008 &  -0.062 &  0.013 &  -0.063 &  0.012 \\
\end{tabular}
\caption{Fiber collision corrected measurements of $\xi_{0,2}$ and
$\hat{\xi}_{0,2}$.  The first column is the logarithmic bin center used in all
plots.  The minimum and maximum redshift space separations in each bin are
listed as $s_{\rm low}$ and $s_{\rm high}$, and  the corresponding maximum
$\mu_{\rm max}$ for $\hat{\xi}_{0,2}$ (see Eq.~\ref{eq:hatmoments}) are listed
as $\mu_{\rm max,low}$ and $\mu_{\rm max,high}$ (recall $\mu_{\rm max}$ is
allowed to vary with $s$).  We use $r_{\sigma} < 0.534$ $h^{-1}$ Mpc to define
$\mu_{\rm max}$, which corresponds to 62'' at $z=0.7$ in the cosmology with
$\Omega_m = 0.274$ used to compute comoving pair separations.  For $\xi_{0,2}$,
$\mu_{max} = 1$ for all $s$ bins.  The latter columns show our fiber-collision
corrected estimates of $\xi_{0,2}$ and $\hat{\xi}_{0,2}$ defined in
Eq.~\ref{eq:moments} and ~\ref{eq:hatmoments} as well as the diagonal elements
of the total (measurements + theory) covariance matrix.  The bolded $\xi_{0,2}$
and $\hat{\xi}_{0,2}$ values are derived using the angular upweighting method
while the rest use the nearest neighbor redshift method.} 
\label{tab:ximeas}
\end{table*}

\subsection{Combined measurement and theory covariance matrix}
\label{sec:covtot}
So far we have accounted for three sources of uncertainty in our measurements:
the standard finite volume sampling, for which we estimate a full measurement
covariance matrix using the 200 bootstrap regions in Fig.~\ref{fig:bootregions};
10\% uncertainty in the angular weights used in the angular upweighting method,
which we propagate to the observables of interest and then add to the diagonal
elements of the covariance matrix; and systematic uncertainty equal to the size
of debiasing correction derived from the mock tiling catalog and added to the
diagonal elements of the covariance matrix.  The combination of these three
terms we call our ``measurement'' uncertainty.  One final source of statistical
error comes from uncertainty in our theoretical prediction.  The total volume
within the DR10 survey mask after applying our redshift cuts is $\sim 2.5$
($h^{-1}$ Gpc)$^3$, though some of this volume is mapped with a low number
density of galaxies.  The HiRes $N$-body simulation box, for which we estimate
the theoretical error, covers only $\sim 1/8$ of that volume.  Our theoretical
calculation averages over all possible realizations of a particlar HOD (for the
given halo catalog), which removes much of the sampling variance from the
theoretical calculation.  Cosmic variance assoicated with the underlying dark
matter realization remains, but the theoretical error will be smaller than the
naive volume comparison suggests. 

\begin{figure*} \includegraphics[width=85mm]{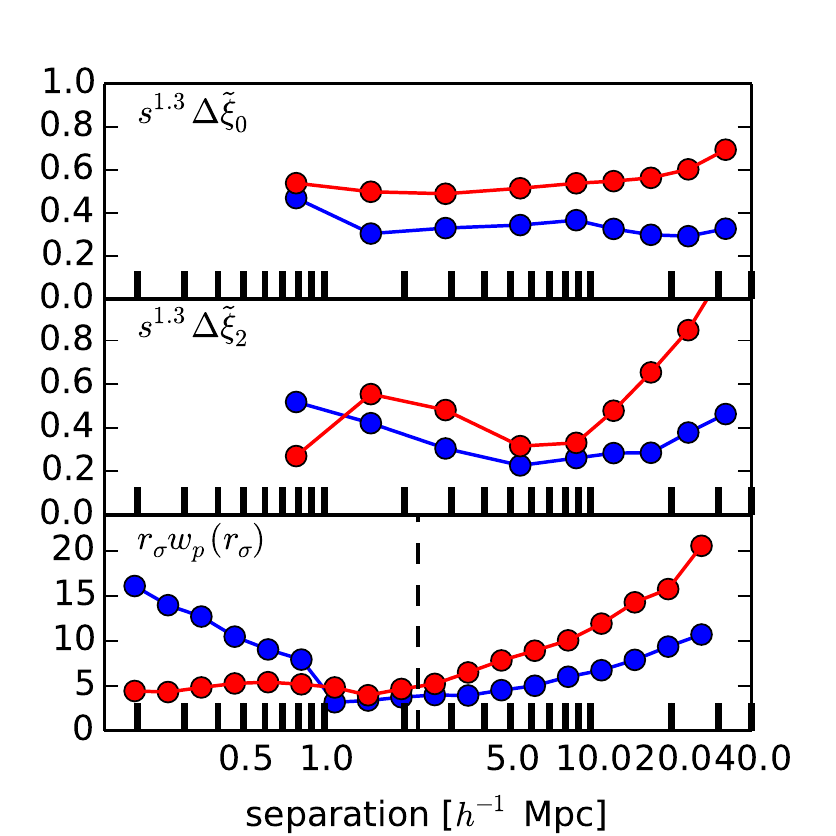}
\includegraphics[width=85mm]{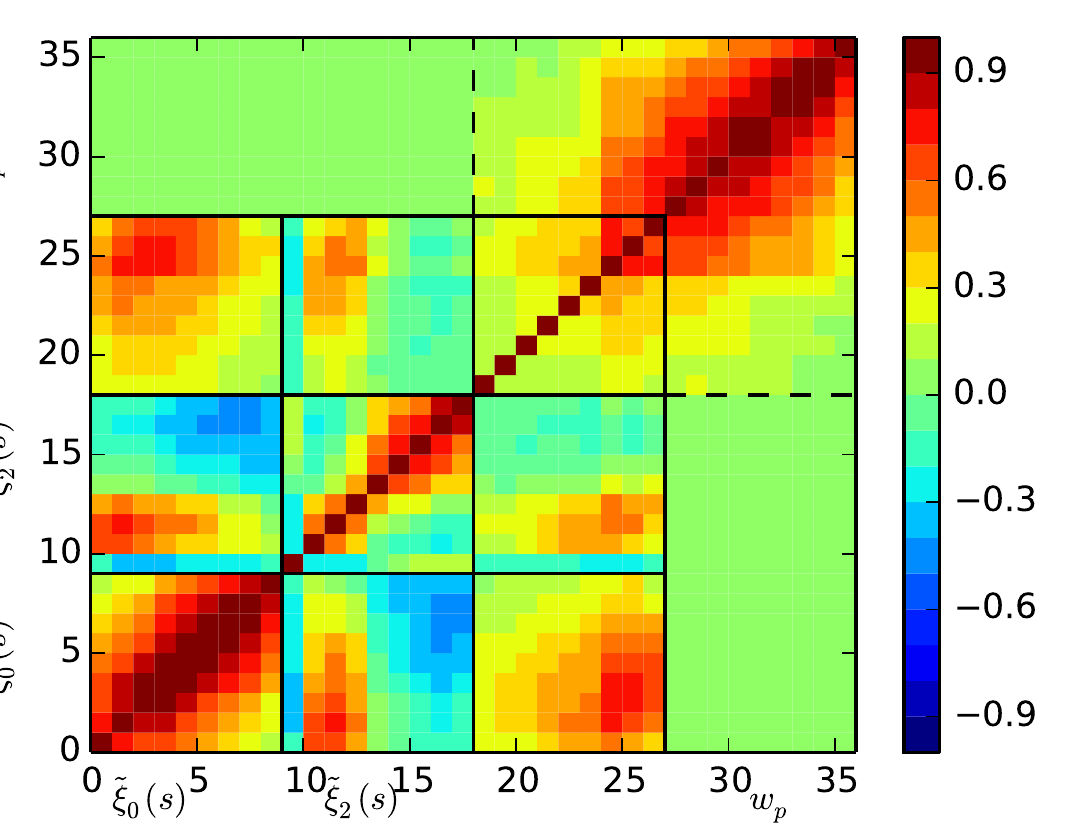} \caption{{\em Left panel:} The diagonal
elements of the measurement covariance matrix (blue) and the theory covariance
matrix (red).  On small scales measurement errors are large due to the 10\%
uncertainty in the angular weights used in the angular upweighting method for
fiber collision corrections.  Theory errors are due to our use of an $N$-body
simulation box smaller than the observational volume and dominate the error
budget on most scales.  Only measurements of $w_p$ below the dashed line are
included in our joint fits to $\hat{\xi}_{0,2}$+$w_p$.  {\em Right panel:} The
reduced total covariance matrix for $\hat{\xi}_{0,2}$+$w_p$ (first 27 elements).
We also show the full $w_p$ covariance out to larger scales, though those data
points are not included in the joint fits.  Off-diagonal elements between
$\hat{\xi}_{0,2}$ and large scale $w_p$ are artificially set to 0 in this plot.
We overlay black lines that divide the $\hat{\xi}_0$, $\hat{\xi}_2$, and $w_p$ 
sections of the covariance matrix into three blocks of nine measurements each.  
Only these points are used to fit the parameters of the model.}
\label{fig:covfinal} \end{figure*}

In order to estimate the theoretical uncertainty, we populate the LowRes
simulation halo catalogs with the same HOD as in Sec.~\ref{sec:mocksoln}.  We
divide the box into $4^3$ subboxes the size of the HiRes box, which allows for a
$10$ $h^{-1}$ Mpc buffer between subboxes in each direction.  We again include
the factor in Eq.~\ref{eq:Hartlapfac} with $n_{\rm boot} = 64$ to unbias the
inverse bootstrap covariance matrix estimate; for $n_{\rm bin} = 27$, the
prefactor is 1.8.  More simulation volume could reduce both the theoretical
uncertainties and covariance, but is only justified if we have not reached a
systematics floor in the theoretical modeling.  The left panel of
Fig.~\ref{fig:covfinal} compares the diagonal elements of the final
``measurement'' uncertainties with the theoretical uncertainties.  We sum the
measurement and theory covariance matrices to arrive at our final covariance
matrix that we will invert and use in a standard $\chi^2$ analysis to do model
parameter fitting.  In the right panel, we show the correlation matrix for the
$C_{\rm tot} = C_{\rm meas} + C_{\rm theory}$.  As is true on large scales as
well, neigboring bins in $\hat{\xi}_0$ are highly correlated, meaning the data
is relatively insensitive to overall changes in the amplitude of clustering, but
more sensitive to spatially-abrupt model signatures (like the BAO feature).  The
$\hat{\xi}_2$ bins are less correlated than $\hat{\xi}_0$ for large separations,
and there is significant covariance between all three observables.  At large
separations, $\hat{\xi}_2$ becomes negative, so a positive correlation in the
amplitude of the multipoles (as we would expect from uncertainty in an overall
bias factor) shows up in that region as an anti-correlation.

\section{Model} \label{sec:HODmodel} The only detailed semi-analytic HOD based
descriptions of galaxy clustering in redshift space available to our knowledge
are given in \citet{Tinker07} and \citet{Zu12}.  The models presented therein
require a description of the probability distribution of pairwise halo
line-of-sight velocities as a function of their real space separation,
orientation with respect to the line of sight, and the two halo masses.  These
distributions have substantial skewness and kurtosis that depends on pair
separation and halo masses.  These semi-analytic models require calibration of
several scaling relations against $N$-body simulations; fine-tuning or extending
it to reach the precision demanded by our measurements would likely be extremely
challenging. 

Throughout the present analysis we therefore resort to deriving our theoretical
predictions directly from mock galaxy catalogs based on $N$-body simulations, as
detailed below.  The disadvantage of this approach is that cosmological
parameter dependences are not easily incorporated, and the theory evaluation
must be fast enough to permit at least a five-dimensional Monte-Carlo HOD
parameter exploration.  Following \citet{Neistein12} we implemented a
pre-calculation of pair counts in fine mass bins; sums over these counts allow
fast evaluation of the theoretical prediction as a function of HOD parameters.
However, parameters that alter the velocity of galaxies change all pair
separations and therefore require recalculation of the pair counts.  We explored
interpolation of the pair counts across the set of three velocity parameters
described in Sec.~\ref{sec:HODp}.  While useful for determining parameter
degeneracies and expected uncertainties, the resulting constraints were not
sufficiently accurate given the coarseness of the velocity parameter sampling.
We therefore resort to varying one or at most two velocity parameters
simultaneously. 

\subsection{Halo and central velocities} \label{sec:halovel} \begin{figure}
\includegraphics[width=85mm]{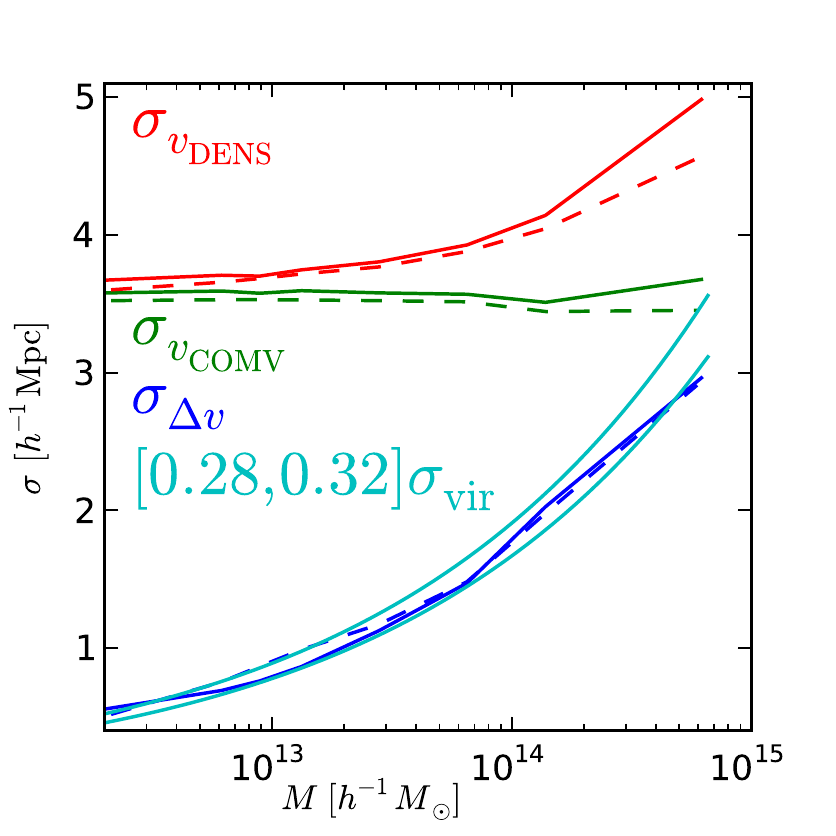} \caption{The rms velocities as a function
of halo mass for the two central velocity definitions (``DENS'' in red and
``COMV'' in green), as well as their difference (blue).  Solid lines are derived
from the HiRes simulation while dashed lines are from the MedRes simulation.
The difference between those vectors has a magnitude consistent with $[0.3 \pm
0.02] \sigma_{\rm vir}$, shown by the cyan curves.} \label{fig:comparevcatsvsM}
\end{figure} \begin{figure*} \includegraphics[width=85mm]{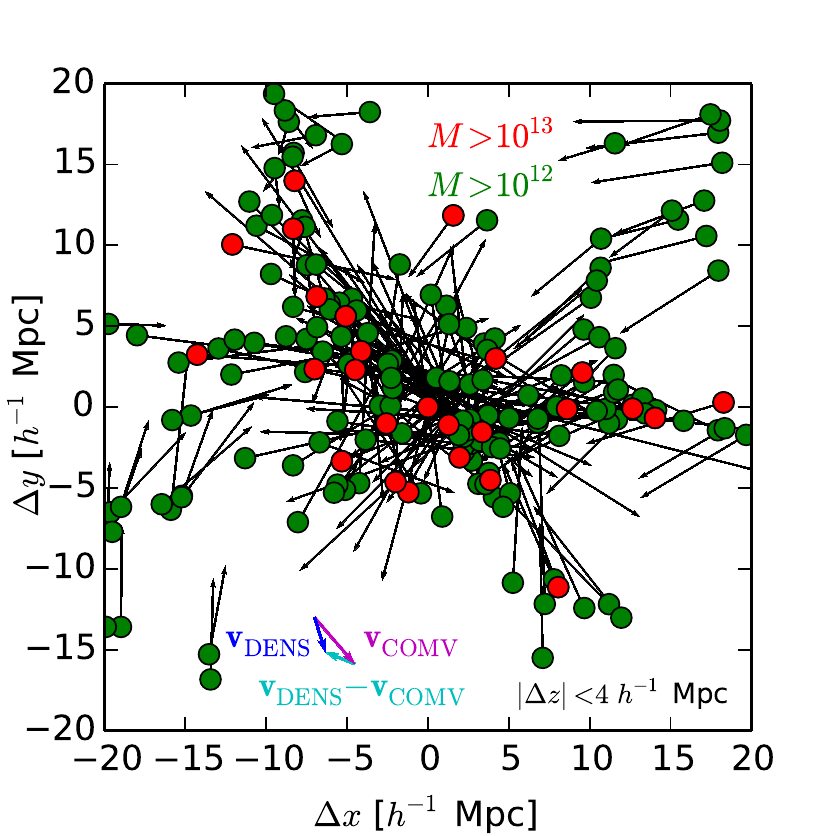}
\includegraphics[width=85mm]{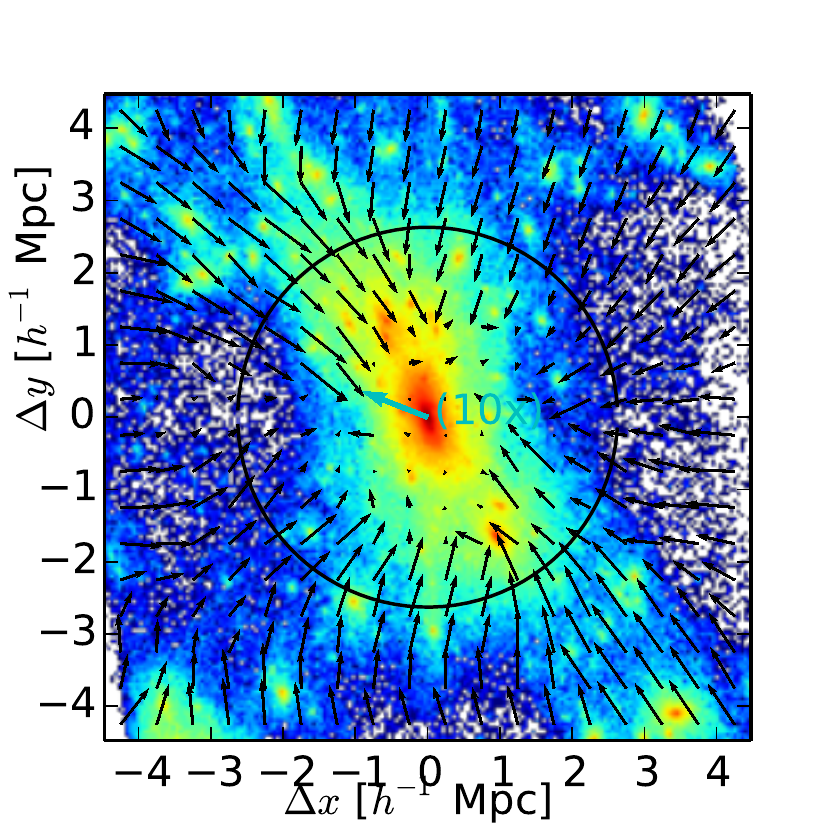} \caption{{\em Left:} A $|\Delta z| < 4$
$h^{-1}$ Mpc slice through the HiRes simulation box, centered on the largest
halo in the box with $M = 1.3 \times 10^{15} \, h^{-1} M_{\odot}$.  Green (red)
dots indicate the positions of halos of mass $M > 10^{12} \, (10^{13})$ $h^{-1}
M_{\odot}$, where all positions have been projected into the plane determined by
the vectors ${\bf v}_{\rm DENS}$ (blue) and ${\bf v}_{\rm COMV}$ (magenta).  The
black arrows indicate the velocity of each halo (in distance units) relative to
${\bf v}_{\rm COMV}$, so that the central halo center of mass is at rest.  {\em
Right:} A zoomed-in version of the left panel with the log of the matter density
over-plotted along with the central halo virial radius $r_{\rm vir} = 2.7$
$h^{-1}$ Mpc.  The matter velocity field is over-plotted in black alongside the
halos; the velocity vectors in this panel were scaled down by a factor of 20 for
visualization purposes.  The central cyan vector shows ${\bf v}_{\rm DENS} -
{\bf v}_{\rm COMV}$, scaled down by a factor of only two (so expanded by a
factor of 10 compared to the other vectors).  The inward flow from the upper
left corner pushes ${\bf v}_{\rm COMV}$ along the $+\hat{e}_x$ compared with
${\bf v}_{\rm DENS}$.  The clear correlation between the density field and
central galaxy velocity will be imprinted differently on $\hat{\xi}_{0,2}$ than
if ${\bf v}_{\rm DENS}$ - ${\bf v}_{\rm COMV}$ were randomly oriented.}
\label{fig:simslice} \end{figure*} For a given SO halo catalog, we consider two
definitions of halo velocities; these velocities are assigned directly to
``central'' galaxies, and the intrahalo velocity component for satellite
galaxies is defined with respect to this halo velocity.  The first choice is to
simply average the velocities of all the halo members, denoted ${\bf v}_{\rm
COMV}$, for center-of-mass velocity.  The dispersion of halo member velocities
around the center-of-mass velocity is \begin{equation} \sigma_{\rm vir} = 2.79
h^{-1} \, {\rm Mpc}\left(\frac{M}{10^{13} h^{-1} M_{\sun}}\right)^{0.331},
\label{eq:vvir} \end{equation} fit to halos in the HiRes box; the HiRes and
MedRes dispersions agree within 2\% with this relation, the LowRes box within
5\%.  The three are in per-cent level agreement above $10^{14} h^{-1} M_{\sun}$.
Therefore, within the range accessible to this study, the intrahalo velocity
dispersions are independent of both cosmology and simulation resolution within a
few per-cent, at fixed SO halo mass.   The green curves in
Fig.~\ref{fig:comparevcatsvsM} show that the rms center-of-mass halo velocity
$\sigma_{\rm COMV}$ is remarkably independent of halo mass (within 2\% of 3.57
$h^{-1} \, {\rm Mpc}$ for $10^{12-15} h^{-1} M_{\sun}$ halos in the HiRes box).
The MedRes ${\sigma}_{\rm COMV}$ is lower by a factor of 1.016, in reasonable
agreement with the linear theory expectation of 1.021 given the ratio of the
values of $f\sigma_8$ for the two boxes.

The second central velocity definition, ${\bf v}_{\rm DENS}$, was defined
precisely in Sec.~\ref{sec:Nbody}, and the sensitivity to this definition 
was explored in more detail in Appendix \ref{sec:vcen}. 
Note that in both catalogs we use that same density peak to define the halo
center, where we place the ``central'' galaxy, so positions in the two halo
catalogs we compare are identical; only the ``central'' galaxy velocities are
different.  Fig.~\ref{fig:comparevcatsvsM} shows that the magnitude of ${\bf
v}_{\rm DENS}$ rises with halo mass.  If we consider the difference vector ${\bf
v}_{\rm COMV} - {\bf v}_{\rm DENS}$, we get the blue curves in
Fig.~\ref{fig:comparevcatsvsM}.  We see that $|{\bf v}_{\rm COMV} - {\bf v}_{\rm
DENS}|$ depends on mass in the same way as the halo virial velocity
(Eq.~\ref{eq:vvir}), but the magnitude is smaller by a factor of 0.3.  

Fig.~\ref{fig:simslice} illustrates these velocity vectors in the local
environment of the largest halo in the HiRes simulation, which has $M_{\rm halo}
= 1.3 \times 10^{15} \, h^{-1} M_{\odot}$.  The real space coordinates have been
shifted to place the halo at the (0,0) and projected into the plane defined by
${\bf v}_{\rm DENS}$ (blue) and ${\bf v}_{\rm COMV}$ (magenta).  In the left
panel we take a $|\Delta z| < 4$ $h^{-1}$ Mpc slice around the central halo and
plot the positions (dots) and velocities (relative to ${\bf v}_{\rm COMV}$ of
the central halo; black vectors) of halos within $\pm 20$ $h^{-1}$ Mpc in
$\Delta x$ and $\Delta y$.  The relative velocity of the dense clump of the
central halo is shown as the cyan vector.  In the right panel we examine the
virial region (marked by the black circle) and surrounding structure with a log
mapping of the density field.  Mean matter velocities are shown with black
arrows, scaled down by a factor of 20 for visualisation purposes.  The net
offset ${\bf v}_{\bf DENS} - {\bf v}_{\bf COMV}$ is shown as the cyan vector,
scaled down only by a factor of two; it is inherently much smaller than the
infall region velocities, and is correlated with a major filamentary structure.
The correlation will be imprinted in $\hat{\xi}_{0,2}$ since the relative
velocity will preferentially move pairs along the filamentary structure and thus
preferentially along their separation vector. 

One final point of interest in comparing these vectors is that the difference
vector contains a component along the ${\bf v}_{\rm COMV}$ direction, such that
the magnitude of ${\bf v}_{\rm DENS}$ is larger in that direction by 1.5\%.
This provides a ballpark upper limit on how much the central galaxy velocity
details may alter the effective $f\sigma_8$ on large scales, if the correlation
is sourced by the quasilinear velocity component driving the large-scale Kaiser
distortions.  We propagate these two velocity choices to galaxy clustering
predictions in Sec.~\ref{sec:vCOMVvDENS}.  Further investigation is warranted
beyond these two choices.  However, given the good agreement between our
``central'' velocity definition and the more detailed phase-space investigation
given in \citet{Behroozi13}, we assert that ${\bf v}_{\rm DENS}$ is the more
physical choice of the two.

\subsection{Number Density Prior and Redshift Evolution} \label{sec:HODnbar} Our
HOD model based on a fixed redshift $N$-body simulation halo catalog can only be
an approximation to the real CMASS galaxy sample, for which the number density
$\bar{n}(z)$ varies considerably across the redshift range of our sample;
potentially the galaxy properties are redshift dependent as well \citep[see
earlier work on this topic in][]{Masters11,Ross12,Ross14}.  Remarkably, in
Appendix \ref{sec:zdep} we find that there is no measurable redshift evolution
in the $\hat{\xi}_{0,2}$ statistic across the sample, even though the number
density drops by a factor of 2.2 in the high redshift sample.  We therefore take
the simple ansatz that galaxies at all redshifts are a random subsample drawn
from a single population.  The observed $\bar{n}(z)$ simply reflects the
fraction of the parent population selected by the CMASS targeting algorithm as a
function of redshift.  While a more complete model would allow all the HOD
parameters to vary with redshift to match the observed $\bar{n}(z)$, the data do
not require it and it cannot be done without considerably increasing the
complexity of our theoretical calculation (i.e., requiring the generation of
light-cone halo catalogs).

\begin{figure} \includegraphics[width=85mm]{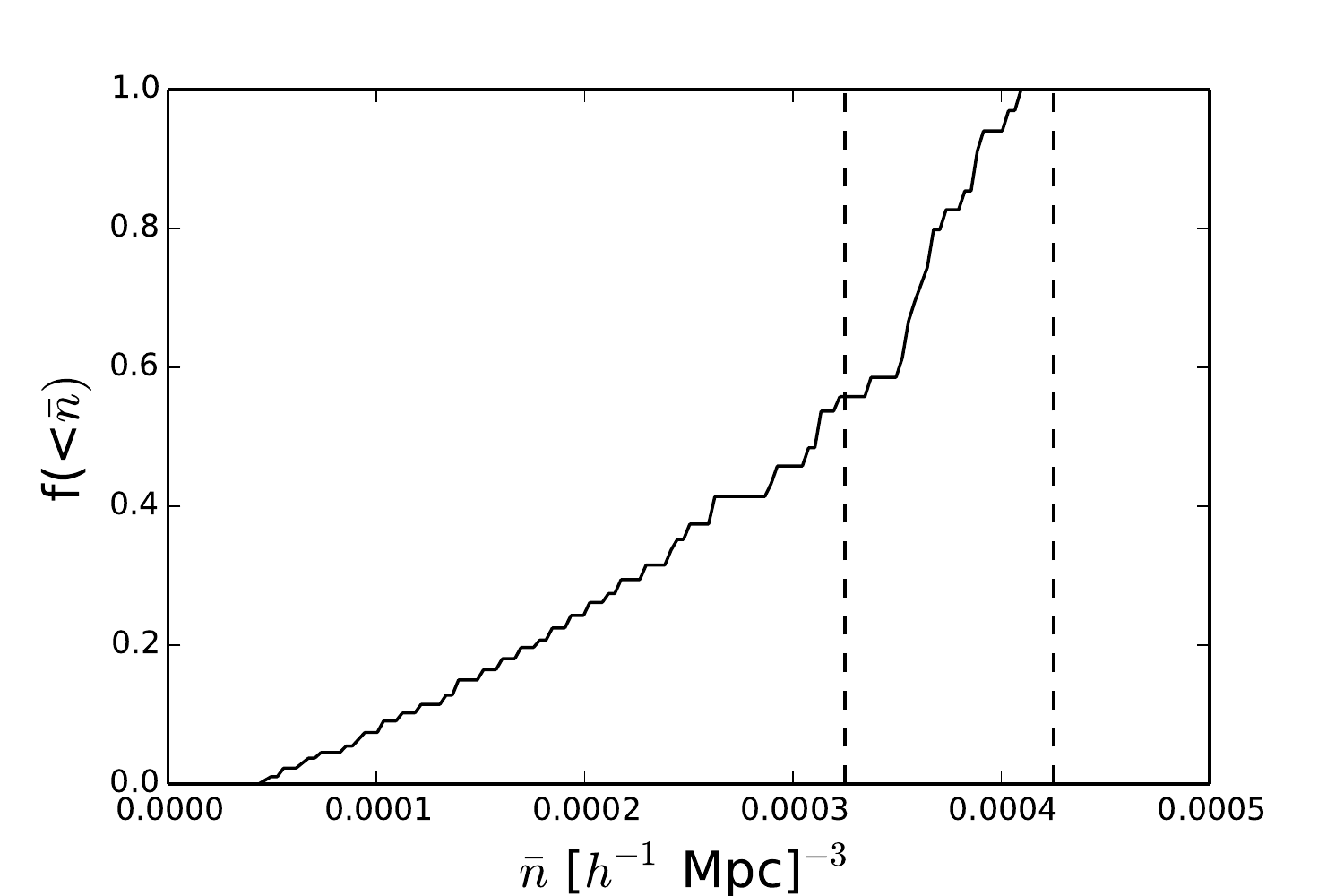} \caption{Cumulative
distribution of $\bar{n}$ for the CMASS sample with redshift cut $0.43 < z <
0.7$.  The vertical dashed line shows the fiducial hard prior adopted when
fitting HOD models, $3.25 < 10^4 \bar{n}_{\rm HOD} < 4.25$ in units of ($h^{-1}$
Mpc)$^{-3}$.} \label{fig:cumnbar} \end{figure}

Fig.~\ref{fig:cumnbar} shows the cumulative probability distribution of
$\bar{n}(z)$ at the redshift of the galaxies in CMASS after applying the
redshift cut $0.43 < z < 0.7$.  The vertical lines show the hard prior we
assumed when fitting the single underlying HOD, $3.25 < 10^4 \bar{n}_{\rm HOD} <
4.25$ in units of ($h^{-1}$ Mpc)$^{-3}$.  The lower bound is set by requiring
the parent population to have higher number density than the typical CMASS
galaxy.  The upper bound depends on the completeness of our target selection,
which in turn depends on the size of the photometric uncertainties in the
imaging data.  We have chosen a value safely above the peak of the number
density distribution for the fiducial case, and will demonstrate in
Sec.~\ref{sec:robustresults} that our constraints on $f\sigma_8$ are insensitive
to this choice.  As discussed in Sec.~\ref{sec:robustresults}, the observed
galaxy clustering amplitude and the abundance of sufficiently highly biased
halos sets a hard upper limit of $\approx 6 \times 10^{-4}$ ($h^{-1}$
Mpc)$^{-3}$.

\subsection{Halo Occupation Distribution (HOD) parameters and implementation}
\label{sec:HODp} The parametrisation of the HOD we adopt follows
\citet{Zheng05}, and has been used in a number of studies focusing on the
SDSS-II Luminous Red Galaxy sample \citep{Reid09}, the SDSS-III CMASS sample
\citep{White11}, and the SDSS-III LOWZ sample \citep{Parejko11}.  We separately
model central and satellite galaxies, assuming that a central galaxy is required
for a given halo to host a satellite galaxy.  We model the probability for a
halo of mass $M$ to host a central galaxy as \begin{equation} \label{eq:Ncen}
N_{\rm cen}(M) = 0.5 \left[1+{\rm erf}\left(\frac{\log_{10} M - \log_{10} M_{\rm
min}}{\sigma_{\log_{10} M}}\right) \right].  \end{equation} In our default
model, the central galaxy is assigned to the position and velocity (${\bf
v}_{\rm DENS}$) of the density peak of its host dark matter halo.  We also test
a model with the same position assignment, but set the central velocity to ${\bf
v}_{\rm COMV}$.  Given that a given halo of mass $M$ hosts a central galaxy, the
number of satellites assigned to the halo is drawn from a Poisson distribution
with mean \begin{equation} N_{\rm sat}(M) = \left(\frac{M - M_{\rm
cut}}{M_1}\right)^{\alpha}.  \end{equation} We set $N_{\rm sat}(M < M_{\rm cut})
= 0$.  The average total number of galaxies in a halo of mass $M$ is then
\begin{equation} \left<N_{\rm gal}(M)\right> = N_{\rm cen}(M) \left(1 + N_{\rm
sat}(M)\right).  \end{equation} For each satellite galaxy we assign the position
and velocity of a randomly chosen dark matter particle member of the host halo.
When fitting for the HOD parameters we sample $\log_{10} M_{\rm min}$,
$\log_{10} M_{1}$, and  $\log_{10} M_{\rm cut}$ so all masses are constrained to
be larger than 0.

We also introduce three new parameters that rescale the galaxy velocities
without altering their positions.  \begin{itemize} \item $\gamma_{\rm HV}$: This
parameter rescales all halo velocities in the simulation.  If linear theory were
accurate on all scales, a fractional change in $\gamma_{\rm HV}$ would be
equivalent to a fractional change in the large-scale peculiar velocity field
amplitude, $f\sigma_8$.  In Sec.~\ref{sec:cosmodep} we demonstrate the validity of
this approximation for relative halo velocities even on non-linear scales.  
Our constraints on $\gamma_{\rm HV}$ are derived by interpolating across the theoretical 
model evaluated between $\gamma_{\rm HV} = 0.9 - 1.1$ in steps of 0.01.  For our fiducial 
fit, the lower bound is $\sim 2.5\sigma$ away from the best fit. 
\item
$\gamma_{\rm IHV}$: This parameter rescales the velocity of satellite galaxies
relative to the host halo.  Conceptually, this amounts to rescaling the virial
velocity of the halo and/or accounting for sub-halo/galaxy velocity bias
effects.  \item $\gamma_{\rm cenv}$: This parameter specifies an additional
random (Gaussian) dispersion for central galaxies in units of the halo virial
velocity.  \end{itemize}

\subsection{Interpreting $\hat{\xi}_{0,2}$ in the halo model}
\label{sec:HODinterp} \begin{figure} \includegraphics[width=85mm]{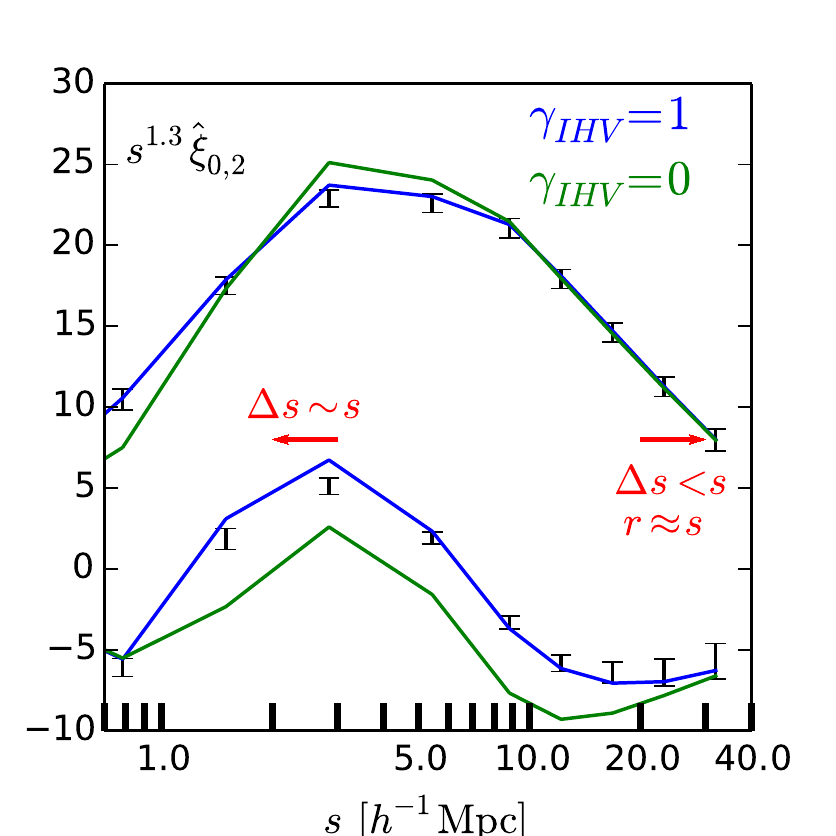}
\caption{Best fit model to the MedRes simulation for $\gamma_{\rm HV} = 1$,
$\gamma_{\rm IHV} = 1$, $\gamma_{\rm cenv} = 0$ (blue) compared with the same
HOD model but with the satellite velocities artificially set to 0 ($\gamma_{\rm
IHV} = 0$; green) and their positions set to the position of the central galaxy.
The satellite velocity dispersion distorts $\hat{\xi}_0$ only below 10 $h^{-1}$
Mpc.  Removing the satellite intrahalo velocities lowers the quadrupole
significantly on all scales of interest.  For separations of $\sim 3$ $h^{-1}$
Mpc and below, the difference between true and apparent (redshift space)
separations is comparable or larger than $s$ and sets the scale of the
transition from positive to negative for the  $\gamma_{\rm IHV}=0$ model.  On
larger scales, true and apparent separations are similar, since $\Delta s$ is
small compared to $s$.  For comparison our measurements are shown with black
error bars.} \label{fig:xi02theory} \end{figure}

One prime advantage of the $w_p$ observable is that the x-axis (projected
separation $r_\sigma$) draws a neat separation between the two physical
components of the halo model; the majority of pairs with $r_\sigma$ smaller than
the virial radius of the typical host halo actually occupy the same host halo
(the ``one-halo'' term); on larger scales, pairs originate from different host
halos (the ``two-halo'' term).  Such a distinction is impossible at small
redshift space separations $s$.  Although an extreme example,
Fig.~\ref{fig:simslice} illustrates the large intrahalo velocities on small
scales, and how they blur the underlying small-scale structure.  Using the
results of \citet{ReiWhi11}, we find that halos of mass $\sim 10^{13.1}$
$h^{-1}$ $M_{\sun}$ have a mean relative pairwise velocity of $\sim -2.7$
$h^{-1}$ Mpc at separations below 10 $h^{-1}$ Mpc and a pairwise velocity
dispersion that increases with scale, also with an rms of $\sim 2.7$ $h^{-1}$
Mpc at true (real space) halo separations of 5 $h^{-1}$ Mpc.  Thus we expect
that below scales of $\sim 3$ $h^{-1}$ Mpc, there is little correlation between
the observed redshift pair separation $s$ and the true one; as shown in
Fig.~\ref{fig:xi02theory}, this is the scale where $\hat{\xi}_2$ transitions
from positive to negative for the underlying halo distribution (see the green
$\gamma_{\rm IHV} = 0$ model curve where satellite velocities have artificially
been set to 0).

Fig.~\ref{fig:xi02theory} also shows that there is a plethora of information in
$\hat{\xi}_2$ on the satellite galaxy velocity dispersion.  The satellite
velocity dispersion distorts $\hat{\xi}_0$ only below 10 $h^{-1}$ Mpc.  Removing
the satellite intrahalo velocities lowers the quadrupole significantly on all
scales of interest.  The $\chi^2$ difference between an HOD model fit to the
data and the same model without satellite velocity distpersions is 400.

\section{Results}
\label{sec:results}
\begin{table*}
\centering
\begin{tabular}{lllllllll}
\hline
  & {\bf fiducial} & HiRes & HiRes & MedRes & COMV & COMV & high $\bar{n}_{\rm HOD}$\\
\hline
$\log_{10} M_{\rm min}$ & ${\bf 13.031 \pm 0.029 }$ & $ 13.055 \pm 0.022 $ & $ 13.089 \pm 0.027 $ & $ 13.004 \pm 0.025 $ & $ 13.152 \pm 0.027 $ & $ 13.027 \pm 0.027 $ & $ 12.926 \pm 0.022 $\\
$\sigma_{\log_{10} M}$ & ${\bf 0.38 \pm 0.06 }$ & $ 0.31 \pm 0.05 $ & $ 0.38 \pm 0.05 $ & $ 0.32 \pm 0.07 $ & $ 0.61 \pm 0.03 $ & $ 0.37 \pm 0.06 $ & $ 0.16 \pm 0.12 $\\
$\log_{10} M_{\rm cut}$ & ${\bf 13.27 \pm 0.13 }$ & $ 13.43 \pm 0.13 $ & $ 13.36 \pm 0.13 $ & $ 13.27 \pm 0.14 $ & $ 13.07 \pm 0.15 $ & $ 13.19 \pm 0.13 $ & $ 13.01 \pm 0.58 $\\
$\log_{10} M_1$ & ${\bf 14.08 \pm 0.06 }$ & $ 14.33 \pm 0.32 $ & $ 14.24 \pm 0.18 $ & $ 14.09 \pm 0.07 $ & $ 14.05 \pm 0.04 $ & $ 14.05 \pm 0.04 $ & $ 14.09 \pm 0.05 $\\
$\alpha$ & ${\bf 0.76 \pm 0.18 }$ & $ 0.40 \pm 0.22 $ & $ 0.53 \pm 0.22 $ & $ 0.73 \pm 0.20 $ & $ 1.03 \pm 0.13 $ & $ 0.90 \pm 0.14 $ & $ 0.93 \pm 0.22 $\\
$\bar{n}_{\rm HOD}$ & ${\bf 4.12 \pm 0.13 }$ & $ 4.14 \pm 0.11 $ & $ 4.08 \pm 0.16 $ & $ 4.16 \pm 0.09 $ & $ 4.05 \pm 0.17 $ & $ 4.14 \pm 0.11 $ & $ 4.64 \pm 0.11 $\\
$f_{\rm sat}$ & ${\bf 0.1016 \pm 0.0069 }$ & $ 0.0997 \pm 0.0068 $ & $ 0.1015 \pm 0.0069 $ & $ 0.1015 \pm 0.0071 $ & $ 0.1038 \pm 0.0065 $ & $ 0.1037 \pm 0.0072 $ & $ 0.1152 \pm 0.0076 $\\
$f\sigma_8$ & ${\bf 0.452 \pm 0.011 }$ & ${\bf 0.482 }$ & $ 0.449 \pm 0.006 $ & ${\bf 0.472 }$ & ${\bf 0.472 }$ & ${\bf 0.472 }$ & ${\bf 0.472 }$\\
$\gamma_{\rm IHV}$ & ${\bf 1.00 }$ & ${\bf 1.00 }$ & ${\bf 1.00 }$ & ${\bf 1.00 }$ & ${\bf 1.00 }$ & ${\bf 1.00 }$ & ${\bf 1.00 }$\\
$\gamma_{\rm cenv}$ & ${\bf 0.00 }$ & ${\bf 0.00 }$ & ${\bf 0.00 }$ & ${\bf 0.00 }$ & ${\bf 0.00 }$ & ${\bf 0.30 }$ & ${\bf 0.00 }$\\
$\chi^2_{w_p}$ (18) & ${\bf 12.4 }$ & $ 9.5 $ & $ 9.7 $ & $ 11.5 $ & $ 28.9 $ & $ 15.5 $ & $ 8.6 $\\
$\chi^2_{\hat{\xi}_{0,2}}$ (18) & ${\bf 27.5 }$ & $ 31.0 $ & $ 24.4 $ & $ 30.6 $ & $ 65.0 $ & $ 49.4 $ & $ 27.1 $\\
$\chi^2_{w_p+\hat{\xi}_{0,2}}$ (27) & ${\bf 32.3 }$ & $ 34.1 $ & $ 26.4 $ & $ 36.8 $ & $ 68.5 $ & $ 50.0 $ & $ 30.0 $\\
\hline
  & MedRes1 & MedRes2 & high $\bar{n}_{\rm HOD}$ & cen/sat test & MedRes0 & MedRes0 & MedRes0\\
\hline
$\log_{10} M_{\rm min}$ & $ 13.035 \pm 0.032 $ & $ 13.037 \pm 0.030 $ & $ 12.951 \pm 0.030 $ & $ 12.983 \pm 0.060 $ & $ 13.034 \pm 0.030 $ & $ 13.017 \pm 0.028 $ & $ 13.024 \pm 0.030 $\\
$\sigma_{\log_{10} M}$ & $ 0.39 \pm 0.06 $ & $ 0.39 \pm 0.06 $ & $ 0.26 \pm 0.10 $ & $ 0.31 \pm 0.11 $ & $ 0.40 \pm 0.07 $ & $ 0.34 \pm 0.06 $ & $ 0.36 \pm 0.06 $\\
$\log_{10} M_{\rm cut}$ & $ 13.26 \pm 0.14 $ & $ 13.28 \pm 0.13 $ & $ 13.08 \pm 0.15 $ & $ 11.89 \pm 0.99 $ & $ 13.24 \pm 0.13 $ & $ 13.24 \pm 0.14 $ & $ 13.25 \pm 0.14 $\\
$\log_{10} M_1$ & $ 14.09 \pm 0.06 $ & $ 14.07 \pm 0.06 $ & $ 14.06 \pm 0.05 $ & $ 14.23 \pm 0.05 $ & $ 14.03 \pm 0.05 $ & $ 14.17 \pm 0.10 $ & $ 14.08 \pm 0.06 $\\
$\alpha$ & $ 0.75 \pm 0.19 $ & $ 0.75 \pm 0.19 $ & $ 0.88 \pm 0.16 $ & $ 1.15 \pm 0.10 $ & $ 0.89 \pm 0.15 $ & $ 0.67 \pm 0.22 $ & $ 0.77 \pm 0.18 $\\
$\bar{n}_{\rm HOD}$ & $ 4.11 \pm 0.14 $ & $ 4.11 \pm 0.13 $ & $ 4.60 \pm 0.13 $ & $ 3.67 \pm 0.28 $ & $ 4.16 \pm 0.09 $ & $ 4.10 \pm 0.14 $ & $ 4.13 \pm 0.12 $\\
$f_{\rm sat}$ & $ 0.1016 \pm 0.0070 $ & $ 0.1017 \pm 0.0068 $ & $ 0.1140 \pm 0.0074 $ & $ 0.1536 \pm 0.0222 $ & $ 0.0998 \pm 0.0069 $ & $ 0.1024 \pm 0.0068 $ & $ 0.1021 \pm 0.0070 $\\
$f\sigma_8$ & $ 0.447 \pm 0.014 $ & $ 0.451 \pm 0.010 $ & $ 0.458 \pm 0.010 $ & $ 0.455 \pm 0.009 $ & $ 0.460 \pm 0.013 $ & $ 0.453 \pm 0.011 $ & $ 0.445 \pm 0.009 $\\
$\gamma_{\rm IHV}$ & ${\bf 1.00 }$ & ${\bf 1.00 }$ & ${\bf 1.00 }$ & ${\bf 1.00 }$ & ${\bf 0.80 }$ & ${\bf 1.20 }$ & ${\bf 1.00 }$\\
$\gamma_{\rm cenv}$ & ${\bf 0.00 }$ & ${\bf 0.00 }$ & ${\bf 0.00 }$ & ${\bf 0.00 }$ & ${\bf 0.00 }$ & ${\bf 0.00 }$ & $ 0.06 \pm 0.05 $\\
$\chi^2_{w_p}$ (18) & $ 10.9 $ & $ 12.5 $ & $ 9.9 $ & $ 8.3 $ & $ 17.4 $ & $ 8.4 $ & $ 13.4 $\\
$\chi^2_{\hat{\xi}_{0,2}}$ (18) & $ 28.2 $ & $ 27.3 $ & $ 27.0 $ & $ 22.4 $ & $ 55.0 $ & $ 21.1 $ & $ 27.2 $\\
$\chi^2_{w_p+\hat{\xi}_{0,2}}$ (27) & $ 31.9 $ & $ 32.1 $ & $ 28.4 $ & $ 22.1 $ & $ 57.3 $ & $ 24.4 $ & $ 32.7 $\\

\end{tabular}
\caption{Marginalized 68\% parameter constraints for under various model
assumptions when fit to our measurements of $w_p(r_{\sigma} < 2.0$ $h^{-1}$ Mpc)
(9 bins) and $\hat{\xi}_{0,2}$ (18 bins).  The ``default'' constraints are shown
in bold in the first column.  We also use bold to indicate parameters that were
held fixed.  We vary the underlying $N$-body simulation (HiRes, MedRes0,
MedRes1, and MedRes2; where not stated, MedRes0 was used), the central galaxy
velocity definition (${\bf v_{\rm DENS}}$ is the default choice, compared with
${\bf v_{\rm COMV}}$ labelled ``COMV''), the prior on $\bar{n}_{\rm HOD}$ ($3.25
< 10^4 \bar{n}_{\rm HOD} (h^{-1} \, {\rm Mpc})^3 < 4.25$ is the default, ``high
$\bar{n}_{\rm HOD}$'' assumes $4.25 < 10^4 \bar{n}_{\rm HOD} (h^{-1} \, {\rm
Mpc})^3 < 4.75$), the relation between central and satellite galaxies (the
default is to assume all halos with satellites also host CMASS centrals, while
Sec.~\ref{sec:robustresults} describes one alternative, labelled ``cen/sat
test''), and whether the velocity parameters $\gamma_{\rm HV}$ ($\propto
f\sigma_8$), $\gamma_{\rm IHV}$, and $\gamma_{\rm cenv}$ are fixed or varied.
The last three rows provide the $\chi^2$ value for the full $w_p$ measurement
(18 bins), $\hat{\xi}_{0,2}$ (18 bins), and our ``default'' data combination (27
bins including $w_p$-$\hat{\xi}_{0,2}$ covariances), evaluated at the best fit
model in the MCMC chain.  Experiments with a direct $\chi^2$ minimization
algorithm indicate the minimum is found within $\sim \Delta \chi^2 = 0.5$.}
\label{tab:hod} \end{table*} In this section we fit the measured $w_p$ and
$\hat{\xi}_{0,2}$ using the covariance matrix presented in Sec.~\ref{sec:covtot}
and Fig.~\ref{fig:covfinal} to the model described in Sec.~\ref{sec:HODmodel}.
Our recommended constraints based on our best guesses regarding modeling choices
(justified further below) are indicated in Table~\ref{tab:hod} as the bold
``fiducial'' column, while many other modeling choices are presented there only for
comparison.  We also indicate parameters held fixed in particular analyses by
bold.  \subsection{Choice of measurement combination $w_p$ and
$\hat{\xi}_{0,2}$} In this analysis we study galaxy clustering on scales below
the typical host halo virial radius, out to the quasi-linear scales used in our
large-scale RSD measurements.  The maximum scale included sets a limitation on
the number of available bootstrap regions from the survey, and that, in turn,
sets a limit on the number of observables for which we can reliably estimate a
covariance matrix.  It was thus our goal to determine a minimal set of
observables that contained most of the available clustering information on the
scales of interest.  Initially we considered fits only to either $\xi_{0,2}$ or
$\hat{\xi}_{0,2}$.  We prefer the latter because of its insensitivity to fiber
collision corrections and smaller uncertainties.  We found that these two
observables preferred distinct regions of HOD parameter space, at least in our
fiducial HOD parametrization: fits to $\hat{\xi}_{0,2}$ alone prefer a low
satellite fraction of $6.5\%$ and did provide a better fit to the small-scale
behavior of $\hat{\xi}_{0,2}$ than presented below.  However, this model was in
strong tension with both $w_p$ and $\xi_{0,2}$ on small scales because of the
low satellite fraction.  We concluded that information relevant to the satellite
HOD parameters was missing from $\hat{\xi}_{0,2}$, and so we decided to jointly
fit $w_p(r_{\sigma} < 2$ $h^{-1}$ Mpc) and $\hat{\xi}_{0,2}$ to search for
models that could fit both adequately.  The number of elements in our data
vector used throughout the rest of the paper is $n_{\rm bin} = 27$: the nine
smallest scale bins in $w_p$ as well as nine bins each for $\hat{\xi}_0$ and
$\hat{\xi}_2$.  The tension between the initial fits to $\xi_{0,2}$ and
$\hat{\xi}_{0,2}$ naively indicates shortcomings in our model; however, as we
show in Sec.~\ref{sec:goodnessfit}, we are able to find a model within our
fiducial parametrisation that adequately fits all three observables.

We planned to use only the HiRes simulation box for our theoretical
calculations, but given the noisiness of the resulting likelihood surface in
$f\sigma_8$, we verified our results by repeating the fits with three independent
MedRes simulation boxes.  Most of the final results we report are based on the
MedRes0 box, but some cases using the HiRes box are presented for comparison.
In general we found excellent agreement between the two. 

\subsection{Comparing ${\bf v}_{\rm COMV}$ and ${\bf v}_{\rm DENS}$ central
galaxy velocity definitions} \label{sec:vCOMVvDENS} \begin{figure}
\includegraphics[width=85mm]{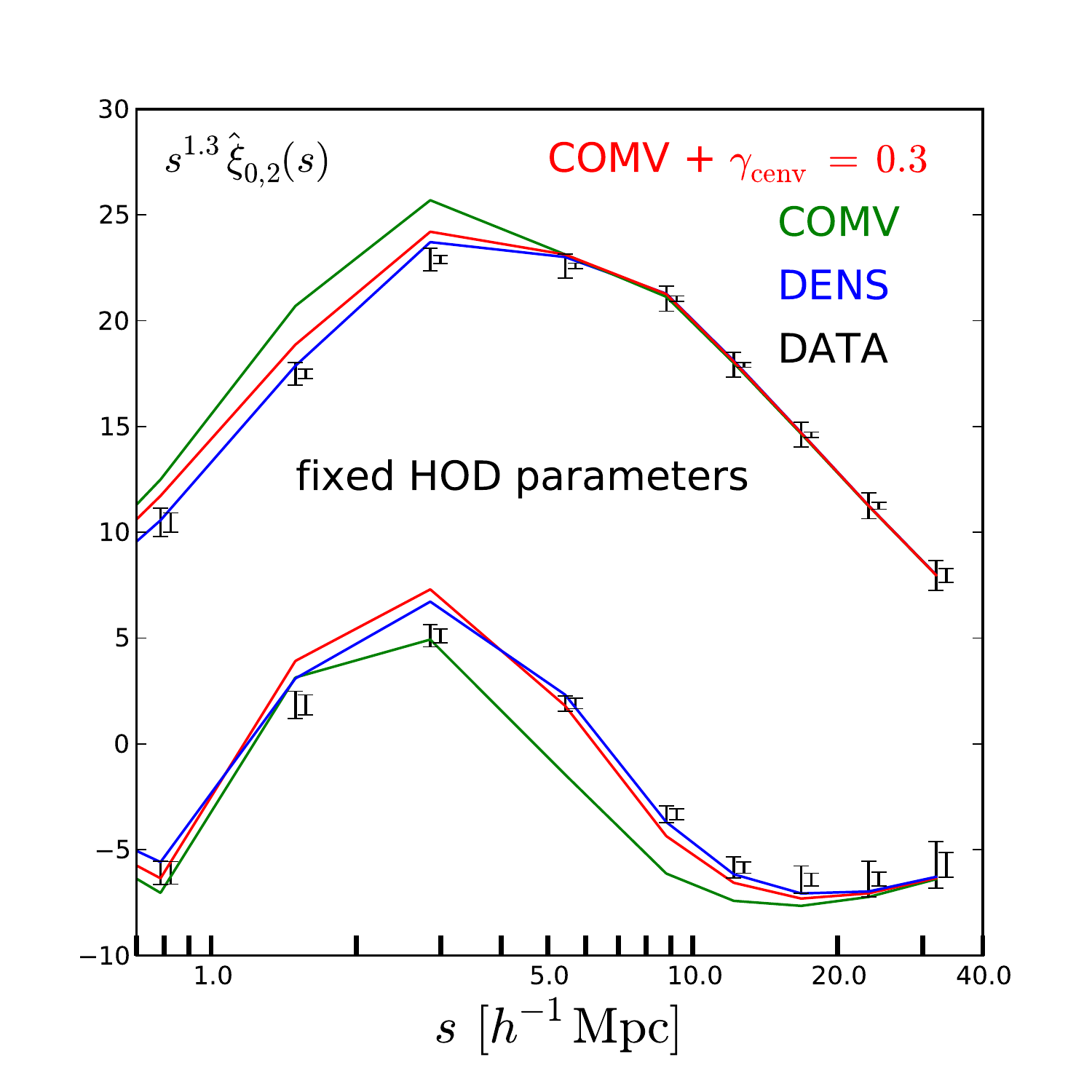} \caption{In order to isolate the effect of
the central galaxy velocity definitions, we fix HOD parameters in this plot to
the best fit values using ${\bf v}_{\rm DENS}$ (fourth upper column in Table
\ref{tab:hod}) and plot the theoretical predictions for the fiducial choice
${\bf v}_{\rm DENS}$ (blue), the halo center-of-mass velocity ${\bf v}_{\rm
COMV}$ (green), and ${\bf v}_{\rm COMV}$ with additional Gaussian dispersion at
0.3$\sigma_{\rm vir}$ (i.e., setting $\gamma_{\rm cenv} = 0.3$; shown in red).
In the other two cases, $\gamma_{\rm cenv} = 0$.  In all three theoretical
curves, $\gamma_{\rm IHV} = 1$ and $f\sigma_8 = 0.472$ is held fixed.  
We show two sets of errors: the larger ones are
the square root of the diagonal elements of the covariance matrix.  There are
strong covariances between the bins, and the smaller error bars attempt to
demonstrate their impact.  The small errors show the change required in a single
bin to change $\chi^2$ by 1, when the data and theory are in perfect agreement
in all other bins.  Though changes between the ${\bf v}_{\rm DENS}$ and ${\bf
v}_{\rm COMV}$ + $\gamma_{\rm cenv} = 0.3$ appear relatively small, fits with
the full covariance matrix disfavor the latter choice by $\Delta \chi^2 = 13$;
see upper columns four through six in Table \ref{tab:hod}.}
\label{fig:comparevcats} \end{figure} We first compare fits using two different
central galaxy velocity choices, ${\bf v}_{\rm DENS}$ (default) and ${\bf
v}_{\rm COMV}$ using the MedRes0 box.  In this comparison central galaxy
positions are fixed and only their velocities are varied.  The prediction for
$\hat{\xi}_{0,2}$ from the best fit HOD model with ${\bf v}_{\rm DENS}$ (upper
fourth column in Table \ref{tab:hod}) is shown in Fig.~\ref{fig:comparevcats} in
blue.  To emphasize the differences caused by the choice of central galaxy
velocity, we also plot the prediction for the same HOD using the center-of-mass
halo velocity for central galaxies (green) and center-of-mass halo velocity plus
a random Gaussian dispersion term consistent with the magnitude of $|{\bf
v}_{\rm DENS} - {\bf v}_{\rm COMV}|$ (red); see Fig.~\ref{fig:comparevcatsvsM}.
The ${\bf v}_{\rm COMV}$ model is clearly a bad fit, which is also true when the
HOD parameters are allowed to vary.  By eye, the difference between the fiducial
model and ${\bf v}_{\rm COMV} + \gamma_{\rm cenv} = 0.3$ does not appear large
compared with the square root of the diagonal elements of the covariance matrix
(larger errors in Fig.~\ref{fig:comparevcats}).  However, the covariance matrix
has very strong correlations; to give an alternate sense of the true
constraining power of our measurements, we also show a second error bar, which
is the size of the change required in a single bin to change $\chi^2$ by 1, when
the model and theory differences are set to 0 in all other bins.  The difference
between these two ``errors'' is largest (a factor of 5) in the 5-17 $h^{-1}$ Mpc
bins of $\hat{\xi}_0$.  Thus the data do show a strong preference for ${\bf
v}_{\rm DENS}$ ($\chi^2 = 36.8$) compared with either ${\bf v}_{\rm COMV}$
($\chi^2 = 68.5$; upper column five of Table \ref{tab:hod}) or ${\bf v}_{\rm
COMV} + \gamma_{\rm cenv} = 0.3$ ($\chi^2 = 50.1$; upper column six), when our
HOD parameters are allowed to vary and assuming the underlying halo clustering
in the simulation is sufficiently similar to that in the real universe.  The
latter comparison indicates that the motion of the dense core of the halo
relative to the center-of-mass of the halo is correlated with the surrounding
cosmic structure, and this correlation propagates into the shape of the
correlation function; Fig.~\ref{fig:simslice} shows the alignment between the
velocity difference vector and the density field around the most massive halo in
the HiRes simulation.  The net effect of this density-velocity correlation is to
increase the redshift separation $s$ between pairs, thus broadening both
$\hat{\xi}_{0}$ and $\hat{\xi}_{2}$ compared to the uncorrelated dispersion
case, ${\bf v}_{\rm COMV} + \gamma_{\rm cenv}=0.3$.  In Fig.~\ref{fig:simslice},
the difference vector ${\bf v}_{\rm DENS}$ - ${\bf v}_{\rm COMV}$ is oriented
such that it will increase the redshift separation between the central halo and
halos falling in along the corresponding filament.  We saw similar levels of
$\chi^2$ differences in the HiRes box compared with the MedRes box when
performing the same test.  These possibilities are certainly not the only
choices for assigning velocities to galaxies and neglect all ``gastrophysical''
effects; however, the investigations in \citet{Behroozi13} do indicate that our
${\bf v}_{\rm DENS}$ has similar properties to a more detailed phase-space based
halo finding algorithm. 

\subsection{Goodness of fit} \label{sec:goodnessfit} For our fiducial choice
using ${\bf v}_{\rm DENS}$ and the MedRes0 box for the underlying halo catalog,
and introducing no free HOD velocity parameters, we find $\chi^2 = 36.8$ for 27
data points and 5 free HOD parameters; a larger $\chi^2$ is expected only 2.5\%
of the time.  This high $\chi^2$ could be an indication of insufficiencies of
our model, non-Gaussianity of our errors, a preference for different
cosmological parameters compared with our simulation parameters, or simply bad
luck.  For further insight, we attempt to fit MedRes and HiRes without the
theoretical error contribution to the covariance matrix.  If our theoretical
predictions were based on single catalog realizations rather than an average
over all possible HOD realizations for a fixed halo catalog, we would expect the
contribution from the measurement and MedRes theory errors to be comparable
because they cover comparable volumes.  We find $\chi^2 = 75$ ($\chi^2 = 60$) to
be compared with the fourth (second) columns in Table~\ref{tab:hod} for the
MedRes0 (HiRes) halo catalogs.  That is, our fiducial model seems adequate,
within at least a factor of $\sim \sqrt{2}$ of the measurement errors.  The fact
that using the HiRes halo catalog, which covers only an eighth of the MedRes box
volume, returns a better $\chi^2$ in this case must indicate that changes to the
observables allowed by our theoretical uncertainties can be mostly absorbed by
tweaking the HOD parameters.  If that were not the case, we would have expected
a much larger contribution from the higher theoretical uncertainty of the HiRes
box to $\chi^2$.  The slight difference could also indicate a preference of the
data for the higher-resolution halo catalog.  In any case, these tests do not
indicate the existence of systematic modeling errors at the level of our total
quoted uncertainty.  Of course, just because the model can fit the data does not
demonstrate that the resulting parameter fits are unbiased.

\subsection{Properties of the Halo Occupation Distribution}
\label{sec:HODresults} \begin{figure*} \includegraphics[width=170mm]{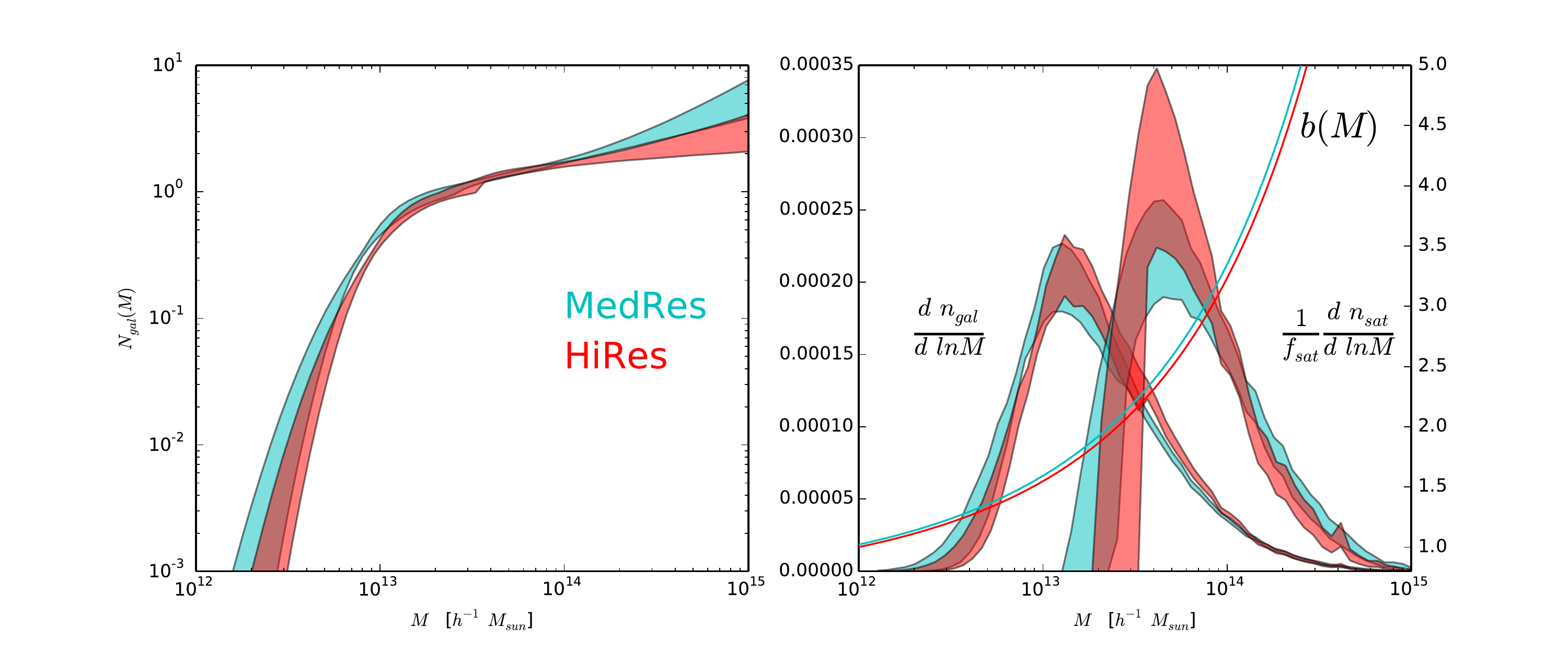}
\caption{Assuming the cosmology for the MedRes (cyan) or HiRes (red) simulation
boxes, we show the halo occupation distribution of CMASS galaxies as a function
of host halo mass.  These HODs have $\bar{n}_{\rm HOD} = 4.15 \pm 0.1$ [$h^{-1}$
Mpc]$^{-3}$, while the typical $\bar{n}$ for the CMASS redshift distribution is
somewhat lower (see Fig.~\ref{fig:cumnbar}).  Our model assumes that the
observed CMASS sample is a random subsample of this HOD. Multiplying by the halo
mass function in the two simulations gives the probability distribution of
galaxies as a function of host halo mass (left curves in right panel).  We also
show the distribution of satellite galaxies separately, rescaling by $f_{sat}$
for visualization purposes.  Finally, we overlay the \citet{Tinker10} $b(M)$
relation for the two cosmologies.  At fixed mass, halos in the HiRes simulation
are slightly less biased since $\sigma_8$ and $\Omega_m$ are slightly larger in
that simulation.} \label{fig:HOD} \end{figure*} Figure \ref{fig:HOD} shows the
halo occupation distribution at the fiducial cosmologies of our HiRes and MedRes
boxes (corresponding to upper columns two and four in Table~\ref{tab:hod}).  We
enforce a hard prior on $0.1 \leq \alpha \leq 2$ which does affect the
constraints from the HiRes box.  Table~\ref{tab:hod} shows that the HOD
parameters are quite stable as we explore different parameter spaces and model
assumptions, with the exception of the ``high $\bar{n}_{\rm HOD}$'' and
``cen/sat test'' cases detailed in the next section.  Within the fiducial
$\bar{n}_{\rm HOD}$ prior discussed in Sec.~\ref{sec:HODnbar}, the data prefer
the largest allowed values of $\bar{n}_{\rm HOD}$; the best fit value is near
the hard prior upper boundary.  Under the fiducial $\bar{n}_{\rm HOD}$ prior,
the fraction of galaxies that are satellites is strongly constrained: $10.2 \pm
0.7$ per-cent.  The data show a strong preference for a non-zero $M_{\rm cut}$
at a value of $\sim 2M_{\rm min}$, which could plausibly be produced by a 1:1
merger of halos of mass $M_{\rm min}$.  The distribution of galaxies across halo
mass is relatively symmetric as a function of $\log_{10} M$ (right panel of
Fig.~\ref{fig:HOD}), which makes the median ($1.7 \times 10^{13}$ $h^{-1}$
$M_{\sun}$) and mean  ($3.3 \times 10^{13}$ $h^{-1}$ $M_{\sun}$) host halo
masses quite different.  For satellite galaxies, the median (mean) host halo
mass is $6 \, (9) \, \times 10^{13}$ $h^{-1}$ $M_{\sun}$.\\

The mean host halo mass is most closely related to the expected amplitude of the
galaxy-galaxy lensing signal.  The amplitude of clustering of CMASS galaxies on
scales substantially larger than a typical host halo virial radius constrains
the product of a linear bias factor $b$ and the overall amplitude of matter
fluctuations $\sigma_8(z_{\rm eff})$ at the effective redshift of the galaxy
sample.  The observed $b\sigma_8$ for the CMASS galaxy sample places it in a
halo mass regime where halo bias depends steeply on mass; $b(M)$ is overlaid in
the right panel of Fig.~\ref{fig:HOD}.  To test the robustness of the mean halo
mass prediction within the context of our HOD model, we allowed a freely varying
spline function to describe $d n_{\rm cen}/d ln M$, constrained by a minimum
$\bar{n}_{\rm HOD}$ set by the observed $\bar{n}(z)$ and constrained to
reproduce the observed $b\sigma_8$.  Adding this freedom to the HOD only
introduced uncertainty in the mean central galaxy halo mass at the $\sim 10\%$
level.\\

The high-mass slope $\alpha$ of the satellite HOD is not well-constrained in our
fits, and in particular, our $\alpha >= 0.1$ prior affects the constraints in
the HiRes case.  However, the satellite galaxy distributions in the right panel
of Fig.~\ref{fig:HOD} are similar, and the corresponding intra-halo velocity
dispersion is well-constrained by our measurements; see
Sec.~\ref{sec:sig2results}.\\

Neglecting slight differences in cosmological parameters, we find excellent
agreement with the analysis of $w_p(0.3 h^{-1}$ Mpc $< r_{\sigma} < 30 h^{-1}$ Mpc$)$ 
in \citet{White11}.  Converting to our HOD
parameter definitions, their Table 2 implies $\sigma_{log_{10} M} = 0.30 \pm
0.07$, $log_{10} M_{\rm min} = 13.08 \pm 0.12$, $\log_{10} M_1 = 14.06 \pm 0.1$,
$M_{\rm cut}/M_{\rm min} = (1.13 \pm 0.38)$, $\alpha = 0.9 \pm 0.19$, and
$f_{sat} = 10 \pm 2$ per-cent.  We find $M_{\rm cut}/M_{\rm min} = (1.8 \pm
0.6)$.  Since \citet{White11} used one-tenth the sky area of the DR10 analysed
in this paper and we have added information from $\hat{\xi}_{0,2}$, it is rather
surprising that our errors on most parameters seem comparable.  Some of the
difference can likely be attributed to our conservative estimate on the angular
upweighting errors that dominate the error budget in the one-halo region of
$w_p$, our wider prior on $\bar{n}_{\rm HOD}$, and our inclusion of a
theoretical error budget.  Our measurements do improve the errors considerably
on the satellite fraction and the $N_{\rm cen}(M)$ mass scale.  The latter is
expected since the larger survey volume and inclusion of $\xi_0$ allow a precise
clustering amplitude measurement on large scales.  We also note that
\citet{Nuza13} showed relatively good agreement between the observed DR9 CMASS
clustering and the predictions of SHAM; their resulting CMASS HOD is also in
broad agreement with the results of \citet{White11} as well as those presented
here.  We note that our estimate of $\xi_0$ derived in Sec.~\ref{sec:FB} and the
\citet{Nuza13} measurement, based on the \citet{Guo12} fiber collision
correction method, significantly disagree on scales between 1 and 8 $h^{-1}$
Mpc.  Finally, we note that \citet{Guo14} have recently used the \citet{Guo12}
fiber-collision correction method to compute the projected clustering $w_p$ of
various luminosity, redshift, and color subsamples of CMASS.  Our results are
not directly comparable because of their cuts, but a cursory examination yields
some interesting differences.  Their $M_0$ parameter (equivalent to our $M_{\rm
cut}$) is constrained to be effectively 0, while our constraints require it to
be at least larger than $M_{\rm min}$.  Their CMASS subsample with $M_i > -21.6$
and $0.48 < z < 0.55$ has the largest $\bar{n}$ among their subsamples (still a
factor of two lower than our best fit HODs), and yet has more satellites per
halo than our HOD for masses above $\sim 10^{14.6}$, despite their lower $f_{\rm
sat} = 7.9$ per-cent.  An examination of the anistropic clustering in their
samples may shed light on this difference.  Their analytic HOD model is
calibrated on FoF halo catalogs, and may therefore require more satellites in
massive halos than our model, since SO halo catalogs have more halo pairs near
the one-halo to two-halo transition. 

\subsection{Fits to $f\sigma_8(z=0.57)$} \label{sec:fs8results} \begin{figure}
\includegraphics[width=85mm]{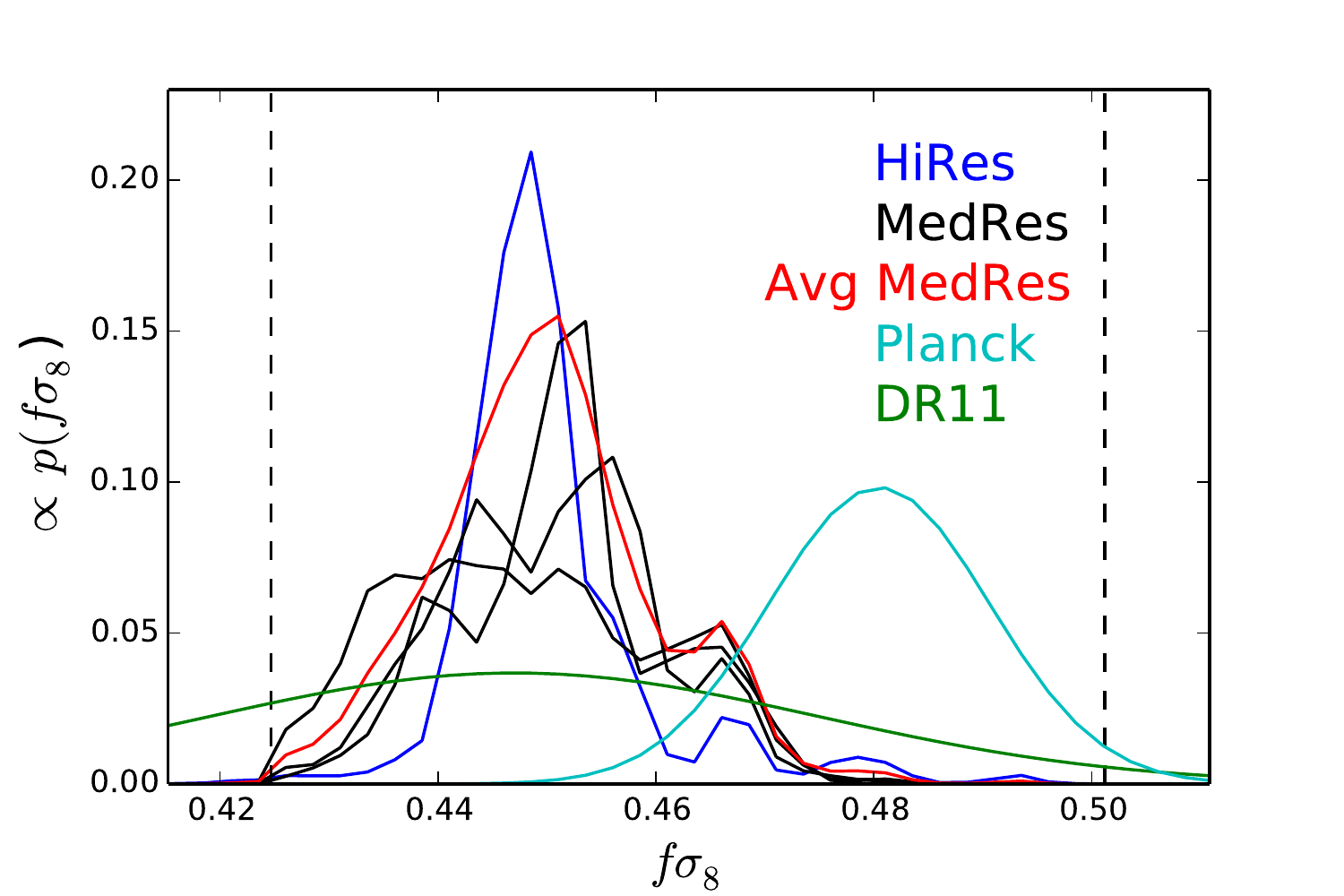} \caption{The marginalized distribution of
$f\sigma_8$ from our HiRes box (blue) and three MedRes boxes (black; red shows
their average).  In this case $\gamma_{\rm cenv} = 0$ and $\gamma_{\rm IHV} = 1$
were held fixed.  We compare this to the constraints from Planck $\Lambda$CDM
fits ($f\sigma_8 = 0.480 \pm 0.010$) and our DR11 analysis of the CMASS galaxy
clustering restricted to large scales $s > 25$ $h^{-1}$ Mpc, where we found
$f\sigma_8 = 0.447 \pm 0.028$.  Vertical dashed lines show our hard prior on the
MedRes box of $\pm 10\%$ of the $f\sigma_8$ value in the MedRes cosmology.}
\label{fig:fs8} \end{figure}
\begin{figure*}
\includegraphics[width=180mm]{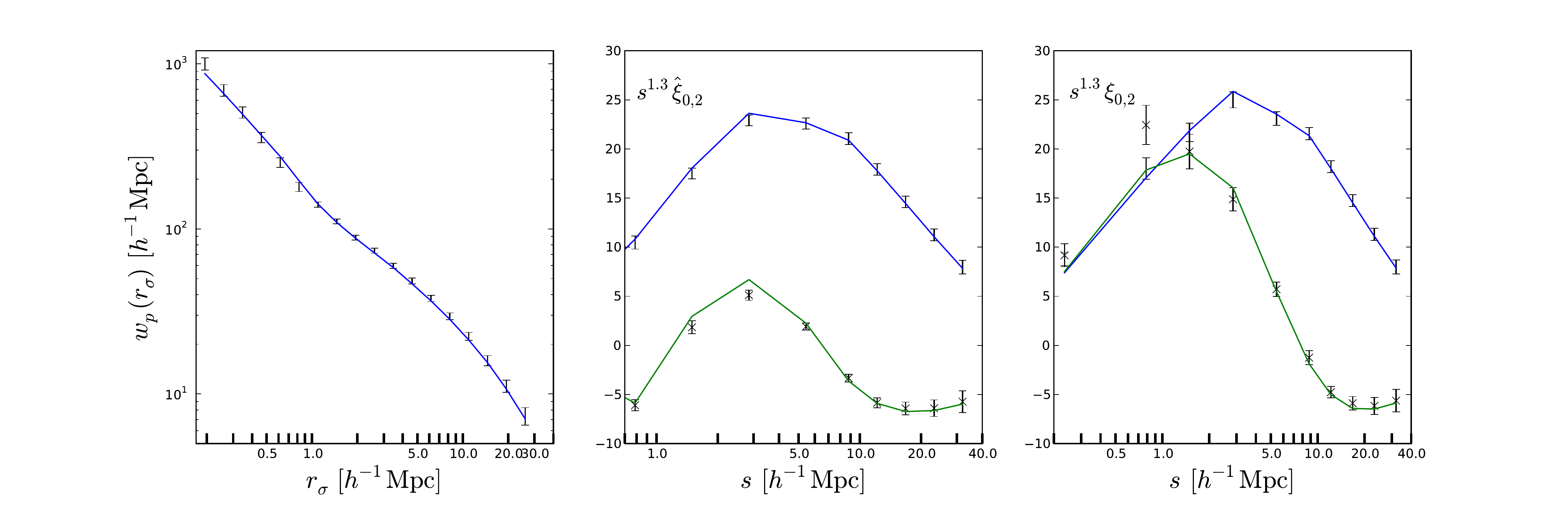} \caption{The best fit model using the
``MedRes0'' box, fixing $\gamma_{\rm IHV}=1$ and $\gamma_{\rm cenv}=0$ compared
to our measurements for $w_p$, $\hat{\xi}_{0,2}$, and $\xi_{0,2}$ (bottom first
column of Table~\ref{tab:hod}).  In the second two panels $\hat{\xi}_2$ and
$\xi_2$ measurements are indicated with an $X$, and the model predictions with
green curves instead of blue, since $\xi_0$ and $\xi_2$ cross on small scales.
The varying parameters took best fit values $f\sigma_8=0.452$, $M_{\rm min} =
10^{13.011}$ $h^{-1}$ $M_{\sun}$, $M_{\rm cut} = 10^{13.159}$ $h^{-1}$
$M_{\sun}$, $M_{1} = 10^{14.068}$ $h^{-1}$ $M_{\sun}$, $\alpha = 0.89$,
$\sigma_{\log_{10} M} = 0.358$.  This HOD has $\bar{n}_{\rm HOD} = 4.23 \times
10^{-4}$ ($h^{-1}$ Mpc)$^{-3}$ and $f_{sat} = 10.4$ per-cent.  The best fit
$\chi^2$ values are listed in the first lower column in Table ~\ref{tab:hod}.
This best fit model is derived by fitting only the first nine bins of $w_p$ and
$\hat{\xi}_{0,2}$, but also provides a good fit to $\xi_{0,2}$, for which
$\chi^2 = 20.9$ for 20 measurement bins.  Compared to the best fit with
$\gamma_{\rm HV} = 1$ ($f\sigma_8 = 0.472$), the fit to the quadrupole on large
scales is improved.  Our last 1.5 bins overlap with the smallest bins in the
large-scale RSD analysis in \citet{Reid12} and \citet{Samushia13} and our best
fit $f\sigma_8$ values are nearly identical.} \label{fig:BF3panel} \end{figure*}

 Next, we consider the effect of linearly varying
the overall amplitude of the peculiar velocity field with the parameter
$\gamma_{\rm HV}$, and interpret the result as a change in the effective
$f\sigma_8$.  We justify this interpretation in Sec.~\ref{sec:cosmodep}.  Here
we consider only the case when the other velocity parameters $\gamma_{\rm cenv}
= 0$ and $\gamma_{\rm IHV} = 1$ are held fixed.  The marginalized distribution
of $f\sigma_8$ shown in Fig.~\ref{fig:fs8} is clearly noisy due to the finite
volume of our $N$-body simulation boxes.  We therefore computed constraints
separately from three independent MedRes boxes (labelled Fiducial=MedRes0,
MedRes1, and MedRes2 in Table \ref{tab:hod}) as well as with the single HiRes
box we had available (top row, third column in the table).  The marginalized
$f\sigma_8$ constraints are consistent across the boxes, despite the
$\sim1\sigma$ shift in fiducial value between the box cosmologies.  Averaging
over the MedRes simulation boxes, we find $f\sigma_8 = 0.450 \pm 0.011$,
consistent with our recent large-scale analysis of DR11 \citep{Samushia13} which
found $f\sigma_8 = 0.447 \pm 0.028$ for a $\Lambda$CDM expansion history.  Our
raw statistical error is equal to Planck's $\Lambda$CDM prediction of $f\sigma_8
= 0.48 \pm 0.010$; the difference between the two independent measurements is
$1.9\sigma$, which we take to be reasonable agreement since we have not included
a modeling systematics error budget.  Despite the dominance of satellite
galaxies on the observed anisotropies (Fig.~\ref{fig:xi02theory}), there is
still ample information on the rate of structure growth on these smaller scales
where the clustering signal is strong and well-measured, resulting in a factor
of 2.5 reduction in uncertainty on $f\sigma_8$ compared with our DR11
large-scale RSD analysis.  In Fig.~\ref{fig:BF3panel} we show the theoretical
prediction from the best fit model using the MedRes0 box.  In this model
$f\sigma_8 = 0.452$ and we have held $\gamma_{\rm IHV} = 1$ and $\gamma_{\rm
cenv} = 0$ fixed.  Compared to the best fit model with $\gamma_{\rm HV} = 1$
($f\sigma_8 = 0.472$) in Fig.~\ref{fig:comparevcats}, the amplitude of $\xi_2$
on large scales provides a better fit to the data.  These are the same scales
dominating the \citet{Samushia13} large-scale RSD measurement of $f\sigma_8$;
the last $\sim 1.5$ bins overlap between the analyses.  The best fit models as a
function of $f\sigma_8$ have nearly identical behavior in the first three bins
$s < 3$ $h^{-1}$ Mpc, and divide on larger scales, indicating that the
constraint on $f\sigma_8$ is driven by the relative amplitudes of $\hat{\xi}_0$
and $\hat{\xi}_2$.  Fig.~\ref{fig:BF3panel} also shows that even though the
model was fit to $w_p(r_{\sigma} < 2$ $h^{-1}$ Mpc) and $\hat{\xi}_{0,2}$, it
provides a good fit to $w_p$ out to 25 $h^{-1}$ Mpc ($\chi^2 = 12.4$ for 18
bins), and correctly models scales below the fiber collision radius, so that
$\xi_{0,2}$ is also fit ($\chi^2 = 20.9$ for 20 bins).  

\subsection{Robustness of the $f\sigma_8$ constraint to model extensions}
\label{sec:robustresults} The basic redshift-independent HOD model we are using
to fit the CMASS clustering assumes that the observed galaxies are a subsample
of objects defined by those HOD parameters.  We enforce only a broad prior on
$\bar{n}_{\rm HOD}$ from the observed CMASS selection function $\bar{n}(z)$.
However, both intrinsic stochasticity in the stellar mass-halo mass relation and
photometric errors in the imaging catalog will broaden the distribution of halo
masses hosting the CMASS sample.  In order to test our sensitivity to the
allowed host halo mass scatter, we refit our measurements with the $\bar{n}_{\rm
HOD}$ prior shifted to higher values: $4.25 < 10^4 \bar{n}_{\rm HOD} (h^{-1} \,
{\rm Mpc})^3 < 4.75$.  The results of fits that fix or vary $f\sigma_8$ are
labelled in Table \ref{tab:hod} as ``high $\bar{n}_{\rm HOD}$.''  This choice is
similar to relaxing our assumption that $N_{\rm cen}(M)$ in Eq.~\ref{eq:Ncen}
approaches one at large halo masses.  Indeed, we find that this region of HOD
parameter space provides a better fit to the observed clustering ($\Delta \chi^2
\sim 4$).  There are small (expected) shifts in the HOD parameters with the
higher $\bar{n}_{\rm HOD}$ prior; most importantly for our conclusions in this
work, the constraint on $f\sigma_8$ shifts by only $\sim 0.5\sigma$.
If we completely remove the $\bar{n}$ prior, the HOD is limited to 
$10^4 \bar{n}_{\rm HOD} (h^{-1} \,{\rm Mpc})^3 < 6$ as $\sigma_{\log_{10} M}$ 
approaches 0, which is an unphysical limit of noisy target selection producing 
a precise mass cut in central galaxy host mass. Given our HOD parametrization, 
models with higher number density are unable to generate sufficiently large 
clustering.  Even in this unrealistic case, the $f\sigma_8$ shifts upward 
compared to our fiducial value by only $1\sigma$.

Both the color selection and photometric errors in the imaging used for target
selection could result in halos where the central galaxy does not pass our
target selection cuts, while one or more satellite galaxies in that halo do
pass.  To test the impact of such cases (labelled ``cen/sat'' test in Table
\ref{tab:hod}), we consider the drastic case where 20\% of centrals in massive
halos are not CMASS selected galaxies, implemented in our model by simply
multiplying $N_{\rm cen}(M)$ by 0.8.  In contrast to the rest of our analyses,
in this test we do not require a central galaxy in order for a particular halo
to host a satellite galaxy, thus lowering the contribution of ``one-halo''
central-satellite pairs at fixed HOD parameters.  This model provides a much
better fit than our fiducial HOD assumptions ($\Delta \chi^2 = 10.2$).  The
satellite fraction is larger, $\bar{n}_{\rm HOD}$ in the model moves closer to
the typical $\bar{n}$ in the sample, and the satellite occupation distribution
steepens.  In future work we hope to explore such model extensions more
generally in concert with a better understanding of the impact of photometric
errors on targeting, as well as redshift evolution and intrinsic diversity in
the CMASS galaxy population.  Again, the important result for the present work
is that a plausible extension of our halo occupation modeling can improve the
fit, but the constraint on $f\sigma_8$ shifts only slightly.  

\begin{figure*} \includegraphics[width=180mm]{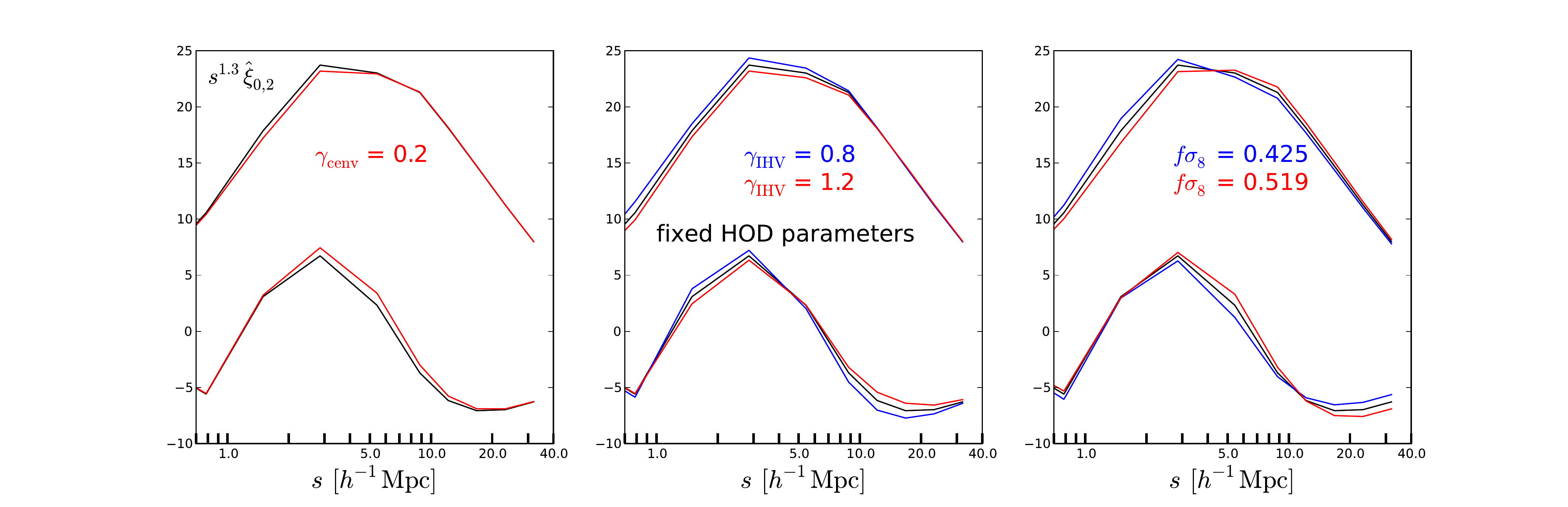}
\caption{The impact on $\hat{\xi}_{0,2}$ of varying the three velocity rescaling
parameters $\gamma_{\rm cenv}$, $\gamma_{\rm IHV}$, and $\gamma_{\rm HV} \propto
f\sigma_8$ at fixed HOD parameters.  } \label{fig:vparams} \end{figure*} Next we
consider the impact of varying the galaxy intrahalo velocities through the
parameters $\gamma_{\rm IHV}$ and $\gamma_{\rm cenv}$ defined in
Sec.~\ref{sec:HODp}.  Their impact on the $\hat{\xi}_{0,2}$ observable is shown
in Fig.~\ref{fig:vparams}, holding the HOD parameters fixed to the best fit
values for $\gamma_{\rm HV} = \gamma_{\rm IHV} = 1.0$ and $\gamma_{\rm cenv} =
0$.  Increasing the intrahalo velocity dispersion lowers the number of pairs at
small $s$ separations, while changing $f\sigma_8$ shifts the peak position in
$s^{1.3} \hat{\xi}_0$.  The impact of $\gamma_{\rm IHV}$ on $\hat{\xi}_2$
extends out to much larger scales than $\gamma_{\rm cenv}$, as expected because
intrahalo satellite velocities have a broader dispersion than centrals.
Changing $f\sigma_8$ has a distinct scale dependence in both $\hat{\xi}_0$ and
$\hat{\xi}_2$, which should be distinguishable from $\gamma_{\rm IHV}$ and
$\gamma_{\rm cenv}$.

We introduce the parameter $\gamma_{\rm IHV}$ to rescale the relative velocity
between satellite galaxies and their host halos.  This parameter is meant to
absorb the effect of galaxy velocity bias as well as variations in the halo mass
function due to cosmological parameter uncertainties.  \citet{White10} examine
in detail the velocity structure of subhalos within group-scale halos at z=0.1,
and suggest a theoretical uncertainty in velocity bias of $\mathcal{O}$(10\%).
\citet{Wu2013} used $N$-body and hydrodynamical simulations to study the
relationship between the galaxy and dark matter intrahalo velocity dispersion in
halos of mass $\sim 10^{14}$ $M_{\sun}$, i.e., well-matched to the typical
satellite galaxy host halo mass according to our HOD model fits.  They found
that averaging over all cluster galaxies, $\sigma_{\rm IHV,gal}/\sigma_{\rm
IHV,DM} = 1.065$, while averaging only over the five brightest satellites yielded a
ratio of 0.868.  The latter is likely more applicable to the massive galaxies
comprising the CMASS sample.  Rather than smoothly varying $\gamma_{\rm IHV}$,
we run separate MCMC chains at $\gamma_{\rm IHV} = 0.8$ and $\gamma_{\rm IHV} =
1.2$; this range incorporates the small velocity biases found in \citet{Wu2013}.
Alternatively, neglecting velocity bias, a $\pm 20$ per-cent variation in
$\gamma_{\rm IHV}$ corresponds to a factor of 0.5-1.7 change in the host halo
mass scale of satellite galaxies.  The fifth and sixth bottom columns of Table
\ref{tab:hod} show the result of these fits; $\gamma_{\rm IHV} = 0.8$ is
strongly disfavored by our data ($\Delta \chi^2 = 25$), but the best fit of
$f\sigma_8$ under this assumption is still in good agreement with our fiducial
case at the $1\sigma$ level.  Our data shows a $\Delta \chi^2 = 8$ preference
for $\gamma_{\rm IHV} = 1.2$, indicating that our fiducial model may not produce
strong enough finger-of-god features.  Again, allowing freedom in $\gamma_{\rm
IHV}$ does not shift or weaken the constraing on $f\sigma_8$.

Finally, we also introduce additional random velocity dispersion for central
galaxies through the parameter $\gamma_{\rm cenv}$ (final bottom column of
Table~\ref{tab:hod}).  The fiducial value $\gamma_{\rm cenv} = 0$ is preferred
by the data.  Allowing $\gamma_{\rm cenv}$ as a free parameter shifts the 68\%
confidence region on $f\sigma_8$ lower by $\sim 0.5\sigma$.  We do note that
preliminary tests using the HiRes box showed that when both $\gamma_{\rm IHV}$
and $\gamma_{\rm cenv}$ are free (and both take large values outside the range
considered here), the best fit value of $f\sigma_8$ is more dramatically
reduced.  The statistical precision of our fiducial $f\sigma_8$ constraint certainly
warrants a further assessment of our uncertainties of the velocity structure of
CMASS-type galaxies relative to their host dark matter halos.  Appendix \ref{sec:vcen} 
showed that the velocity of the halo center is not well-defined, but depends on the 
averaging scale.  We have chosen a scale that roughly matches the size of the typical 
CMASS galaxy, but ideally the uncertainty associated with this choice 
(and the potential impact of baryonic effects) should  be accounted for in future work.

\subsection{Non-linear velocities, cosmology dependence and light-cone effects}
\label{sec:cosmodep} \begin{figure} \includegraphics[width=85mm]{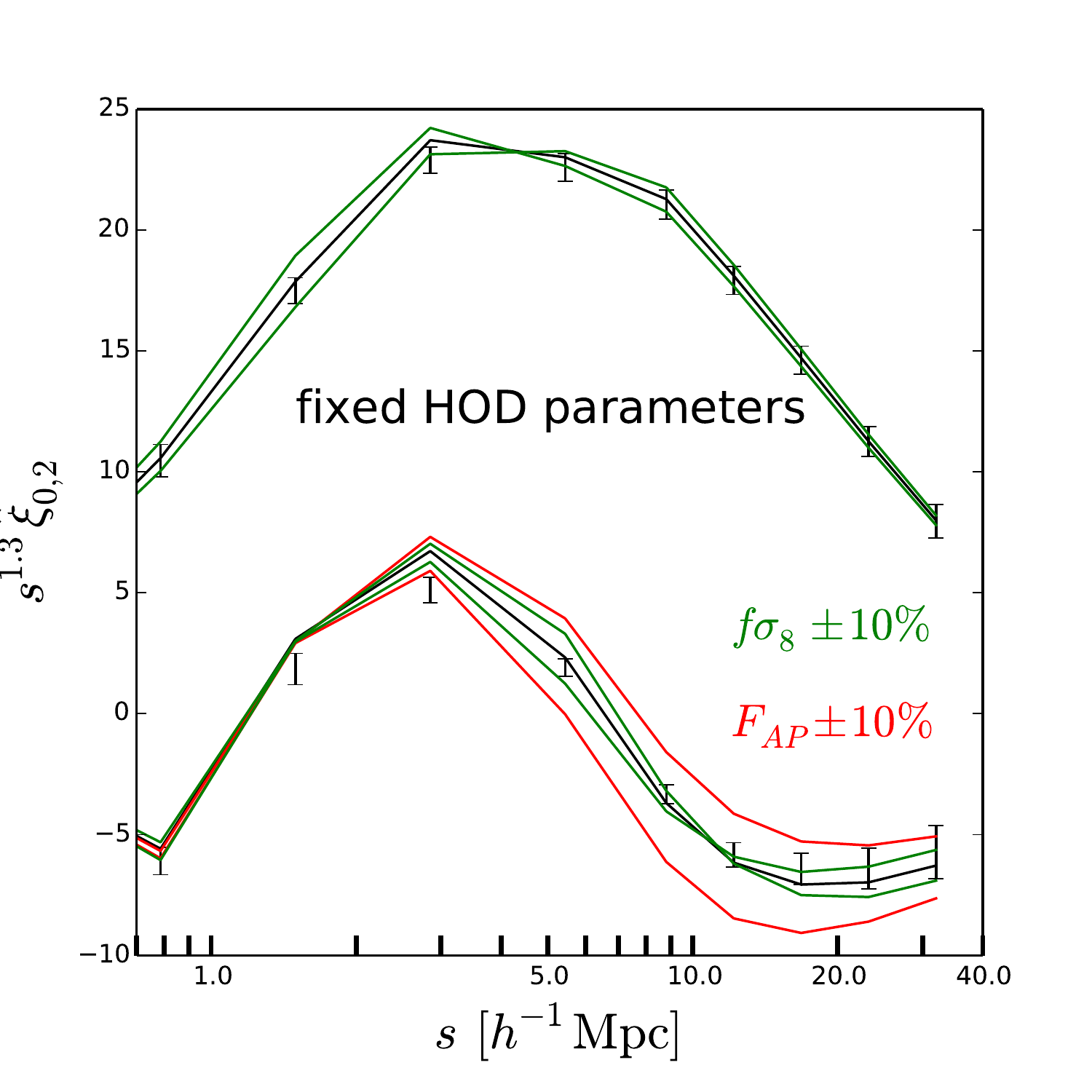}
\caption{Predictions for $\hat{\xi}_{0,2}$ when the growth rate of structure
$f\sigma_8$ or the geometric Alcock-Paczynski parameter $F_{AP}$ is varied by
10\%, at fixed HOD parameters.  In the latter case, we also held fixed the
isotropic BAO scale $\propto D_A^{2/3} H^{-1/3}$, so that the prediction for
$\hat{\xi}_0$ is unchanged and thus not shown.  $F_{AP}$ is tightly constrained
assuming a flat $\Lambda$CDM expansion history, but more general dark energy
models will modify both $f\sigma_8$ and $F_{AP}$.  Our current results therefore
cannot yet be applied to constraining dark energy.} \label{fig:APtest}
\end{figure} With the single exception of the cosmological parameter combination
determining $f\sigma_8$, we have not explored how our constraints depend on
cosmological parameters.  In this section we merely discuss where the largest
sensitivities lie.  Under the assumption of adiabatic fluctuations and the
standard three species of massless neutrinos, the CMB observations tightly
constrain the power spectrum of matter fluctuations \citep{PlanckXVI} with $k$
in units of Mpc; under these assumptions, $P(k)$ depends only on physical
densities $\Omega_{c,b,\gamma} h^2$ and $n_s$ \citep[see Section 5.1.1 of
][]{Reid12}.  These tight constraints on the linear matter power spectrum should
translate to strong constraints on the scale-dependence of halo clustering as
well.  As in other analyses simultaneously fitting cosmology and HOD-like
parameters \citep{Tinker11,Cacciato13,Mandelbaum13}, we naively expect most of
our sensitivity to cosmological parameters to be through some combination of
$\sigma_8$ and $\Omega_m$. 

We have allowed the overall amplitude of the halo peculiar velocity field in our
simulations to vary, and in Sec.~\ref{sec:fs8results} have interpreted this
amplitude as a constraint on $f\sigma_8$.  This linear scaling is expected to
break down in the non-linear regime; recall, for instance, that perturbative
corrections to the power spectrum are proportional to powers of the linear
growth factor ($D^2$, $D^4$, ...) times different functions of $k$.  To check
the impact of both light cone effects and the $f\sigma_8$ scaling approximation,
we examine halo catalogs from the MedRes0 box at neighboring redshift outputs:
$z=0.45,0.55,0.65$, for which $\sigma_8(z) = 0.59, 0.62, 0.65$.  Thus the edges
span a 10\% change in the large-scale amplitude of matter fluctuations.  We
divide our fiducial halo catalog at $z=0.55$ into four mass bins split on the
cumulative mass distribution from our best fit HOD model with boundaries edges
at $[10\%,30\%,50\%,70\%,90\%]$.  Motivated by our observation in Appendix 
\ref{sec:zdep} that there is no measurable redshift evolution of the CMASS clustering,  
we shift these mass bins slightly at $z=0.45$
and $z=0.65$, with the bin centers shifted to match the large scale value of
$b(M) \sigma_8(z)$ in the original bins.  The difference in the corresponding
mass bin centers between the $z=0.45$ and $z=0.65$ outputs was at most 20\%, for
a 10\% change in $\sigma_8$ across this redshift range.  Across this same
redshift range, we measure the normalization of the $\sigma_{\rm vir}(M)$
relation to decrease by 7\%.  These two effects nearly cancel each other, so for
HODs selecting halos with the same distribution of $b\sigma_8$, the effective
$\gamma_{\rm IHV}$ will remain $\approx 1$ at the-percent level, well within 
the range $\pm 20\%$ explored in the previous section.  
We expect small changes in other cosmological parameters
to be within this prior as well.   With these mass bins we can also compare the
clustering and velocity statistics at different redshifts.  By design, our bins
have the large-scale clustering amplitude matched, so we can isolate the impact
of non-linear growth on the underlying halo clustering.  We first compare the
matched real space correlation functions and pairwise infall velocities across
redshift.  Changes in non-linear growth to $\xi(r)$ and $v_{12}(r)$ at fixed
large scale $b\sigma_8$ is not well-detected in this measurement using a single
simulation box, but is constrained to be smaller than $\pm 2\%$, except in the
largest halo mass bin ($10^{13.43} - 10^{13.82}$ $h^{-1} M_{\sun}$), where the
real and redshift-space monopoles change by $\sim 10$ and $\sim 15$ per-cent
respectively below 1 $h^{-1}$ Mpc in the expected directions. We detect no
significant trends in $\hat{\xi}_2$.  We conclude that at fixed cosmological
parameters (other than $\sigma_8$), non-linear corrections to the theoretical
template can be neglected when inferring $f\sigma_8$.  Since the relevant halo
clustering is so similar at the three redshifts, we infer that light-cone
effects should have a negligible impact on our theoretical template.  That is,
our theoretical template from a fixed redshift output should be nearly the same
as if we had generated it from a light cone (after perhaps small shifts in HOD
parameters).  

We have tested this assertion by refitting the measurements using the halo
catalog from the MedRes box output at $z=0.65$, where $\sigma_8$ is 5\% lower
than at our fiducial $z=0.55$.  The matter clustering in this redshift slice
should be closer to the clustering in a model with the overall amplitude of
fluctuations lowered to match our best fit $f\sigma_8$.  However, the value of
$f\sigma_8$ in the higher redshift slice is only smaller by 1\% since
$\Omega_m(z)$ increases with redshift.  Accounting for this difference, we find
$f\sigma_8 = 0.449 \pm 0.008$, in excellent agreement with our fiducial fit
using the $z=0.55$ halo catalog. 

One additional impact of cosmological parameters is in the conversion of angles
and redshifts into comoving coordinates.  As in \citet{Anderson13} and
\citet{Samushia13}, we assume a cosmology with $\Omega_m = 0.274$ and $h=0.7$
for this conversion.  We did not account for the difference between the fiducial
cosmology and simulation cosmology in the theoretical model, but the error on
the angle-averaged distance scale $D_V \propto D_A^{2/3}(z_{\rm eff})
H^{-1/3}(z_{\rm eff})$ in $h^{-1}$ Mpc units is only 1.0\%.  We checked that
this error has a negligible impact on our theoretical predictions: the
difference amounts to $\Delta \chi^2 = 0.9$ at fixed HOD parameters.  The
Alcock-Paczynski \citep{AP} parameter $F_{AP} \propto D_A(z_{eff}) H(z_{eff})$
distorts line-of-sight distances relative to transverse distances and at fixed
$D_V$, a change in $F_{AP}$ alters $\xi_2$ while holding $\xi_0$ basically
fixed.  Fig.~\ref{fig:APtest} compares the impact of 10\% changes in $f\sigma_8$
and $F_{AP}$ on $\hat{\xi}_2$ at fixed HOD parameters; changes caused by the two
parameters are distinguishable because of their differing scale dependence.
Allowing uncertainty in $F_{AP}$ will however degrade our constraints on
$f\sigma_8$; in \citet{Reid12} and \citet{Samushia13} we report joint
constraints on $f\sigma_8$-$F_{AP}$.  Since geometric parameters are ``slow''
variables in our theoretical calculation (they alter the separation between halo
pairs), we defer a joint $F_{AP}-f\sigma_8$ constraint to future work.  Note
that non-cosmological constant dark energy affects both geometric and growth of
structure parameters, so our measurement of $f\sigma_8$ cannot be used to
constrain dark energy without accounting for this degeneracy.  The current work
can be considered only a consistency test of the $\Lambda$CDM + general
relativity model, where $F_{AP}$ is constrained to within 0.6\%
\citep{PlanckXVI}.  At fixed cosmological and HOD parameters in our model,
varying $F_{AP}$ by 1.2\% produces a change in the theoretical prediction of
$\Delta \chi^2 = 1$, so we can safely neglect this uncertainty when testing
models that assume a flat $\Lambda$CDM expansion history.

\subsection{Sensitivity of $f\sigma_8$ constraint to small scales}
Finally, we assess the sensitivity of our $f\sigma_8$ constraint to the
small-scale velocity distribution probed by $\hat{\xi}_{0,2}$ by performing fits
to $w_p(r_{\sigma} < 2$ $h^{-1}$ Mpc) + $\hat{\xi}_{0,2}(s < 10.3$ $h^{-1}$ Mpc)
and  $w_p(r_{\sigma} < 2$ $h^{-1}$ Mpc) + $\hat{\xi}_{0,2}(s > 10.3$ $h^{-1}$
Mpc); that is, the first (second) choice combines our fiducial $w_p$
measurements with the first five (last four) $s$ bins of our $\hat{\xi}_{0,2}$
measurement.  The first fit essentially recovers the results of our fiducial fit
including all $s$ bins to nearly the same precision, implying that essentially
all of our $f\sigma_8$ information comes from these non-linear scales and
therefore potentially sensitively depends on the accuracy of our HOD modeling
approach.  Table \ref{tab:hod} shows that for a reasonable range of extensions
to our fiducial model, the $f\sigma_8$ constraint is stable.

The fit restricted to larger $s$ bins has considerable shifts in HOD parameters.
While the satellite fraction is still well-constrained with the same central
value, the distribution of the satellites shifts to lower halo masses.  Rather
than a constraint on $M_{\rm cut}$, this fit prefers $log_{10} M_{\rm cut} <
13.04$ with 95\% confidence, $log_{10} M_1$ increases by 0.2, and $\alpha = 1.07
\pm 0.12$.  This HOD model has weaker FOG features, which  therefore lowers
$f\sigma_8$ to $0.435 \pm 0.033$ but also alters $\hat{\xi}_{0,2}$ on small
scales.  The best fit to $w_p(r_{\sigma} < 2$ $h^{-1}$ Mpc) + $\hat{\xi}_{0,2}(s
> 10.3$ $h^{-1}$ Mpc) is strongly disfavored using our fiducial set of
measurements ($\chi^2 = 70$).

\subsection{Predicting $\sigma^2_{\rm FOG}$} \label{sec:sig2results}
\begin{figure} \includegraphics[width=85mm]{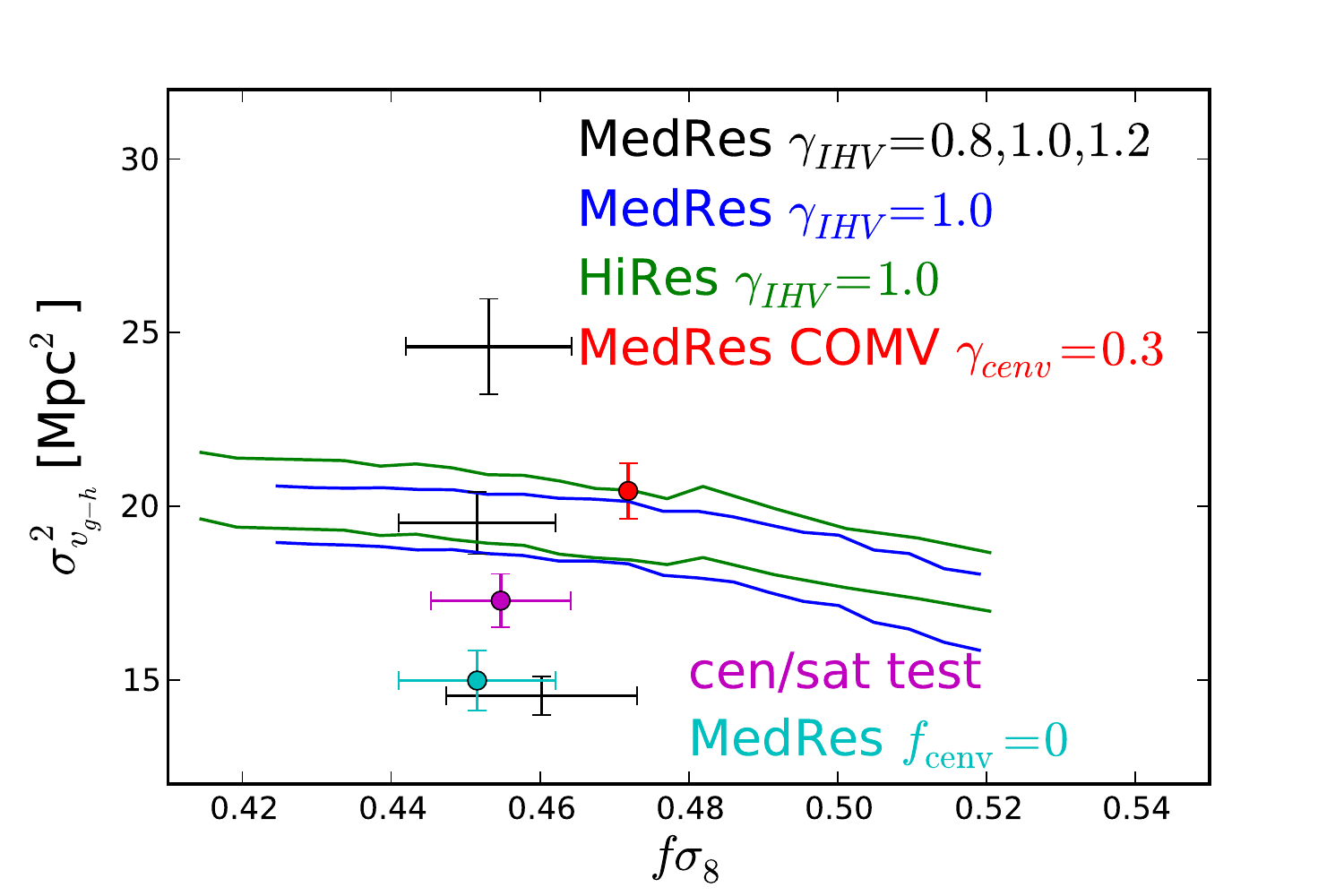} \caption{Second moment of
the velocity dispersion of centrals and satellites relative to the
center-of-mass velocity of their host dark matter halos computed from the HOD
constraints presented in Table \ref{tab:hod}.  Ignoring the goodness-of-fit, we
find the dependence of $\sigma_{v_{g-h}}$ as a function of $f\sigma_8$ using
several chains at fixed $f\sigma_8$ values.  The MedRes (blue) and Hires (green)
fits give similar results.  The black points show the constraints from the
chains that vary $f\sigma_8$ at fixed $\gamma_{\rm IHV} = 0.8, 1.0, 1.2$ using
the MedRes0 simulation box.  Larger $\gamma_{\rm IHV}$ corresponds to a larger
$\sigma^2_{v_{g-h}}$.   Neglecting the fiducial central galaxy dispersion moves
the central black constraint to the cyan one, demonstrating that central galaxy
intra-halo velocities are non-negligible.  Finally, we obtain a similar value
for $\sigma^2_{v_{g-h}}$ when we assign central galaxies the center-of-mass
velocity of their halo and then add velocity dispersion with $\gamma_{\rm cenv}
= 0.3$ (red) or when using the model described in Sec.~\ref{sec:robustresults}
as the ``cen/sat'' test (magenta).  $\sigma_{v_{g-h}}$ should be directly
related to the nuisance parameter $\sigma^2_{\rm FOG}$ used in \citet{Reid12}
and \citet{Samushia13}.} \label{fig:sig2vgh} \end{figure} \begin{figure}
\includegraphics[width=85mm]{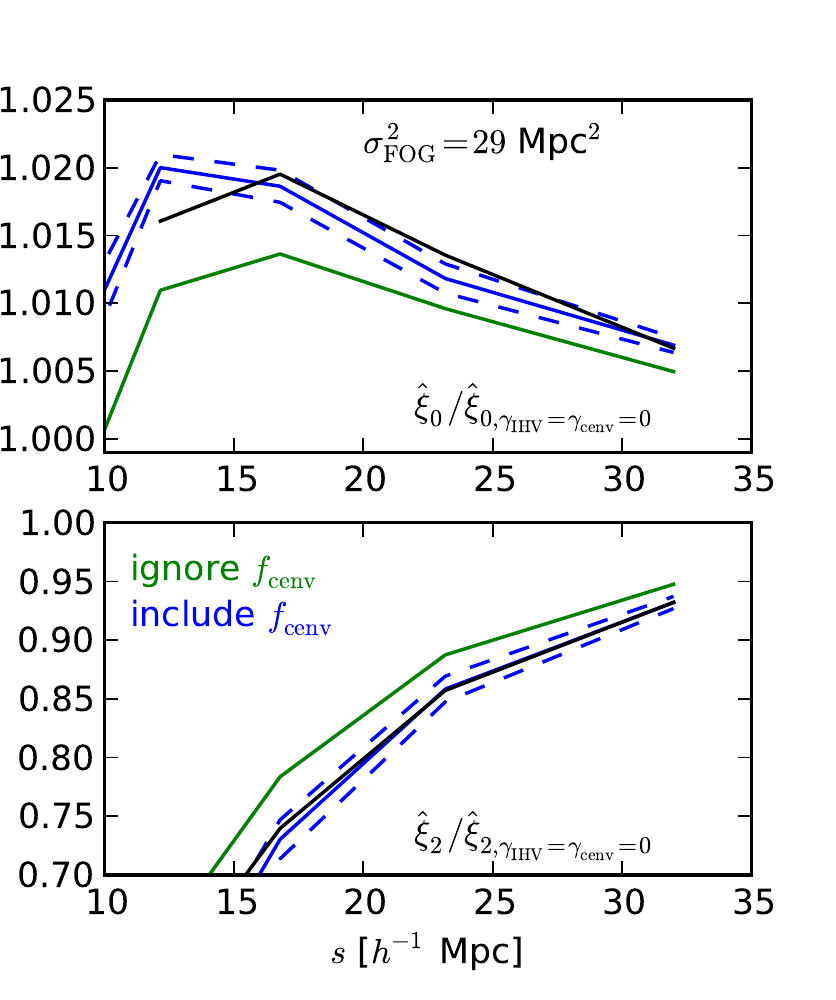} \caption{We use the ratios
$\hat{\xi}_{\ell}/\hat{\xi}_{\ell,\gamma_{\rm IHV} = \gamma_{\rm cenv}= 0}$ for
$\ell=0,2$ as a proxy for the nuisance parameter $\sigma^2_{\rm FOG}$.  These
ratios are tightly constrained by our measurements (blue bands).  For this
purpose we compute the theoretical prediction using the center-of-mass velocity
${\bf v}_{\rm COMV}$ for the central galaxies to account for the dispersion in
${\bf v}_{DENS} - {\bf v}_{\rm COMV}$; neglecting this term produces the green
curves.  We compute the analogous quantity using {\em cosmoxi2d} as a function
of nuisance parameter $\sigma^2_{\rm FOG}$, recovering 29 Mpc$^2$ as the best
fit for the case where $\gamma_{\rm IHV} = 1.0$ and $f\sigma_8$ is allowed to
vary.  Reassuringly, the {\em cosmoxi2d} model reproduces the scale-dependence
of this ratio as measured from our HOD model predictions.} \label{fig:sig2rat}
\end{figure}

In \citet{Reid12} and \citet{Samushia13}, we analysed the large scale
$\xi_{0,2}$ with an analytic Gaussian streaming model to constrain $f\sigma_8$
along with geometric parameters $D_A(z_{\rm eff})$ and $H(z_{\rm eff})$.  The
model has a single parameter, $\sigma^2_{\rm FOG}$, to account for the effect of
small-scale motions, like satellite galaxies within their host halos
(traditional ``fingers-of-god'').  The model convolves the predicted halo
auto-correlation function with a Gaussian, approximating the velocities of
galaxies relative to their host halos as uncorrelated with the quasilinear flows
of interest.  We can estimate the probability distribution function of those
velocities, assuming the central galaxies also have some residual motion
specified by a fraction $f_{\rm cenv}$ of the halo virial velocity.  Using the
halo model and assuming a central is required for a halo to host a satellite, we
can estimate the probability distribution of galaxy velocities relative to their
host halos as \begin{eqnarray} p(\Delta v_{g-h}) = n_{tot}^{-1} \int dM n(M)
N_{\rm cen}(M) \times \nonumber \\ \left[G\left(\Delta v_{g-h}, f_{\rm cenv}
\sigma_{\rm vir}^2(M)\right) + N_{sat}(M) G\left(\Delta v_{g-h}, \sigma_{\rm
vir}^2(M)\right) \right], \label{eq:pdv} \end{eqnarray} where $G(\Delta v_{g-h},
\sigma^2)$ is a Gaussian probability distribution with variance $\sigma^2$ and
mean 0, and $n(M)$ is the halo mass function.  To get the pairwise velocity
distribution component due to galaxy motions relative to the center-of-mass
velocity of their host halos for pairs in different host halos, we convolve
Eq.~\ref{eq:pdv} with itself.  The result has a narrow distribution about
$\Delta v_{g-h} = 0$ from central galaxy pairs, and an exponential tail due to
the much larger satellite galaxy velocities.  We can evaluate the second moment
of this distribution, $\sigma^2_{v_{g-h}}$,  at each point in our chains to
determine its mean and uncertainty.  68\% confidence intervals are shown in
Fig.~\ref{fig:sig2vgh} as a function of $f\sigma_8$ for both the HiRes (green)
and MedRes (blue) simulations over a broader range of $f\sigma_8$ values than is
preferred by our fiducial MedRes fits (central black point, with 68\% confidence
in $f\sigma_8$ also shown).  The plot shows that $\sigma^2_{v_{g-h}}$ is very
well constrained by our measurements and has only a modest degeneracy with
$f\sigma_8$.  Varying the satellite galaxy relative velocities by $\gamma_{\rm
IHV}$ by $\pm 20\%$ (upper and lower black points) also changes
$\sigma^2_{v_{g-h}}$ by 25\%; note from Table \ref{tab:hod} that $\gamma_{\rm
IHV} = 0.8$ provides a poor fit to our measurements of $\hat{\xi}_{0,2}$.  Also
shown in red is the constraint derived from a MedRes chain adopting
center-of-mass velocities for central galaxies, but adding a random dispersion
to central galaxies of magnitude $0.3 \sigma_{\rm vir}$ and holding $\gamma_{\rm
HV}=1$ fixed.  The ``cen/sat test'' model (magenta point) described in
Sec.~\ref{sec:robustresults} relaxes the assumption that halos hosting satellite
CMASS galaxies also always host centrals; this model produces a similar value of
$\sigma^2_{v_{g-h}}$ as our fiducial one.  Finally, we can also separate the
contributions from central and satellites to the dispersion in Eq.~\ref{eq:pdv}.
In our default chain, we find the central term to contribute $4.5 \pm 0.1$ and
the satellite term to contribute $15.0 \pm 0.9$ (shown in cyan on the figure).
It is therefore imperative to allow for some dispersion in both the central and
satellite galaxies, relative to the bulk halo motion.  To incorporate a
reasonable uncertainty on $\gamma_{\rm IHV}$ and $\gamma_{\rm cenv}$, we suggest
a conservative Gaussian prior on $\sigma^2_{v_{g-h}}$ centered at 19.5 Mpc$^2$
and with uncertainty of $\sqrt{2} \times 5$ Mpc$^2$ to account both for
$1\sigma$ uncertainty corresponding to $\gamma_{\rm IHV}$ uncertain by 20\% and
central galaxy dispersion uncertain at the 100\% level.

Before we can apply this prior to our analysis of large scale clustering, we
need to understand the relation between the nuisance parameter $\sigma^2_{\rm
FOG}$ and $\sigma^2_{v_{g-h}}$, which we estimate from the HOD constraints from
small scale clustering; the relation is non-trivial due to the non-Gaussianity
of $p(\Delta v_{g-h})$.  Unfortunately, constraints on $\sigma^2_{\rm FOG}$ by
fitting mock catalogs directly with {\em cosmoxi2d} (the theoretical prediction
software used in \citealt{Reid12} and \citealt{Samushia13}) on the same large
scales as the data is extremely noisy, even at known $f\sigma_8$ and geometric
parameters.  We remove much of the cosmic variance in the inference of
$\sigma^2_{\rm FOG}$ from mock catalogs by considering the ratios
$\hat{\xi}_{0,2}/\hat{\xi}_{0,2,\gamma_{\rm IHV} = \gamma_{\rm cenv}= 0}$.  In
particular, we assign center-of-mass halo velocities to the central galaxies to
compute the denominator but use ${\bf v}_{\rm DENS}$ in the numerator, since we
found the first term in Eq.~\ref{eq:pdv} is not negligible, and the majority of
perturbation theory models (including {\em cosmoxi2d}) are validated using halo
catalogs containing ${\bf v}_{\rm COMV}$.  The resulting ratio is shown in blue
in Fig.~\ref{fig:sig2rat} for the MedRes box.  A 68\% error band shown as blue
dashed curves are derived from the uncertainty on the HOD parameters using the
default chain with $\gamma_{\rm IHV} = 1.0$, $\gamma_{\rm cenv} = 0$, and
$\gamma_{\rm HV}$ free; comparison with two additional MedRes boxes indicates
that cosmic variance is a subdominant contribution to the uncertainty for the
ratio.  

We compute an analogous ratio using {\em cosmoxi2d}, varying $\sigma^2_{\rm
FOG}$ in the numerator and setting it to 1 Mpc$^2$ in the denominator,
consistent with the expected value for halos reported in \citet{ReiWhi11}.  {\em
cosmoxi2d} is based on perturbation theory, and the underlying model for halo
clustering breaks down at $\sim 25 \, h^{-1}$ Mpc.  We therefore determine the
best fit $\sigma^2_{\rm FOG}$ using only the last bin in our small-scale
measurements, 27-38 $h^{-1}$ Mpc and find $\sigma^2_{\rm FOG} = 29 \pm 2$
Mpc$^2$.  Fig.~\ref{fig:sig2rat} shows that the scale-dependent distortions to
$\xi_{0,2}$ caused by the relative velocities between halos and galaxies is
described well by the nuisance parameter $\sigma^2_{\rm FOG}$, even to smaller
scales than included in the {\em cosmoxi2d} analysis.  Experimenting with a few
other cases, we find the mapping $\sigma^2_{\rm FOG} = 1.5 \sigma^2_{v_{g-h}}$
to be a good predictor for the best fit {\em cosmoxi2d} nuisance parameter.
Thus we adopt a Gaussian prior on $\sigma^2_{\rm FOG} = 29 \pm 10$ Mpc$^2$ and
recompute DR11 large scale constraints in the case assuming a $\Lambda$CDM
expansion history, where \citet{Samushia13} originally found $f\sigma_8 = 0.447
\pm 0.028$.  With the new $\sigma^2_{\rm FOG}$ prior, the constraint shifts to
$f\sigma_8 = 0.457 \pm 0.025$.  The modest reduction in the uncertainty is not
surprising, since the original hard prior on $\sigma^2_{\rm FOG}$ (0-50 Mpc$^2$)
was only slightly broader than the new prior.  The small shift in the central
value comes from the new prior eliminating the lowest end of the originally
allowed range of $f\sigma_8$; the small-scale clustering requires satellite
galaxies, and thus non-zero $\sigma^2_{\rm FOG}$.  One important difference
between the small-scale and large scale analyses is the use of FKP weights in
the latter; we therefore verified that we recover very similar small-scale
clustering when applying the FKP weights, so that our derived prior on
$\sigma^2_{\rm FOG}$ remains valid for an FKP weighted correlation function used
to analyse large scales.

\section{Implications for modified gravity models} \label{sec:modgrav}
Substantial research efforts have recently been devoted to exploring
modifications of gravity as a means to explain the apparent cosmic acceleration
\citep{Clifton12}.  \citet{Schmidt09} showed that for the $f(R)$ model,
constraints on non-linear scales, in this case from clusters, are $10^{4}$ times
stronger than those obtained on quasi-linear scales.  Recently, the velocity
structure around massive halos on 1-30 Mpc scales has been identified as a
promising observational probe of modified gravity models, both for $f(R)$ models
with the chameleon screening mechanism \citep{Lam12,Lam13,Zu13} and a galileon
model with the Vainshtein screening mechanism \citep{Zu13}.  In both the $f(R)$
and galileon models studied by \citet{Zu13}, the infall velocity around
$10^{14}$ $h^{-1}$ $M_{\sun}$ halos at $z=0.25$ was enhanced by $\sim 20-40\%$
on scales of 5 $h^{-1}$ Mpc, the real space halo-matter cross-correlation
function showed a scale-dependent enhancement, peaking at $\sim 40\%$ at 2
$h^{-1}$ Mpc, and velocity dispersions increased as well.  All three effects
would propagate to our $\hat{\xi}_{0,2}$ observable, and we would expect
modifications of the same order.  \citet{Lam12,Lam13} frame this gravity test in
combination with weak lensing, used to measure the mass of the central halos;
\citet{Zu13} showed that similar deviations persist in abundance matched samples
of halos.  While we plan to incorporate galaxy-galaxy lensing constraints on
halo masses in future work \citep{Leauthaud0X}, one can still search for the
signatures of modified gravity using our measurements, but comparing to
clustering-amplitude matched halo samples.  The overall amplitude of galaxy
clustering observed in our sample constrains the product of the mean halo bias
(determined by the galaxy HOD) and the amplitude of matter fluctuations at the
effective redshift of the galaxy sample, $b(M, z_{\rm eff}) \sigma_8(z_{\rm
eff})$.  Thus, for a given modified gravity model (realized with an $N$-body
simulation), the HOD would be constrained by the same procedure as we have
implemented here for the case of general relativity.  With the overall amplitude
of clustering matched on $\sim 30$ Mpc scales, the modifications to the pairwise
infall velocities and dispersions will propagate to scale-dependent changes in
our $\hat{\xi}_{0,2}$ observables.  While we are unable to provide any
quantitative constraints without halo catalogs derived from modified gravity
simulations, the $\sim 2.5\%$ precision of our GR-based $f\sigma_8$ constraint
should severely limit the types of modifications described above. 

\section{Conclusions and Future Prospects} \label{sec:conc} We have made the
most precise comparison to date between the observed anisotropic clustering of
galaxies at relatively small separations ($\sim 0.8 - 32$ $h^{-1}$ Mpc) and the
predictions of a standard halo model in the context of $\Lambda$CDM.  We found
good agreement between our simplified, redshift-independent HOD model and our
measurements of both the projected  and anistropic clustering on small scales.
Our fits constrain the growth rate of cosmic structure at the effective redshift
of our galaxy sample: $f\sigma_8(z_{\rm eff} = 0.57) = 0.450 \pm 0.011$.  This
constraint is consistent with but improves on our DR11 analysis of large scale
anistropy \citep{Samushia13} by a factor of 2.5.  Intriguingly, our result has
the same statistical power but is $\sim 1.9\sigma$ low compared with Planck's
$\Lambda$CDM prediction, $f\sigma_8 = 0.480 \pm 0.010$ \citep{PlanckXVI}. 

The competitive statistical precision of our measurement warrants a systematic
evaluation of the observational and modeling systematics.  For the former, we
introduced a new anisotropic clustering statistic $\hat{\xi}_{0,2}$ that does
not include information below the fiber collision scale, but approaches the
usual multipoles on large scales.  We carefully assessed the systematic and
observational uncertainties from the angular upweighting method to correct fiber
collisions to order to estimate the projected correlation function
$w_p(r_{\sigma})$.  We combined these measurements to obtain robust joint
constraints on the halo occupation distribution and growth rate of cosmic
structure probed by CMASS galaxies. 

To assess the robustness of our modeling assumptions, we investigated several
generalizations to our HOD assumptions and particularly how we assign velocities
to the mock galaxies from which we draw our theoretical predictions; the results
are summarized in Table \ref{tab:hod}.  The variations we examined caused at
most $\sim 0.5\sigma$ shifts in the $f\sigma_8$ constraints.   However, given
the statistical precison of our reported constraint, further investigation with
more sophisicated modeling of the galaxy-halo connection is warranted.  Of the
possibilities we explored, a model that relaxes the assumption that halos
hosting satellite galaxies also host centrals (labelled ``cen/sat test'')
improved the fit to our measurements by $\Delta \chi^2 = 10$ but did not shift
$f\sigma_8$ constraints appreciably.  Such a model is well-motivated by our
target selection process -- both color cuts and photometric errors cause massive
galaxies to scatter in and out of the sample.  Alternatively, we can also
improve the model fit by increasing the satellite galaxy velocity dispersion at
fixed halo mass. 

At least within the cosmological parameter space we explored, we found that for
two different definitions of the central galaxy velocity, the data prefer ${\bf
v}_{\rm DENS}$, the motion of the densest $\sim 0.2r_{\rm vir}$ clump, over
${\bf v}_{\rm COMV}$, the halo center-of-mass velocity averaged over all
particles within $\Delta_m = 200$.  A comparison of these two velocity fields
also indicates a possible shift of $\sim 1.5\%$ in the effective large scale
$f\sigma_8$, and should therefore be considered when this level of precision is
reached.

While we have not tested any explicit modified gravity models, we have shown
that the clustering of few $\times 10^{13}$ $h^{-1}$ Mpc halos are consistent
with the expectations of $\Lambda$CDM and a simple picture of galaxy formation
in which halo mass is the only relevant variable determining the probability of
hosting a CMASS galaxy.  To quantify the precision of this test, our best fit
model matches the observed $\hat{\xi}_{0}$ at the 3 per-cent level from 0.8 - 32
$h^{-1}$ Mpc, and 15 to 5 per-cent level from 5 to 32 $h^{-1}$ Mpc for
$\hat{\xi}_2$, with reasonable agreement compared to our uncertainties on
smaller scales as well.

As the example of $f(R)$ shows, modified gravity could potentially
dramatically alter structure growth on these scales, and our analysis should be
used to constrain such models.  In addition to the $f\sigma_8$ constraint
afforded by our measurements, more precise galaxy velocity bias predictions in
$\Lambda$CDM would allow our joint constraints on $\gamma_{\rm IHV}$ and
$b\sigma_8$ to be interpreted as an additional consistency test between the halo
mass inferred from clustering amplitude $b\sigma_8$, and from the halo virial
velocities probed by $\gamma_{\rm IHV}$.

Finally, even ignoring the information of the small-scale clustering on
$f\sigma_8$, our data tightly constrain the impact of the intra-halo motions of
galaxies on clustering at relatively large scales.  We derive a prior on the
``finger-of-god'' nuisance parameter that is tighter but consistent with the
prior adopted in \citet{Reid12} and \citet{Samushia13}.  Moreover, our detailed
study of the clustering on small scales also allowed us to validate that
$\sigma^2_{\rm FOG}$ as defined in those works can precisely describe the impact
of intra-halo velocities of CMASS galaxies on quasi-linear scales. 

\section{Acknowledgements}
We thank Cameron McBride for help in creating the tiled mock, as well as Zheng
Zheng, Hong Guo, and the anonymous referee for useful discussions/suggestions.

Funding for SDSS-III has been provided by the Alfred P. Sloan
Foundation, the Participating Institutions, the National Science
Foundation, and the U.S. Department of Energy Office of Science.
The SDSS-III web site is http://www.sdss3.org/.

\noindent SDSS-III is managed by the Astrophysical Research Consortium for the
Participating Institutions of the SDSS-III Collaboration including the
University of Arizona,
the Brazilian Participation Group,
Brookhaven National Laboratory,
University of Cambridge,
Carnegie Mellon University,
University of Florida,
the French Participation Group,
the German Participation Group,
Harvard University,
the Instituto de Astrofisica de Canarias,
the Michigan State/Notre Dame/JINA Participation Group,
Johns Hopkins University,
Lawrence Berkeley National Laboratory,
Max Planck Institute for Astrophysics,
Max Planck Institute for Extraterrestrial Physics,
New Mexico State University,
New York University,
Ohio State University,
Pennsylvania State University,
University of Portsmouth,
Princeton University,
the Spanish Participation Group,
University of Tokyo,
University of Utah,
Vanderbilt University,
University of Virginia,
University of Washington,
and Yale University.

\noindent The simulations used in this paper were analysed at the
National Energy Research Scientific Computing Center, the Shared Research
Computing Services Pilot of the University of California and the
Laboratory Research Computing project at Lawrence Berkeley National Laboratory.

\noindent BAR gratefully acknowledges support provided by NASA through Hubble
Fellowship grant 51280 awarded by the Space Telescope Science Institute, which
is operated by the Association of Universities for Research in Astronomy, Inc.,
for NASA, under contract NAS 5-26555.

\noindent This work was supported in part by the National Science Foundation
under Grant No. PHYS-1066293 and the hospitality of the Aspen Center for
Physics.

\noindent AL was supported by the World Premier International Research Center
Initiative (WPI Initiative), MEXT, Japan.

\appendix
\section{Redshift evolution of $\hat{\xi}_{0,2}$} 
\label{sec:zdep}
\begin{figure*} \includegraphics[width=140mm]{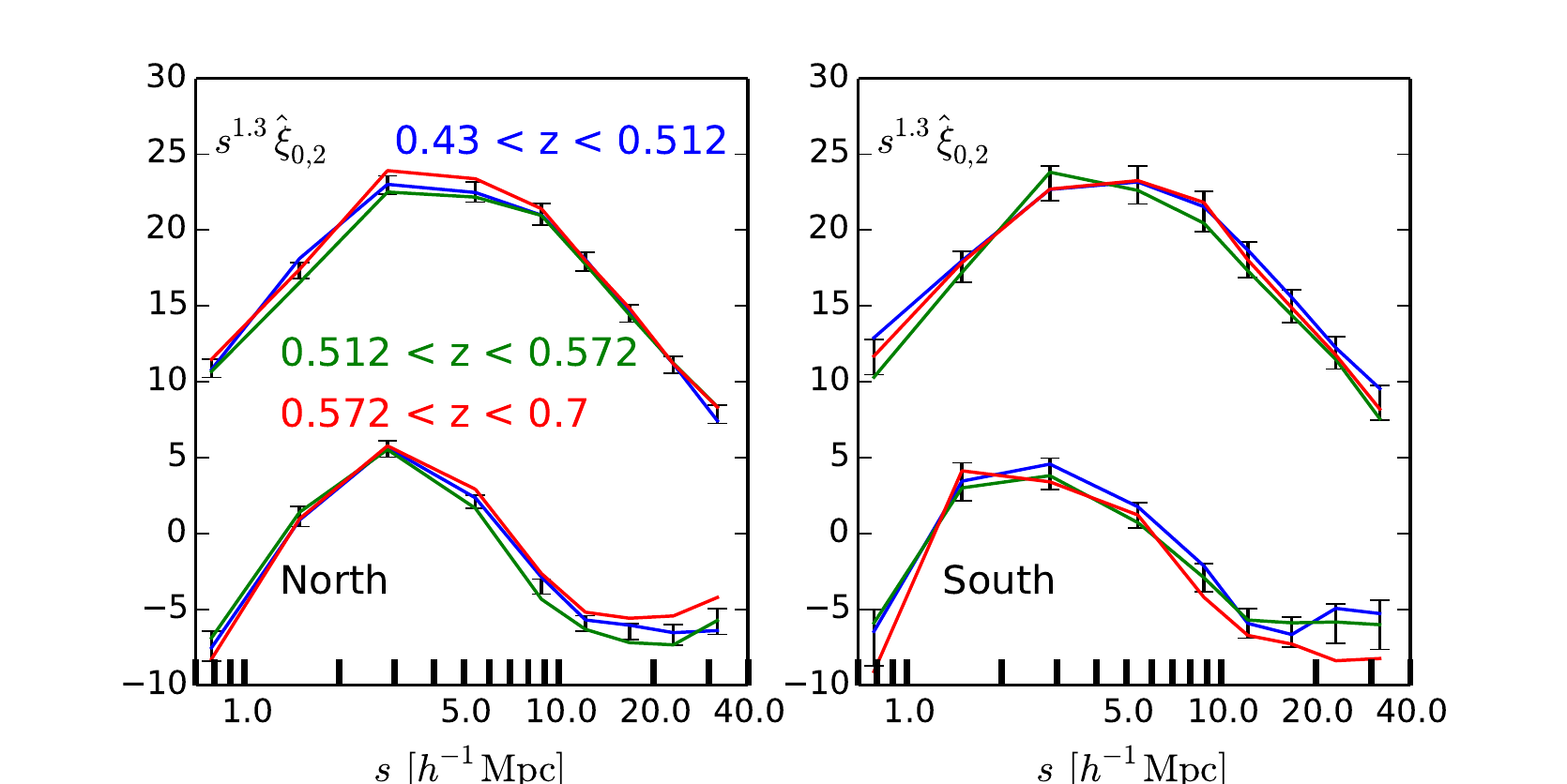}
\caption{$\hat{\xi}_{0,2}$ measured using the nearest neighbor fiber collision
correction method for three redshift subsamples with equal numbers of galaxies,
computed separately in the north and the south.  Remarkably, our target
selection algorithm selects galaxies with very similar clustering across the
redshift range of our sample, even though the linear growth factor increases by
14\% in our fiducial cosmology between $z=0.7$ and $z=0.43$.  The black error
bars indicate our measurement of $\hat{\xi}_{0,2}$ for the full sample; the
error bars shown are the square root of the diagonal elements of the measurement
covariance matrix for the full sample, after rescaling it by the fraction of
galaxies in the north or south and multiplying by a factor of 3 to approximately
indicate the level of scatter expected between the redshift slices.  Using the
same rescaled covariance matrix to assess consistency between the subsamples and
full sample, we find $\chi^2 \approx 12 \pm 2$ in each redshift except the high
redshift bin in the north ($\chi^2 \approx 33$).} \label{fig:zdep}
\end{figure*}

Throughout this analysis we
have treated the CMASS sample within $0.43 < z < 0.7$ as a single population.
One reason for this is that fiber collision corrections using the angular
upweighting become substantially more uncertain as the spectrosopic sample
becomes a smaller subset of the target sample.  However, the nearest neighbor
correction method should be valid in arbitrary redshift bins.  Fig.~\ref{fig:zdep}
shows our measurement of $\hat{\xi}_{0,2}$ using the nearest neighbor redshift correction 
for redshift cuts that split the
sample equally into three bins in both the north and south.  These bins also
have different galaxy weighted number densities: $10^4\bar{n} = $ 3.4, 3.4, and
1.5 $(h^{-1} {\rm Mpc})^{-3}$.  Rescaling the bootstrap covariance matrix
derived for the full sample by $N_{\rm tot}/N_{\rm subsample}$ to compute an
approximate $\chi^2$ difference between the subsamples and the full $0.43 < z < 0.7$
sample, we find these subsample clustering measurements to be consistent with
being drawn from the full sample ($\chi^2 \approx 12 \pm 2$), with the exception
of the highest redshift bin in the north ($\chi^2 \approx 33$ for 18 bins);
comparison with the same bin in the south suggests most of the difference could
be cosmic variance rather than a systematic change from the clustering of the
full sample.  The covariance matrix used for this comparison is approximate and
does not include any of our theoretical error budget.  Remarkably, our target
selection algorithm selects galaxies with very similar clustering across the
redshift range of our sample, even though the linear growth factor increases by
14\% in our fiducial cosmology between $z=0.7$ and $z=0.43$.

\section{Sensitivity to the halo central velocity definition}
\label{sec:vcen}
\begin{figure*} \includegraphics[width=150mm]{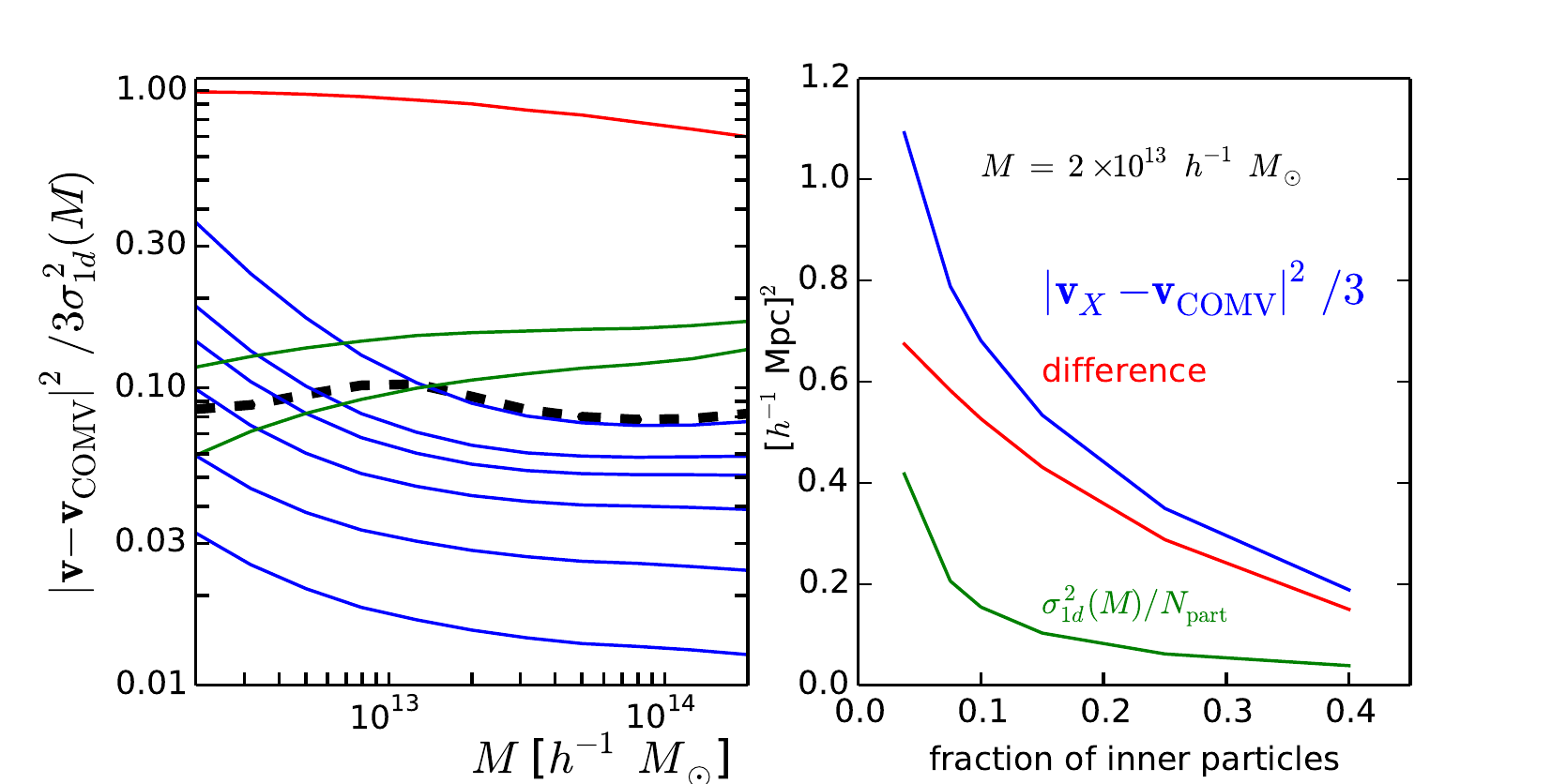}
\caption{Square differences between ${\bf v}_{\rm COMV}$, the halo center of
mass velocity, and various other definitions of the ``central'' velocity
measured from our MedRes simulation.  In the left panel we normalize by
$3\sigma^2_{1d}(M)$, the three-dimensional halo velocity dispersion averaged
over all halo members.  The thick dashed black curve shows the fiducial
``central'' definition detailed in Sec.~\ref{sec:Nbody} and adopted throughout
for our analysis.  For ease of comparison with other work, for all other curves
we define the center by the minimum of the potential; we verified that the two
centers have negligible offsets (0.01-0.02 $h^{-1}$ Mpc).  To compute the red
curve we simply take the velocity of the particle at the potential minimum; the
dispersion of this particle is the same as the average halo particle at low
masses and slightly lower in high mass halos.  The green curves are computed by
averaging over the nearest 10 and 20 particles.  The blue curves average over a
fixed fraction of the innermost halo members: 0.0375, 0.075, 0.1, 0.15, 0.25,
and 0.4 from top to bottom.  The right panel shows how the one-dimensional
square difference depends on that fraction (blue curve) for the halo mass bin $M
= 2 \times 10^{13}$ $h^{-1}$ $M_{\odot}$.  If velocities of the innermost
particles were uncorrelated (i.e., there were no bulk motion) and they had the
same dispersion as the full halo member population ($\sigma^2(M)_{1d}$), then we
would expect a much smaller value for this measurement (green line).  The red
curve is the difference of the two, indicating a significant bulk velocity
relative to the center-of-mass.  The amplitude of the central bulk velocity
depends on the averaging scale, with no apparent convergence on small scales.
Results are similar for other mass bins, with the noise contribution getting
smaller at larger halo masses.} \label{fig:vcenapp} \end{figure*}

Using the MedRes simulation box, we explore in the left panel of
Fig.~\ref{fig:vcenapp} the mean square velocity difference between the halo
center of mass velocity and various central velocity definitions, in units of
the halo velocity dispersion: the velocity of the particle at the potential
minimum (red), the velocity averaged over the innermost 10 and 20 particles
(green), the velocity averaged over a fixed fraction of the innermost halo
members, ranging from 3.75\% to 40\% (blue), and the fiducial choice adopted in
this work (thick black dashed line) and detailed in Sec.~\ref{sec:Nbody}.  The
right panel demonstrates that the mean square dispersion of the central velocity
depends on the fraction of innermost halo particles used to determine the
central velocity (blue curve).  For uncorrelated intrahalo motions, we would expect the green
curve; we interpret the difference (red curve) as evidence for bulk motion of the central
galaxy. This naive noise estimate suggests that our fiducial central velocity
definition has a sizeable contribution from particle noise.  We integrate the
square velocity difference in the central velocity dispersion between our MedRes
and HiRes simulations (lower dashed and solid curves in
Fig.~\ref{fig:comparevcatsvsM}) over the best fit HOD, and find only a 13\%
excess in the MedRes simulation.  Since the HiRes mass resolution is eight times
larger, we are therefore confident that our central galaxy velocities are
negligibly affected by resolution in the mass range of interest for this
analysis.

However, since their is no apparent convergence of the central velocity
dispersion at small smoothing scales, we expect our predictions for galaxy
clustering to be sensitive to this choice.  In Sec.~\ref{sec:Nbody} we argued
that our fiducial choice corresponds to the typical galaxy size for the
population we are modeling, at least over the halo mass range that dominates the
clustering signal.  Moreover, our fiducial choice provides approximately the
correct amount of central velocity dispersion to match the observed
$\hat{\xi}_{0,2}$; slight modifications to the central velocity definition may
improve the fit on small scales.  We have not explored this possibility further
here, but hope to in future work.  We note that the final test in
Sec.~\ref{sec:robustresults} showed that our measurements do not favor
additional random central velocity dispersion.  Examination of hydrodynamic
galaxy formation simulations would shed light on both the impact of baryonic
effects and could determine the best algorithm to estimate the central galaxy
velocity from dark matter-only simulations.  

Finally, we note that a similar effect was discussed in \citet{Behroozi13};
their Figure 11 includes the {\em median} three-dimensional velocity offset at
$z=0.53$ in halo mass bins, but averaged in spherical shells rather than
including all particles within a given radius.  Nonetheless, from their plots we
would expect $\sim 55 (125)$ km $s^{-1}$ for $10^{13}$ and $10^{14}$ $M_{\odot}$
halos, averaged over 0.06 (0.12) Mpc, while our measurements shown in
Fig.~\ref{fig:comparevcatsvsM} predicts $\sim$110 (237) km $s^{-1}$ for the rms
three-dimensional rms velocity at this redshift.  In addition, we find the
one-dimensional velocity distribution of $v_{x,\rm DENS} - v_{x,\rm COMV}$ to be
approximately exponential, for which we expect the rms to be larger than the
median by a factor of $\sqrt{2}/\ln{2} \approx 2.04$.  Therefore the amplitude
of the bulk velocities in this work seems consistent with that found in
\citet{Behroozi13} using the ROCKSTAR halo finder with slightly higher mass
resolution.  


\begin{thebibliography}{1999}
\bibitem[{{Planck Collaboration}(2013)}]{PlanckXVI}
  Ade P. et al., 2013, arXiv:1303.5076

\bibitem[{{Ahn et al.}(2013)}]{Ahn14}
  Ahn C., et al., 2014, ApJS, 211, 17 

\bibitem[{{Aihara et al.}(2011)}]{DR8} 
  Aihara H., et al., 2011, ApJS, 193, 29

\bibitem[{Alcock \& Paczynski}(1979)]{AP}
  Alcock C., Paczynski B., 1979, Nature, 281, 358

\bibitem[{{Anderson et al.}(2013)}]{Anderson13}
  Anderson L., et al., 2013, arXiv:1312.4877

\bibitem[{{Behroozi, Wechsler, \& Wu}(2013)}]{Behroozi13}
  Behroozi P., Wechsler R., Wu H., 2013, ApJ, 762, 109

\bibitem[{{Benson et al.}(2000)}]{Benson00}
  Benson A., Cole S., Frenk C., Baugh C., Lacey C., 2000, MNRAS, 311, 793

\bibitem[{{Berlind \& Weinberg}(2002)}]{Berlind02}
  Berlind A., Weinberg D., 2002, ApJ, 575, 587

\bibitem[{{Beutler et al.}(2013)}]{Beutler13}
  Beutler F., et al., 2013, arXiv:1312.4611

\bibitem[{{Blanton et al.}(2003)}]{Tiling}
  Blanton M., et al., 2003, AJ, 125, 2276

\bibitem[{{Bolton et al.}(2012)}]{Bolton12}
  Bolton A., et al., 2012, AJ, 144, 144 

\bibitem[{Cacciato et al.}(2013)]{Cacciato13}
  Cacciato M., van den Bosch F., More S., Mo H., Yang X., 2013, MNRAS, 340, 767

\bibitem[{{Chuang et al.}(2013)}]{Chuang13}
  Chuang et al., 2013, arXiv:1312.4889

\bibitem[{{Clifton et al.}(2012)}]{Clifton12}
  Clifton T., Ferreira P., Padilla A., Skordis C., 2012, Physics Reports, 513, 1

\bibitem[{{Cohn \& White}(2013)}]{Cohn13}
  Cohn J., White M., 2013, arXiv:1311.0850

\bibitem[{{Conroy, Wechsler, \& Kravtsov}(2006)}]{Conroy06}
  Conroy C., Wechsler R., Kravtsov A., 2006, ApJ, 647, 201

\bibitem[{{Cooray \& Sheth}(2002)}]{CooraySheth02}
  Cooray A., Sheth R., 2002 Physics Reports, 372, 1

\bibitem[{{Davis et al.}(1985)}]{Davis85}
  Davis M., Efstathiou G., Frenk C., White S., 1985, ApJ, 292, 371

\bibitem[{{Dawson et al.}(2012)}]{Daw13}
  Dawson K., et al., 2013, AJ, 145, 10

\bibitem[{{de la Torre et al.}(2013)}]{delaTorre13}
  de al Torre S., 2013, A\&A, 557, 54

\bibitem[{{Eifler, Kilbinger, \& Schneider}(2008)}]{Eifler08}
  Eifler T., Kilbinger M., Schneider P., 2008, A\&A, 482, 9

\bibitem[{{Eisenstein et al.}(2011)}]{Eis11}
  Eisenstein D.J., et al., 2011, AJ, 142, 72

\bibitem[{{Fisher}(1995)}]{Fisher95}
  Fisher K.B., 1995, ApJ, 448, 494

\bibitem[{{Gao et al.}(2005)}]{Gao05}
  Gao L., Springel V., White S., 2005, MNRAS, 363, 66

\bibitem[{{Gunn et al.}(1998)}]{Gunn98}
  Gunn J.E., et al., 1998, AJ, 116, 3040

\bibitem[{{Gunn et al.}(2006)}]{Gunn06}
  Gunn J.E., et al., 2006, AJ, 131, 2332

\bibitem[{{Guo et al.}(2014)}]{Guo14}
  Guo H., et al., 2014, arXiv:1401.3009

\bibitem[{{Guo, Zehavi, \& Zheng}(2012)}]{Guo12}
  Guo H., Zehavi I., Zheng Z., 2012, ApJ, 756, 127

\bibitem[{{Hartlap, Simon \& Schneider}(2007)}]{Hartlap07}
  Hartlap J., Simon P., Schneider P., 2007, A\&A,464,399

\bibitem[{{Hawkins et al.}(2003)}]{Hawkins03}
  Hawkins E., et al., 2003, MNRAS, 346, 78

\bibitem[{{Hikage}(2014)}]{Hikage14}
  Hikage C., 2014, arXiv:1401.1246

\bibitem[{{Kaiser}(1987)}]{Kaiser87}
  Kaiser N., 1987, MNRAS, 227, 1

\bibitem[{{Keisler}(2013)}]{Keisler13}
  Keisler R., Schmidt F., 2013, ApJ, 765, 32

\bibitem[{{Komatsu et al.}(2011)}]{WMAP7}
  Komatsu E., et al., 2011, ApJS, 192, 18

\bibitem[{{Jackson}(1972)}]{Jackson72}
  Jackson J., 1972, MNRAS, 156, 1P

\bibitem[{{Krause et al.}(2013)}]{Krause13}
  Krause E., 2013, MNRAS, 428, 2548

\bibitem[{{Lam et al.}(2012)}]{Lam12}
  Lam T., Nishimichi T., Schmidt F., Takada M., 2012, PhRvL, 109, 1301

\bibitem[{{Lam et al.}(2013)}]{Lam13}
  Lam T., Schmidt F., Nishimichi T., Takada M., 2013, PhRvD, 88, 023012

\bibitem[{{Landy \& Szalay}(1993)}]{LanSza93}
  Landy S.D., Szalay A.S., 1993, ApJ 412, 64

\bibitem[{{Leauthaud et al.}(prep)}]{Leauthaud0X}
  Leauthaud A., et al., in prep.

\bibitem[{{Limber}(1954)}]{Limber54}
  Limber D., 1954, ApJ, 119, 655

\bibitem[{{Li et al.}(2006)}]{Li06}
  Li C., Kauffmann G., Jing Y., White S., B\''{o}rner G., Cheng F., 2006, MNRAS, 368, 21

\bibitem[{Mandelbaum et al.}(2013)]{Mandelbaum13}
  Mandelbaum R., Slosar A., Baldauf T., Seljak U., Hirata C., Nakajima R., Reyes R., Smith R., 2013, MNRAS, 432, 1544

\bibitem[{{Marin}(2011)}]{Marin11}
  Marin F., et al., 2011, ApJ, 737, 97

\bibitem[{{Marinoni \& Hudson}(2002)}]{Marinoni02}
  Marinoni C., Hudson M., 2002, ApJ, 569, 101

\bibitem[{{Masters et al.}(2011)}]{Masters11}
  Masters K. et al., 2011, MNRAS, 418, 1055

\bibitem[{{Neistein \& Khochfar}(2012)}]{Neistein12}
  Neistein E., Khochfar S., 2012, arXiv:1209.0463

\bibitem[{{Norberg et al.}(2009)}]{Norberg09}
  Norberg P., Baugh C., Gazta\~{n}aga E., Croton D., 2009, MNRAS, 396, 19

\bibitem[{{Nuza et al.}(2013)}]{Nuza13}
  Nuza et al., 2013, MNRAS, 432, 743

\bibitem[{{Parejko et al.}(2013)}]{Parejko11}
  Parejko J., et al., 2013, MNRAS, 429, 98

\bibitem[{Peacock \& Smith}(2000)]{Peacock00}
  Peacock J., Smith R., 2000, MNRAS, 318, 1144

\bibitem[{{Piloyan et al.}(2014)}]{Piloyan14}
  Piloyan A., Marra V., Baldi M., Amendola L., arXiv:1401.2656

\bibitem[{{Reid et al.}(2012)}]{Reid12}
  Reid B., et al., 2012, MNRAS, 426, 2719

\bibitem[{{Reid \& Spergel}(2009)}]{Reid09}
  Reid B., Spergel D., 2009, ApJ, 702, 249

\bibitem[{{Reid \& White}(2011)}]{ReiWhi11}
  Reid B., White M., 2011, MNRAS,  417, 1913

\bibitem[{{Reyes}(2010)}]{Reyes10}
  Reyes R., 2010, Nature, 464, 256

\bibitem[{{Ross et al.}(2012)}]{Ross12}
  Ross A. et al., 2012, MNRAS, 424, 564

\bibitem[{{Ross et al.}(2014)}]{Ross14}
  Ross A. et al., 2014, MNRAS, 437, 1109

\bibitem[{Ross et al.}(2007)]{Ross07}
  Ross N., et al., 2007, MNRAS, 381, 573

\bibitem[{{Samushia et al.}(2013)}]{Samushia13}
  Samushia L., et al., 2013, arXiv:1312.4899

\bibitem[{{Sanchez et al.}(2013)}]{Sanchez13}
  Sanchez A., et al., 2013, arXiv:1312.4854

\bibitem[{{Schmidt, Vikhlinin, \& Hu}(2009)}]{Schmidt09}
  Schmidt F., Vikhlinin A., Hu W., 2009, PhRvD, 80, 083505

\bibitem[{{Sheth}(2005)}]{Sheth05}
  Sheth R., 2005, MNRAS, 364, 796

\bibitem[{Seljak}(2000)]{Seljak00}
  Seljak U., 2000 MNRAS, 318, 203

\bibitem[{{Seo, Eisenstein, \& Zehavi}(2008)}]{Seo08}
  Seo H.~J., Eisenstein D., Zehavi I., 2008, ApJ, 681, 998

\bibitem[{{Tinker}(2007)}]{Tinker07}
  Tinker J., 2007, MNRAS 374, 477

\bibitem[{{Tinker et al.}(2011)}]{Tinker11}
  Tinker J. et al., 2012, ApJ, 745, 16

\bibitem[{{Tinker et al.}(2008)}]{TinkerSO}
  Tinker J.,  Kravtsov A., Klypin A., Abazajian K., Warren M., Yepes G., Gottl\''{o}ber S., Holz D., 2008, ApJ, 688, 709

\bibitem[{{Tinker et al.}(2010)}]{Tinker10}
  Tinker J., Robertson B., Kravtsov A., Klypin A., Warren M., Yepes G., Gottl\''{o}ber S., 2010, ApJ, 724, 878

\bibitem[{{White et al.}(2011)}]{White11}
  White M., et al., 2011, ApJ, 728, 126

\bibitem[{{White Cohn, \& Smit}(2010)}]{White10}
  White M., Cohn J., Smit R., 2010, MNRAS, 408, 1818

\bibitem[{{White, Hernquist, \& Springel}(2001)}]{White01}
  White M., Hernquist L., Springel V., 2001, ApJ, 550, 129

\bibitem[{{White \& Padmanabhan}(2009)}]{White09}
  White M., Padmanabhan N., 2009, MNRAS, 395, 2381

\bibitem[{{Vale \& Ostriker}(2006)}]{Vale06}
  Vale A., Ostriker J., 2006, MNRAS, 371, 1173

\bibitem[{{van den Bosch et al.}(2013)}]{vdB13}
  van den Bosch F., More S., Cacciato M., Mo H., Yang X., 2013, MNRAS, 430, 725

\bibitem[{{van den Bosch et al.}(2005)}]{vdB05}
  van den Bosch F., Weinmann S., Yang X., Mo H., Li C., Jing Y., 2005, MNRAS, 361, 1203

\bibitem[{{Wang et al.}(2013)}]{Wang13}
  Wang L., Weinmann S., De Lucia G., Yang X., 2013, MNRAS, 433, 515

\bibitem[{{Wu et al.}(2013)}]{Wu2013}
  Wu H., Hahn O., Evrard A., Wechsler R., Dolag K., 2013, MNRAS, 436, 460

\bibitem[{{Yang et al.}(2004)}]{Yang04}
  Yang X., Mo H., Jing Y., van den Bosch F., 2004, MNRAS, 350, 1153

\bibitem[{{Yang, Mo, \& van den Bosch}(2003)}]{Yang03}
  Yang X., Mo H., van den Bosch F., MNRAS, 339, 1057

\bibitem[{{Yang, Mo, \& van den Bosch}(2009)}]{Yang09}
  Yang X., Mo H., van den Bosch F., 2009, ApJ, 695, 900

\bibitem[{{York et al.}(2000)}]{Yor00}
  York D.G., et al., 2000, AJ, 120, 1579

\bibitem[{{Zehavi et al.}(2011)}]{Zehavi11}
  Zehavi I. et al., 2011, ApJ, 736, 59

\bibitem[{{Zentner, Hearin, van den Bosch}(2013)}]{Zentner13}
  Zentner A., Hearin A., van den Bosch F., 2013, arXiv:1311.1818

\bibitem[{{Zheng et al.}(2005)}]{Zheng05}
  Zheng Z., et al., 2005, ApJ, 633, 791

\bibitem[{{Zu et al.}(2013)}]{Zu13}
  Zu Y., Weinberg D., Jennings E., Baojiu L., Wyman M., 2013, arXiv:1310.6768

\bibitem[{{Zu \& Weinberg}(2012)}]{Zu12}
  Zu Y., Weinberg D., 2013, MNRAS, 431, 3319

\end{thebibliography}
\end{document}